\documentclass[aps,floatfix,superscriptaddress,reprint,10pt,prb]{revtex4-1}
\usepackage{bm}
\usepackage{amsmath}
\usepackage{amssymb}
\usepackage{multirow}
\usepackage{empheq}
\usepackage{graphicx}
\usepackage{mathrsfs}
\usepackage{amsfonts}
\usepackage{amsthm}
\usepackage{color}
\usepackage{bigints}
\usepackage{txfonts}
\usepackage{hyperref}
\hypersetup{
     unicode=false,          % non-Latin characters in Acrobat?s bookmarks
     pdftoolbar=true,        % show Acrobat?s toolbar?
     pdfmenubar=true,        % show Acrobat?s menu?
     pdffitwindow=false,     % window fit to page when opened
     pdfstartview={FitH},    % fits the width of the page to the window
     pdftitle={My title},    % title
     pdfauthor={Author},     % author
     pdfsubject={Subject},   % subject of the document
     pdfcreator={Creator},   % creator of the document
     pdfproducer={Producer}, % producer of the document
     pdfkeywords={keyword1} {key2} {key3}, % list of keywords
     pdfnewwindow=true,      % links in new PDF window
     colorlinks=false,       % false: boxed links; true: colored links
     linkcolor=red,          % color of internal links (change box color with linkbordercolor)
     citecolor=green,        % color of links to bibliography
     filecolor=magenta,      % color of file links
     urlcolor=cyan           % color of external links
}
\makeatletter \tolerance = 10000 \tolerance = 10000
\usepackage{color}

\makeatother
\begin{document}

%\title{The static and dynamic shear and Hall viscosity in noninteracting graphene}
%\title{Viscous Dirac electron fluid}
\title{Viscosity Enhancement by Electron-Hole Collisions in Dirac Electron Fluid}

\author{Weiwei Chen}
\affiliation{School of Science, Westlake University, 18 Shilongshan Road, Hangzhou 310024,  China}
\affiliation{Institute of Natural Sciences, Westlake Institute for Advanced Study, 18 Shilongshan Road, Hangzhou 310024, China}

\author{W. Zhu}
%\thanks{zhuwei@westlake.edu.cn}
%\{zhuwei@westlake.edu.cn}
\affiliation{School of Science, Westlake University, 18 Shilongshan Road, Hangzhou 310024, China}
\affiliation{Institute of Natural Sciences, Westlake Institute for Advanced Study, 18 Shilongshan Road, Hangzhou 310024,  China}

\date{\today}
\begin{abstract}
Rejuvenation of hydrodynamic transport in solids provides a new window to study collective motion of electrons, where electrons behave like a viscous fluid akin to classical liquids. Experimental observations of such exotic states have not been realized until recent years, and an on-going quest is to amplify the hydrodynamic effect in electron fluids. Here we investigate the hydrodynamic properties of Dirac electron fluid in graphene from a microscopic viewpoint, and elucidate a novel way to enhance  electron hydrodynamics. In particular, we present strong evidences that the shear viscosity of Dirac electrons can be enhanced by frequent electron-hole collisions, through three distinct aspects: promoting electrons and holes around the Dirac point by disorder, creating electron-hole shared zeroth Landau level by external magnetic field, and inducing electron-hole excitations by dynamic deformation. We also study Hall viscosity, which is closely related to the geometric topology and exhibits quantum behavior analogous to Hall conductivity. Therefore, our work demonstrates the exotic landscape of  hydrodynamic electronics in graphene, and presents experimentally relevant responses to quantify the effects of electronic viscosity.

%So far the discussion on the fate of hydrodynamic electronics in the presence of disorder scattering is largely phenomenological, and conventional wisdom always expects that the influence of disorder suppresses collective motions by introducing a momentum-relaxing collisions. Here we investigate the hydrodynamic properties of Dirac electron fluid in graphene from a microscopic viewpoint. We present strong evidences that, due to frequent electron-hole collisions, the shear viscosity of Dirac electrons can be enhanced by electron-disorder scattering, which is in sharp contrast to the common sense of traditional Fermi liquids. Moreover, we clarify that the anomalous behavior of Dirac electron fluid can be also revealed by the dynamic and magneto-hydrodynamics. Therefore, our work demonstrates the exotic landscape of  hydrodynamic electronics in graphene, and presents experimentally relevant responses to quantify the effects of electronic viscosity. 
\end{abstract}
\maketitle

\section{Introduction}
\label{sec1}

Hydrodynamic behavior of electrons in solids has been predicted for decades \cite{Gurzhi1963}, but only in the recent years has it become a reality in two-dimensional materials, where many signatures of viscous electron flows have been probed, such as negative local resistance \cite{Bandurin2016,Bandurin2018}, superballistic flow \cite{Guo2017,Kumar2017,Keser2021}, current vortices \cite{Levitov2016,Mayzel2019,Falkovich2017}, Poiseuille flow \cite{Sulpizio2019,Moll2016}, Hall viscosity \cite{Berdyugin2019} and violation of the Wiedemann-Franz law \cite{Gooth2018,Crossno2016}. Generally speaking, to maximize hydrodynamic effects in experiment, so far there are two main roads. One is to seek a material in which momentum relaxing effects (e.g. electron-phonon scattering) are greatly suppressed. 
%{\color{red}so that the collective behavior dominated by electron collisions tends to be pronounced and the Navier-Stokes equation departures from a Drude-like form}. 
In this respect, graphene becomes a promising candidate to observe hydrodynamic phenomena, due to its extremely stiffness. The other way is, to create inhomogeneous electron flows moving through artificial constriction geometry such as a narrow slit \cite{Bandurin2016}, where hydrodynamic behavior is generally expected. Beyond the above two ways,  an ongoing quest is to find more mechanism that can amplify the hydrodynamic effect in electron fluids, which is not only conceptually important but also practically relevant for experiment.

Most of existing studies on hydrodynamic effect focus on the electron-electron collisions \cite{Briskot2015,Muller2009,Narozhny2019}, and theoretical modeling of viscous electron flow is usually based on the simulation of Navier-Stokes equations \cite{Torre2015,Pellegrino2016prb,Narozhny2019}, where the prevalence of imperfection like disorder is usually overlooked. One plausible reason is that disorder is always expected to disrupt collective motions by introducing a momentum-relaxing collisions \cite{Lucas2015,Lucas2018}.
To date a systematic study of disorder effect on hydrodynamic phenomena in experimentally relevant two-dimensional materials (e.g. graphene) is still lacking.
%{\color{red} However, the hydrodynamic electron system can only reach the steady state by means of disorder scattering \cite{Narozhny2019,Briskot2015,Narozhny2015}.}
On the other hand, it has been known that disorder profoundly impacts carrier transport in graphene. One novel example is the observation of minimum conductivity in graphene \cite{Novoselov2005}, originating from impurity assisted resonant tunneling of massless Dirac fermions, where the impurity induces significant potential barriers and the systems are likely to split into a random distribution of p-n junctions. \cite{Katsnelson2006,Titov2007,Nomura2007,Adam2007}.
Hence, with this fact on hand, it is natural to ask to what extent hydrodynamic phenomena could be promoted or symbiotic coexistence in disordered Dirac electron fluids. We study the shear viscosity ($\eta_s$) in disordered graphene using analytic and numerical methods. Counterintuitively, we discover an enhancement of viscosity for Dirac electron fluid in the regime of low carrier concentrations, which endows the graphene system with interesting, yet mostly unexplored, static and dynmaic hydrodynamic behavior.
%the fate of hydrodynamic phenomena of disordered electron fluids, and ask under which conditions signatures of hydrodynamics survive.

Inspired from this peculiar behavior, we further investigate the viscosity in graphene in the presence of a perpendicular magnetic field, since the zeroth Landau level around the neutral point is shared by the electrons and holes. As expected, the shear viscosity around the Dirac point can be further enhanced by an external magnetic field. These findings are in stark contrast to those found in the conventional two-dimensional ordinary electron gas (2DEG)~\cite{Burmistrov2019}, suggesting the importance of the inherent electron-hole coherence around the Dirac point of graphene. Away from charge neutrality, the viscosity coefficients tend to agree with those of Fermi liquids. (For a detailed comparison please see Tab. \ref{tab1})

%In this paper, we investigate the disorder effect on Dirac electron liquid in graphene.
%Counterintuitively, we discover an enhancement of viscosity for Dirac electron fluid, which endows the graphene system with interesting, yet mostly unexplored, static and dynmaic hydrodynamic behavior.

Going even further, we consider the dynamic situation. Dynamic deformation is introduced, which will result in transitions between energy levels with an interval of dynamic frequency. Thus, when the Fermi energy is lower than the dynamic frequency, the interband (electron-hole) transitions are induced. The results in this regime show that shear viscosity is positively related to frequency, in other words, shear viscosity increases with more electron-hole collisions.

\begin{table*}
	\normalsize
	\caption{\label{tab1}
		Analytic results of static shear ($\eta_s$) and Hall ($\eta_H$) viscosity of disordered graphene in the absence of magnetic field $B=0$ and presence of perpendicular magnetic field $\bm{B}=B\hat{z}$. $E$ is the Fermi energy, $\rho$ is the density of states, $\tau$ is the quasiparticle relaxation time, $N$ is index of Landau level, $A$ is disorder scattering parameter, and $\varepsilon$ is the distance between the Fermi energy and the nearest Landau level center. In 2DEG, $\tilde{\omega}_c=\frac{eB}{m}$ is the cyclotron frequency. In graphene, $\tilde{\omega}_c=\frac{eB}{|E|/v_f^2}$ is effective cyclotron frequency. They are equivalent by introducing effective mass $m=|E|/v_f^2$ in graphene. ``Overlapped" and ``Well Separated" are the regions divided by $\tilde{\omega}_c\tau\lesssim1$ and $\tilde{\omega}_c\tau\gg1$.}
	\begin{ruledtabular}
		\begin{tabular}{cccccc}
			& $\eta_s$ in the case $B=0$&\multicolumn{2}{c}{$\eta_s$ in the case $\bm{B}=B\hat{z}$}&\multicolumn{2}{c}{$\eta_H$ in the case of $\bm{B}=B\hat{z}$} \\
			%		\hline
			& & Overlapped &  Well Separated&Overlapped &  Well Separated\\
			\hline
			\multirow{2}*{2DEG} 
			&\multirow{2}*{$\frac{1}{2}E^2\rho\tau$\cite{Pellegrino2017,Burmistrov2019}} 
			&\multirow{2}*{$\frac{1}{2}\frac{E^2\rho\tau}{1+4(\tilde{\omega}_c\tau)^2}$\cite{Burmistrov2019,Alekseev2016}}  
			&\multirow{2}*{$\frac{N^2\hbar}{8\pi^2l_B^2}(1-2A\varepsilon^2)$\cite{Burmistrov2019}} 
			&\multirow{2}*{$\frac{\rho \tilde{\omega}_c\tau^2E^2}{1+4\tilde{\omega}_c^2\tau^2}$\cite{Pellegrino2017,Burmistrov2019,Alekseev2016}}
			&\multirow{2}*{$\frac{N^2\hbar}{8\pi l_B^2}-\frac{1}{4\tilde{\omega}_c}\frac{\rho E^2}{1+4\tilde{\omega}_c^2\tau^2}$\cite{Burmistrov2019}}\\
			&&&&&\\
			\multirow{2}*{Graphene} 
			&\multirow{2}*{$\frac{1}{8}E^2\rho\tau+\frac{3}{32}\frac{\hbar^2\rho}{\tau}$} 
			&\multirow{2}*{$\frac{1}{8}\frac{E^2\rho\tau}{1+4\tilde{\omega}_c^2\tau^2}+\frac{\hbar^2}{32\tau}\rho\frac{3+16\tilde{\omega}_c^2\tau^2}{1+4\tilde{\omega}_c^2\tau^2}$}
			&\multirow{2}*{$\frac{(N^2+\delta_{N,0})\hbar}{2\pi^2l_B^2}(1-2A\varepsilon^2)$} 
			&\multirow{2}*{$\frac{1}{4}\frac{\rho\tilde{\omega}_c\tau^2E^2}{1+4\tilde{\omega}_c^2\tau^2}$}
			&\multirow{2}*{$\frac{{\rm sgn}(E)(2N^2+2N+1)\hbar}{4\pi l^2_B}-\frac{1}{16\tilde{\omega}_c}\frac{\rho E^2}{1+4\tilde{\omega}_c^2\tau^2}$}\\
			&&&&
		\end{tabular}
	\end{ruledtabular}
\end{table*}

To fully understand the hydrodynamic properties of Dirac electronic fluids, we also study the Hall viscosity ($\eta_H$), which is closely related to the geometric topology and exhibits quantum behavior analogous to Hall conductivity, in both static and dynamic situations, with or without an external magnetic field. In the presence of magnetic field, assuming good separation of Landau levels, we find the contributions from the interband transition counterparts $E_{n,-}\to E_{n+2,+}$ and $E_{n,+}\to E_{n+2,-}$, which are same to dynamic shear viscosity, cancel each other in dynamic Hall viscosity. This implies the electron-hole collision effect is different for shear and Hall viscosity.

All these findings are highly relevant to the on-going experiments on graphene, calling for detailed studies of hydrodynamic electronics of Dirac fluid.

\section{Results}
For systematicness, in the following, we present our results with the order: A. static shear viscosity with and without magnetic field; B. static Hall viscosity; C. dynamic shear viscosity with and without magnetic field; D. dynamic Hall viscosity.
\subsection{Static Shear Viscosity $\eta_S$}
\label{sec-dc-shear}

We start our discussion from evaluating the static shear viscosity $\eta_S$ in graphene using the linear response theory (see Appendix I): 
%Based on the relations: ${\rm Re}[{\rm Tr}(T_{xy}G^A_{\omega+\Omega}T_{xy}G^A_{\omega})]={\rm Re}[{\rm Tr}(T_{xy}G^R_{\omega+\Omega}T_{xy}G^R_{\omega})]$ and $\lim\limits_{\Omega\to0}\frac{f_{\omega+\Omega}-f_{\omega}}{\Omega}=-\delta(\omega-E)$, the Kubo formula of $\eta_S$ is simplified as follows
\begin{equation}\label{eq:kubo-xx-b0-1}
	\eta_s(E)={\rm Re}\left[\eta_s^{RA}(E)-\eta_s^{RR}(E)\right]
\end{equation}
with
\begin{equation}\label{eq:kubo-xx-b0-2}
	\eta_s^{LM}(E)=4\frac{\hbar}{2\pi\mathcal{V}}{\rm Tr}\left[G^L(E)T_{xy}G^M(E)T_{xy}\right]
\end{equation}
where $G^{L,M}$ denotes the retarded and advanced Green's function, the stress tensor components are chosen as $T_{ij}=T_{kl}=T_{xy}$, and the factor 4 denotes the degeneracy of spin and valley. In the Boltzmann transport theory, the contribution from the retarded-retarded (RR) channel is usually discarded. %In the studies of quantum conductivity of graphene, however, this term becomes as much relevant as the RA one in proximity to Dirac point \cite{Chen2021}. 
In this work, we find $\eta_s^{RR}$  becomes as much relevant as the $\eta_s^{RA}$ one in proximity to Dirac point.

In the following, we will calculate and analyze the shear viscosity in the two cases: (1) in the absence of magnetic field, $\bm{B}=0$; (2) in the presence of a magnetic field perpendicular to the $xy$-plane, $\bm{B}=B\hat{z}$. The vertex corrections of the Kubo formula for viscosity in both cases $\bm{B}=0$ and $\bm{B}=B\hat{z}$ are proved to be zero due to short-range disorder scattering in Appendix  IV. A plausible explanation is given by analogizing the vertex correction in Kubo formula to the transport relaxation time correction in Boltzmann transport theory, which is a scheme to distinguish the contributions of forward scattering and back scattering. The Kubo formula for viscosity is in nature a stress-stress correlation function and the stress is an outer product of velocity and momentum. Therefore, the effect of short-range disorder consisting of uniformly distributed plane waves in the momentum space on viscosity isotropic due to a nontrivial compensation between the velocity direction and momentum direction.

%of the conductivity.  for conductivity and a stress-stress correlation function for viscosity. The vertex correction of conductivity in Kubo formula, approximately equivalent to the transport relaxation time correction in Boltzmann transport theory,  In the case of viscosity, however, the stress is a outer product of velocity and momentum so that there is no deviation from anisotropy when the short range disorder is consisted of uniformly distributed plane waves in momentum space.

\begin{figure}
	%\centering
	\flushleft
	\includegraphics[width=0.45\textwidth]{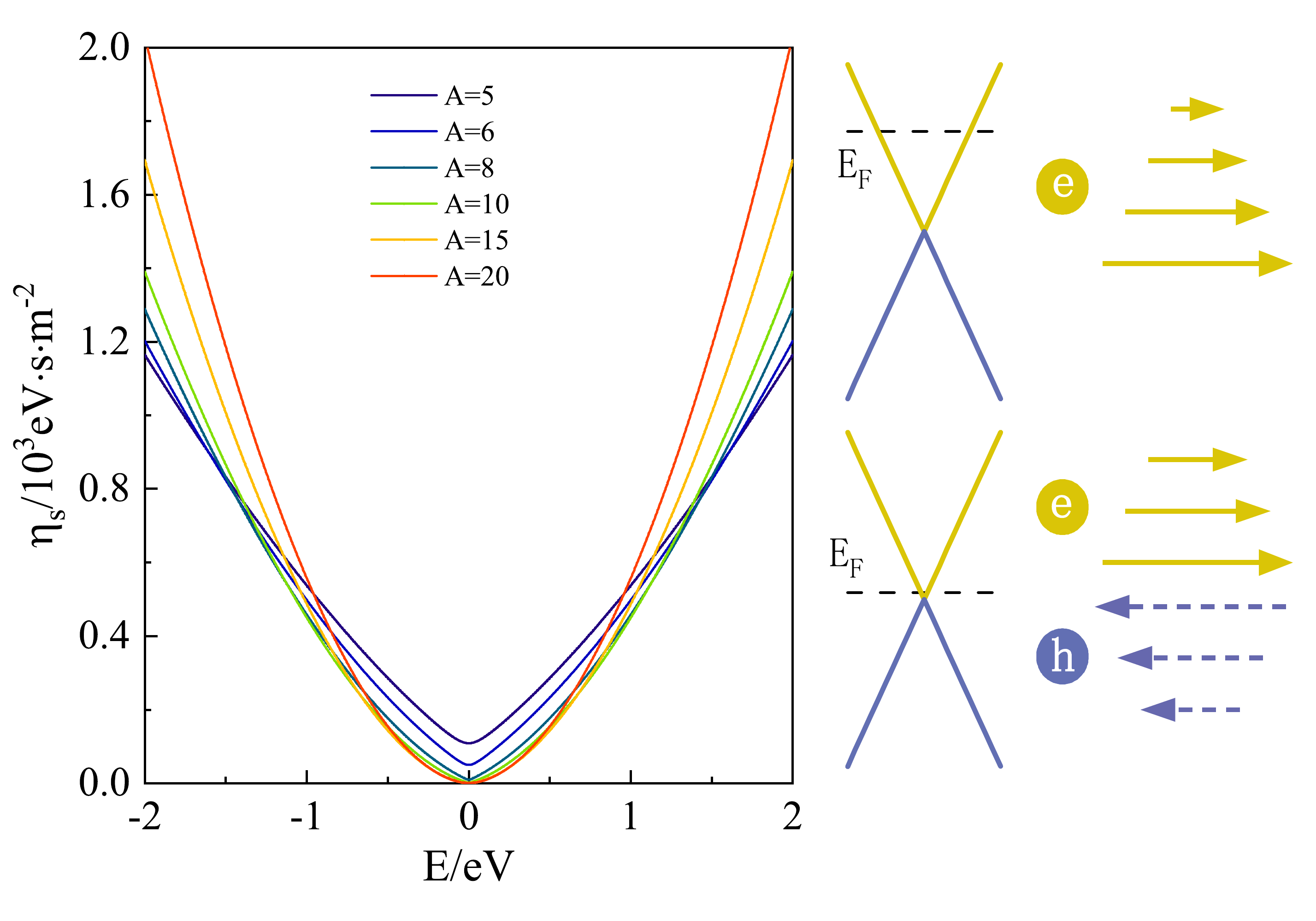}
	\caption{(color online) (Left) Numerical result for the shear viscosity $\eta_s$ as a function of Fermi energy in the absence of magnetic field. $A= \frac{4\pi (\hbar v_f)^2}{n_i V_0^2} $ is a dimensionless parameter characterizing the scattering strength (see main text). 
	(Right) Cartoon picture for hydrodynamic flow for doped and undoped graphene. }
	\label{fig:dcshearb0}
\end{figure}

\subsubsection{$B=0$}
In the absence of magnetic field, the $\eta^{LM}_s$ in Eq.(\ref{eq:kubo-xx-b0-2}) can be analytically derived (see Appendix V).
%can be expanded as
%\begin{equation}\label{eq:kubo-xx-b0-3}
%	\eta_s^{LM}(E)=\frac{\hbar^3v_f^2}{8\pi^2}\int dkk^3(g_+^L+g_-^L)(g_+^M+g_-^M)
%\end{equation}
%where $g_{\pm}^{R/A}(E)=(E\mp\hbar v_fk-\Sigma^{R/A})^{-1}$ is the full Green's function  with self-energy $\Sigma^{R/A}$ approximately obtained from the SCBA equation (see Appendix III). The detail of analytic derivation of $\eta_s^{LM}$ is performed in Appendix V. 
$\eta_s^{RA}$ finally arrives at $
%\begin{equation}
	\eta_s^{RA}(E)=\frac{1}{8}E^2\rho\tau+\frac{\hbar^2}{32\tau}\rho
%\end{equation}
$, where $\tau=\frac{\hbar}{2|{\rm Im}\Sigma(E)|}$ is the quasiparticle relaxation time. The real part of $\eta_s^{RR}$ approaches to $
%\begin{equation}
	{\rm Re}\eta_s^{RR}(E)=-\frac{\hbar^2}{16\tau}\rho
%\end{equation}
$, which is as much relevant as $\eta_s^{RA}$ in the limit $E\to0$. The total real part of shear viscosity, therefore, is obtained as
\begin{equation}\label{eq:shear-b0-total}
	\eta_s(E)=\frac{1}{8}E^2\rho\tau+\frac{3}{32}\frac{\hbar^2\rho}{\tau}
\end{equation}
If we plug the expressions of self-energy  and density of states (see Appendix III) into the above result, it can be rewritten as
\begin{equation}\label{eq:shear-b0-total-2}
	\eta_s(E)=\frac{\hbar}{8\pi^2(\hbar v_f)^2}\left[AE^2+\frac{3}{A}(\pi|E|+E_cAe^{-A/2})^2\right]
\end{equation}

The first term in Eq.~(\ref{eq:shear-b0-total}) and Eq.~(\ref{eq:shear-b0-total-2}) is in accordance with that in 2DEG, and can also be written as $\frac{1}{4}m\bar{n}v_f l$, where $m$ is the effective mass, $\bar{n}=\int\rho dE\approx\frac{1}{2}\rho E$ is the average charge density and $l=v_f\tau$ is mean free path. The second term, which does not exist in 2DEG, is inversely proportional to the quasiparticle relaxation time and implies the viscosity will be enhanced by disorder potential near the Dirac point. This is inconsistent with ordinary metals and classical fluids, where impurities can pack to eliminate any sign of collective behavior of particles. Similar to the minimal conductivity of graphene due to impurity assisted resonant tunneling through the electron-hole coherence and retarded-retarded ($RR$) channel \cite{Chen2021}, this anomalous term in static shear viscosity also mainly comes from the electron-hole coherence and $RR$ term. However, in distinction to the charge current, the momentum current also contains single-particle contributions. Thus, the disorder-assisted transport is significantly larger in viscosity compared to conductivity.

In order to clearly see the effect of the anomalous term in $\eta_s$, we also performed the numerical calculation of the Eq.~(\ref{eq:kubo-xx-b0-1}). The results for several different disorder strengths are illustrated in Fig.~\ref{fig:dcshearb0}. The relationship between $\eta_s$ and disorder scattering can be divided into two regions with opposite behaviors. In the high energy region that we call the ``normal region", $\eta_s$ decays rapidly with the increasing of disorder scattering. This is consistent with the 2DEG and classical fluids. In the low energy region roughly $[-1.1eV,1.1eV]$, which we call ``anomalous region", $\eta_s$ has a peculiar enhancement induced by disorder. It is well agree with the analytic prediction in Eq.~(\ref{eq:shear-b0-total}). Furthermore, the range of this ``anomalous region" is quite large, not just the interval extremely close to the Dirac point, which is consistent with the hydrodynamic transport region detected in experiments \cite{Kumar2017,Bandurin2016,Bandurin2018}. Intuitively, the different observation in doped and undoped graphene can be understood by the pictures as illustrated in Fig. ~\ref{fig:dcshearb0}: In the updoped case in the vicinity of Dirac point, electrons flow in the direction opposite to the holes, and frequent electron-hole collisions leads to enhancement of hydrodynamic behavior. While in the doped case with a large Fermi surface, only one type of carriers contributes so the hydrodynamic behavior should be similar to the case of normal Fermi liquids.

\begin{figure}
	\centering
	%\flushleft
	\includegraphics[width=0.4\textwidth]{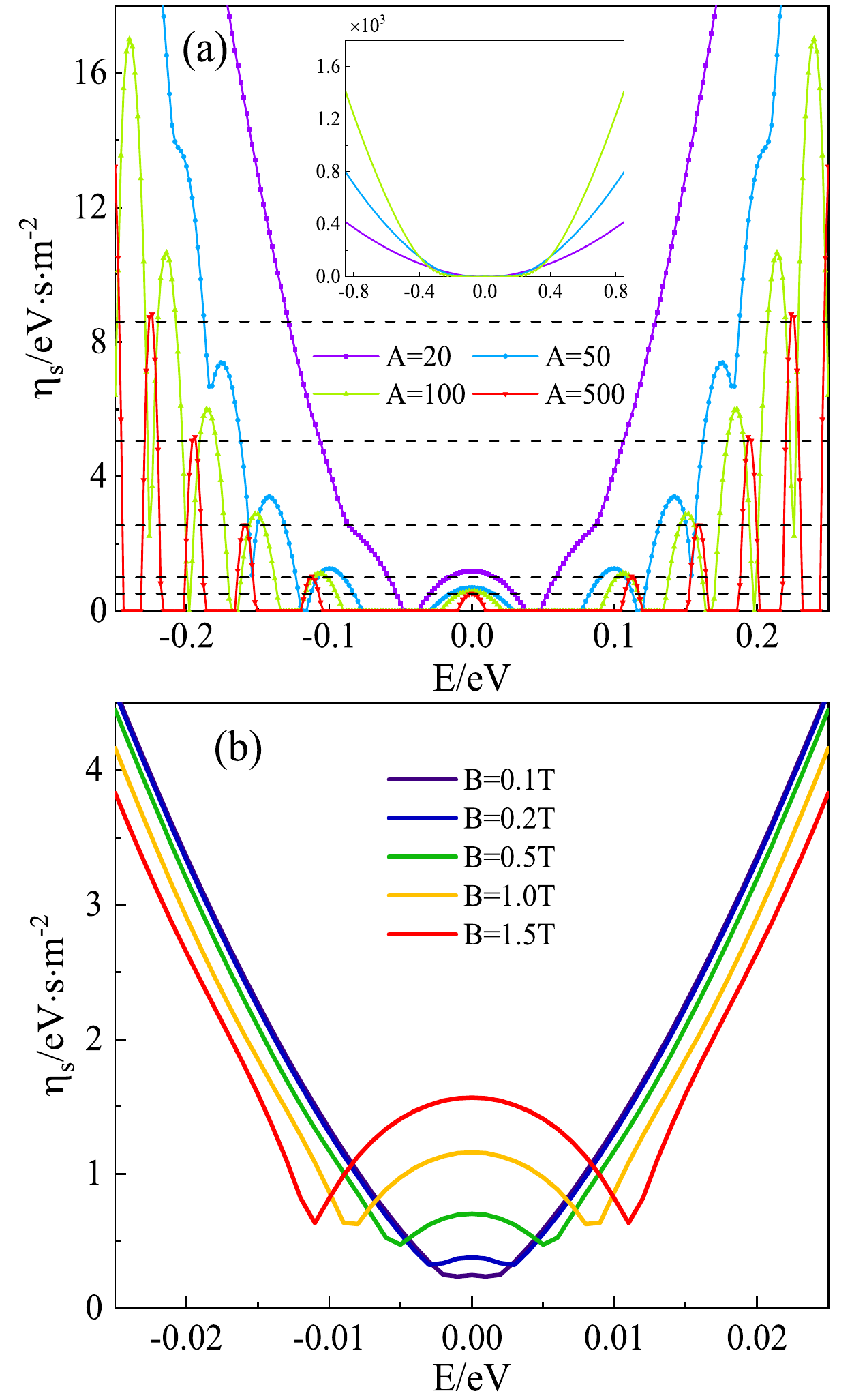}\\
	\caption{(color online) (a)Shear viscosity $\eta_s$ as a function of Fermi energy, fixed magnetic field strength $B=10T$ and different disorder scattering $A= \frac{4\pi (\hbar v_f)^2}{n_i V_0^2}=20$, $50$, $100$, $500$. Dashed lines in top pane denote the predicted values of static shear viscosity of each Landau level in the absence of disorder: $\eta_s=(N^2+\delta_{N,0})\frac{\hbar}{2\pi^2l_B^2}$. The inset is the same plot with a larger energy scope. (b) Static $\eta_s$ vs $E$ for fixed $A=15$ and $B=$0.1T, 0.2T, 0.5T, 1.0T, 1.5T. }
	\label{fig:dcshearb10}
\end{figure}

\subsubsection{$B\neq0$}
In the presence of a magnetic field perpendicular to the graphene, 
we find the anomalous response of shear viscosity remains.
%we get
%\begin{equation}\begin{aligned}\label{eq:kubo-xx-bneq0-1}
%		\eta_s^{LM}(E)=&\frac{(\hbar\omega_c)^2}{4\pi^2l_B^2}\sum_{n}(n+1){\rm Re}(g^L_{n}g^M_{n+2}+g^M_{n}g^L_{n+2})
%\end{aligned}\end{equation}
%where 
%$%\begin{equation}\begin{aligned}
%		g^{R/A}_n=\frac{1}{2}(G^{R/A}_{n,+}+G^{R/A}_{n,-})=\frac{E-\Sigma^{R/A}}{(E-\Sigma^{R/A})^2-n(\hbar\omega_c)^2}.
%$%\end{aligned}\end{equation} 
%is full Green's function under external magnetic field.
Here the discussion of the analytic expression of $\eta_s$ for $B\neq0$ is separated into two parts. 
 When the Landau levels are well separated, i.e. $\tilde{\omega}_c\tau\gg1$, we get
\begin{equation}
	\eta_s(E)=(N^2+\delta_{N,0})\frac{\hbar}{2\pi^2l_B^2}(1-2A\varepsilon^2)
\end{equation}
where $\varepsilon=(E-E_{NS})/2\hbar\omega_c$ characterizes the distance between the Fermi energy and the center of nearest Landau level. The predicted values of static shear viscosity at the center of Landau levels are quantized as $\eta_s=(N^2+\delta_{N,0})\frac{\hbar}{2\pi^2l_B^2}$. When the Landau levels are overlapped ($\tilde{\omega}_c\tau\lesssim1$), $\eta_s$ is approximately given by
\begin{equation}\label{dc-shear-bn0}
	\eta_s=\frac{1}{8}\frac{E^2\rho\tau}{1+4\tilde{\omega}_c^2\tau^2}+\frac{\hbar^2}{32\tau}\rho\frac{3+16\tilde{\omega}_c^2\tau^2}{1+4\tilde{\omega}_c^2\tau^2}
\end{equation}

%where $\tilde{\omega}_c=\frac{\hbar\omega_c^2}{2E}$ is an effective cyclotron frequency, which equals to the cyclotron frequency in 2DEG, $\tilde{\omega}_c=\frac{eB}{mc}$, by substituting the effective mass $m=\frac{2E}{v_f^2}$. This effective mass $m$ is twice the effective mass $m^*$ introduced above through the ratio of Fermi momentum and Fermi velocity. 
The first term in this result is consistent with the result for 2DEG \cite{Burmistrov2019}, except for the cyclotron frequency $\tilde{\omega}_c$, which depends on both magnetic field and Fermi energy in graphene but only depends on magnetic field in 2DEG. Since the well separated and overlapped regions are determined by the value of $\tilde{\omega}_c\tau$, the relation $\tilde{\omega}_c\propto \frac{1}{|E|}$ means that the separated region in graphene is close to the Dirac point. The second term, which can reduce to the anomalous term in Eq.~(\ref{eq:shear-b0-total}) when $\tilde{\omega}_c\to0$, is positively related to both disorder scattering and magnetic field strength according to the self-energy. After plugging the self-energy function  and density of states into the Eq.(\ref{dc-shear-bn0}), the static shear viscosity can be further evaluated by separated the states into $|E|/\hbar\omega_c>1$ and $|E|/\hbar\omega_c\lesssim1$. For the states $|E|/\hbar\omega_c>1$, the static shear viscosity is obtained as
\begin{equation}
	\eta_s(E)=\frac{\hbar}{4\pi^2l_B^2}\frac{AE^2}{(\hbar\omega_c)^2(1+4\alpha^2)}\left[1+\frac{4\alpha^2\delta}{1+4\alpha^2}\cos\frac{\pi E}{\hbar\tilde{\omega}_c}\right]
\end{equation}
where $\alpha=\frac{A}{\pi}\frac{\hbar\tilde{\omega}_c}{E}$ and $\delta=e^{-\frac{4\pi^2E^2}{A(\hbar\omega_c)^2}}$. The second term in the above expression exhibits Shubnikov-de Haas-type oscillations, but the contribution of this term will be suppressed as $E$ increases since the parameters $\alpha$ and $\delta$ both decrease with $E$.

For the states $|E|/\hbar\omega_c\lesssim1$, which are in the vicinity of Dirac point, the static shear viscosity can be approximately given by
\begin{equation}\begin{aligned}
	\eta_s(E)
	%=&\frac{3}{4\pi^2l_B^2}\frac{A}{(\hbar\omega_c)^2}\left[E_ce^{-A/2}+\frac{(\hbar\omega_c)^2}{2E_ce^{-A/2}}\right]^2\\
	=&\frac{3A}{8\pi^2\hbar^2v_f^2}\left[E_ce^{-A/2}+\frac{\hbar v_f^2}{E_ce^{-A/2}}eB\right]^2
\end{aligned}\end{equation}
which is positively related to the strength of magnetic field. 

In Figure~\ref{fig:dcshearb10} we show the numerical results of $\eta_s$ in the presence of magnetic field, which are obtained by numerically solving Kubo formula. As the Fig.~\ref{fig:dcshearb10}(a) shown, the curves of $\eta_s(E)$ with different disorder strength change from separated peaks to Shubnikov-de Haas-type oscillations and then to smooth, which is consistent with the prediction of analytic solutions. From the inset in Fig.~\ref{fig:dcshearb10}(a), one can also find an anomalous disorder-induced shear viscosity enhancement behavior in the low energy region, similar to the case in the absence of a magnetic field. Furthermore, for the states near the Dirac point, the shear viscosity can also be enhanced by magnitude of the applied magnetic field $B$ as shown in Fig.~\ref{fig:dcshearb10}(b). This behavior is contract to the 2DEG, where the static shear viscosity decays with $B$ as a function: $\eta_s\propto\frac{B^2_0}{B_0^2+B^2}$ \cite{Pellegrino2017}.

\begin{figure}
	%\centering
	\flushleft
	\includegraphics[width=0.45\textwidth]{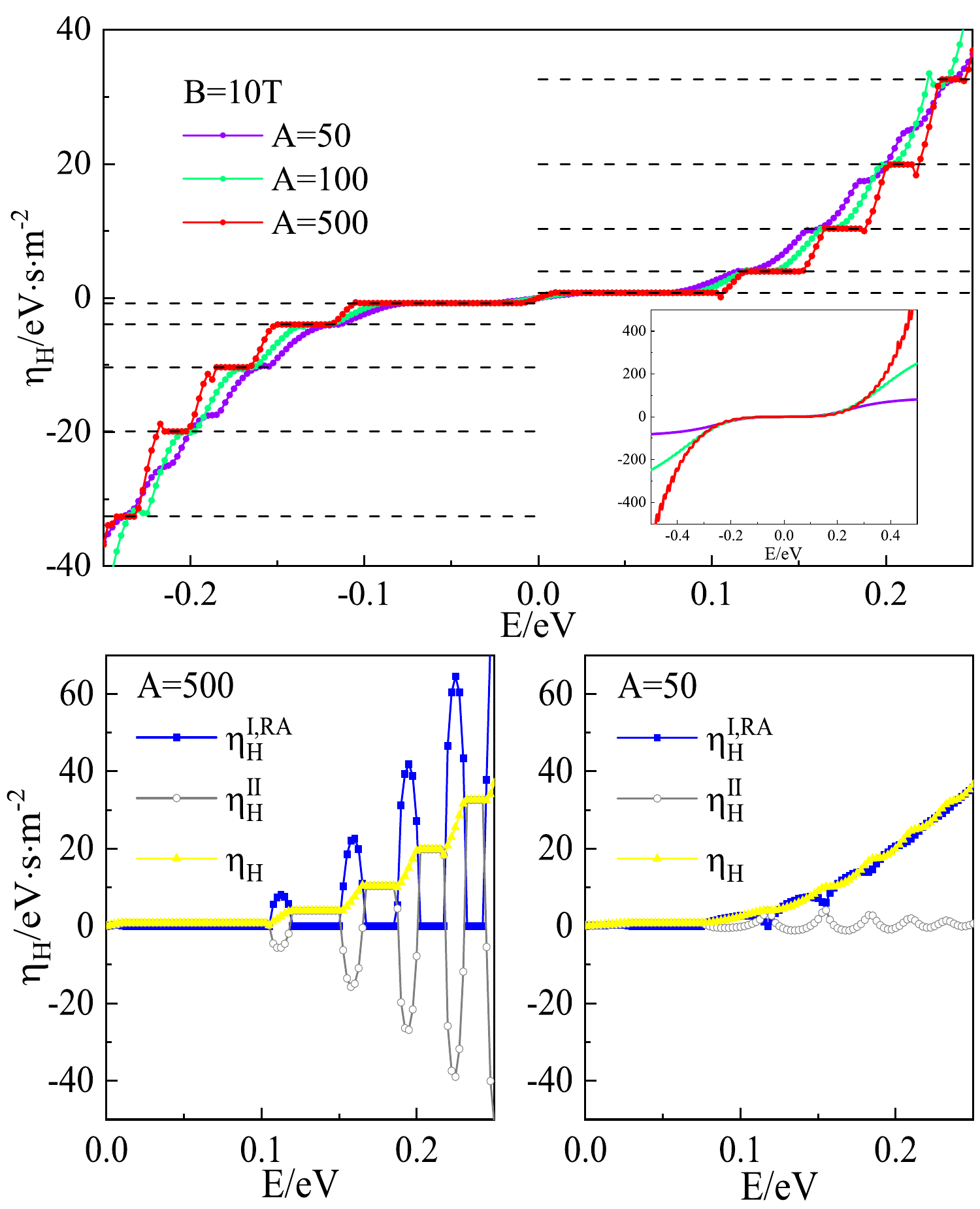}\\
	\caption{(color online) Top: dc Hall viscosity $\eta_H$ as a function of Fermi energy $E$ for fixed magnetic field strength $B=10T$ and different disorder scattering $A= \frac{4\pi (\hbar v_f)^2}{n_i V_0^2} =50$, $100$, $500$. Dashed lines denote the predicted values of Hall viscosity of each Landau level in the absence of disorder: $\eta_H=\frac{{\rm sgn}(E)(2N^2+2N+1)\hbar}{4\pi l^2_B}$. Bottom: different contributions of $\eta_H$ (see main text).}
	%\oim{reduce the height of figures; use open symbols for $\eta_H^{II}$}
	\label{fig:dchallb10}
\end{figure}

\subsection{Static Hall Viscosity $\eta_H$}
\label{sec-dc-Hall}
The Hall viscosity is rooted in the gravitational anomaly, which can  be represented by the strain-induced Berry curvature of wave function and therefore exhibit the striking phenomenon of topology  \cite{Hoyos2012,Sherafati2016,Hu2021}.
Here, we focus on the static Hall viscosity $\eta_H$ in the presence of magnetic field perpendicular to the graphene plane. It can be calculated by the real part of Kubo formula  with $T_{ij}=T_{xy}$, $T_{kl}=\frac{1}{2}(T_{xx}-T_{yy})$: 
%Based on the relations: $-\frac{1}{\Omega}{\rm Re}\left[{\rm Tr}(T_{ij}G^R_{\omega+\Omega}T_{kl}G^R_{\omega}-T_{ij}G^A_{\omega+\Omega}T_{kl}G^A_{\omega})\right]={\rm Re}\left[{\rm Tr}(G^R_{\omega}T_{kl}\frac{G^R_{\omega+\Omega}-G^R_{\omega}}{\Omega}T_{ij}-\frac{G^R_{\omega+\Omega}-G^R_{\omega}}{\Omega}T_{kl}G^R_{\omega+\Omega}T_{ij})\right]$ and $\lim\limits_{\Omega\to0}\frac{G^R_{\omega+\Omega}-G^R_{\omega}}{\Omega}=\frac{dG^R_{\omega}}{d\omega}$, the Kubo formula of $\eta_H$ can be separated into three parts as
\begin{equation}\label{eq:kubo-hall-1}
	\eta_H(E)={\rm Re}\left[\eta_H^{I,RA}(E)-\eta_H^{I,RR}(E)+\eta_H^{II}(E)\right]
\end{equation}
where
\begin{equation}\label{eq:kubo-hall-2}
	\eta_H^{I,LM}(E)=4\frac{\hbar}{4\pi\mathcal{V}}{\rm Tr}\left[G^L(E)(T_{xy}-T_{yy})G^M(E)T_{xy}\right]
\end{equation}
\begin{equation}\begin{aligned}\label{eq:kubo-hall-3}
	\eta_H^{II}(E)=&4\frac{\hbar}{4\pi\mathcal{V}}\int d\omega f_{\omega}{\rm Tr}\left[G^R(\omega)(T_{xy}-T_{yy})\frac{dG^R(\omega)}{d\omega}T_{xy}\right.\\
	&\left.-\frac{dG^R(\omega)}{d\omega}(T_{xy}-T_{yy})G^R(\omega)T_{xy}\right]
\end{aligned}\end{equation}
which is analogous to the Kubo-Streda formula of the Hall conductivity. Compared with $\eta_s$, the Kubo formula of $\eta_H$ has an extra term $\eta^{II}$. $\eta^{II}$ is strikingly different from terms $\eta^{I,LM}$ in that it contains contributions from the entire Fermi sea, while $\eta^{I,LM}$ only from electron states at the Fermi surface. 
Plugging in the form of Green's function, the analytic expression of the total static Hall viscosity $\eta_H$ is
\begin{equation}\label{eq:dc-Hall-total}
	\eta_{H}(E)=\left\{\begin{array}{ll}\frac{{\rm sgn}(E)(2N^2+2N+1)\hbar}{4\pi l^2_B}-\frac{1}{16\tilde{\omega}_c}\frac{\rho E^2}{1+4\tilde{\omega}_c^2\tau^2};&\tilde{\omega}_c\tau\gg1\\
		\frac{1}{4}\frac{\rho\tilde{\omega}_c\tau^2E^2}{1+4\tilde{\omega}_c^2\tau^2};&\tilde{\omega}_c\tau\ll1\end{array}\right.
\end{equation}

From this result, we find the behavior of static Hall viscosity is somewhat similar to the Hall conductivity. It is also quantized in the Landau level gaps where the density of states vanishes, and has opposite signs for the electron states and hole states. Additionally, $\eta_H$ is not integral quantized but has an additional $1/4$, which is analogy to the additional $1/2$ in Hall conductivity of graphene, both of which are the hallmark of the chiral nature of graphene \cite{Sarma2011}. Besides, we find that the quantized value  is mainly contributed from the Fermi sea states, i.e. $\eta_H^{II}$. With the overlap of the Landau levels, the contribution from Fermi sea fades and the contribution from Fermi surface states increases. Therefore, when the Landau level gaps are smoothed by the decrease of $\tilde{\omega}_c\tau$, the quantization behavior of $\eta_H$  disappears.

In Figure~\ref{fig:dchallb10}, we show the numerical results of $\eta_H$ for a fixed magnetic field strength $B=10T$ and different disorder scattering strengths: $A=50$, $100$, and $500$ (top pane), and compare the different contributions from $\eta_H^{I,RA}$ and $\eta_H^{II}$ at $A=50$ and $500$ (bottom pane).  
%In the top pane of Figure~\ref{fig:dchallb10}, we display the numerical results of $\eta_H$ for a fixed magnetic field strength $B=10T$ and different disorder scattering strengths: $A=50$, $100$, and $500$. 
For the extremely small scattering $A=500$ close to pure graphene, it can be clearly seen that $\eta_H$ has a step structure. The height of the step in the Landau level gap is well consistent with the analytically prediction: $\eta_H=\frac{{\rm sgn}(E)(2N^2+2N+1)\hbar}{4\pi l^2_B}$. When the disorder scattering is strengthened, the plateau of the step structure shrinks due to the broaden of Landau levels, and the center of plateau slightly moves to zero energy due to the shift of Landau levels. Furthermore, the contribution from Fermi sea $\eta^{II}$ is fading away with the overlap of Landau levels.

%\section{Dynamic Viscosity}
%\label{sec-dynamic-viscosity}
\subsection{Dynamic Shear Viscosity $\eta_s(\Omega)$}
\label{sec-ac-shear}
Next we turn to the dynamic shear viscosity $\eta_S(\Omega)$ in graphene, where $\Omega$ is the frequency. 
%It is expressed by the Kubo formula Eq.~(\ref{eq:kubo-2}) with the stress tensor components $T_{ij}=T_{kl}=T_{xy}$, which is similar as the static shear viscosity except the finite frequency $\Omega$. 
%In this case, the Kubo formula of $\eta_s$ is simplified as 
%$%\begin{equation}\label{eq:kubo-ac-shear-b0-1}
%	\eta_s(E)={\rm Re}\left[\eta_s^{RA}(E)-\eta_s^{RR}(E)\right]
%$%\end{equation}
%with
%\begin{equation}\label{eq:kubo-ac-shear-b0-2}
%	\eta_s^{LM}(\Omega)=-4\frac{\hbar}{2\pi\mathcal{V}}\int_{-\infty}^{\infty}d\omega\frac{f_{\omega+\Omega}-f_{\omega}}{\Omega}{\rm Tr}\left[G^L_{\omega+\Omega}T_{xy}G^M_{\omega}T_{xy}\right]
%\end{equation}
Similar to the static shear viscosity, we separate the calculation of dynamic shear viscosity into two conditions: $\bm{B}=0$ and $\bm{B}=B\hat{z}$.

\begin{figure}
	%\centering
	\flushleft
	\includegraphics[width=0.45\textwidth]{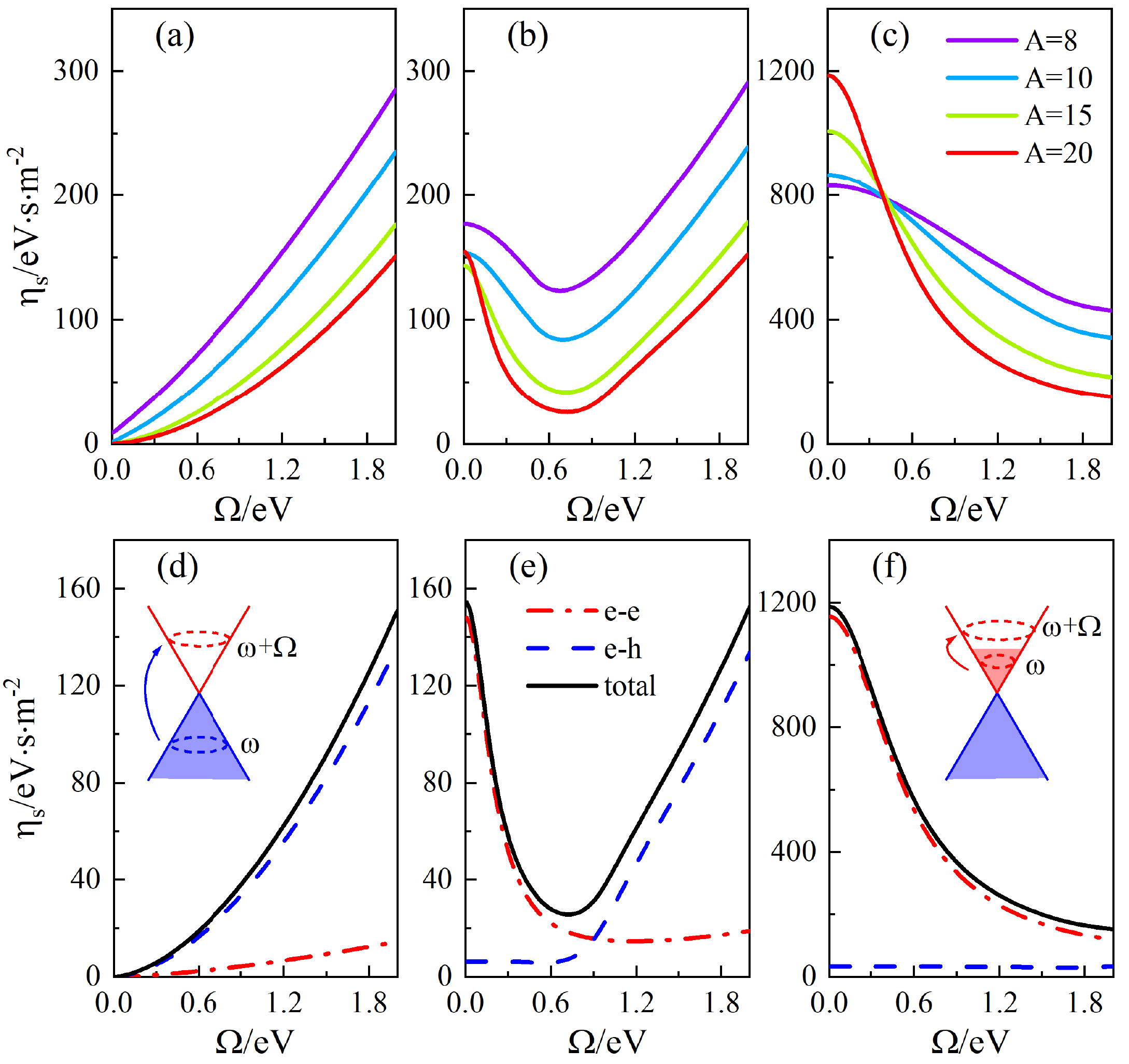}\\
	\caption{(color online) (a-c) Numerical results of as shear viscosity $\eta_s$ versus frequency $\hbar\Omega$ in the disordered graphene at zero temperature with Fermi energy: (a) $E=0eV$; (b) $E=0.5eV$; and (c) $E=1.5eV$. (d-e) Comparison of the contributions of electron-electron coherence and electron-hole coherence for disorder scattering $A=20$ and Fermi energy corresponding to (a-c). } 
	%\oim{reduce the height of figures}
	\label{fig:acshearb0}
\end{figure}

\subsubsection{$B=0$}
In the absence of magnetic field, the analytic derivation of $\eta_s(\Omega)$ is displayed in Appendix. VIII.
%we rewrite $\eta_s^{LM}(\Omega)$ as
%\begin{equation}\begin{aligned}\label{eq:ac-shear-b0}
%		\eta_s^{LM}(\Omega)=&-\frac{\hbar^3v_f^2}{8\pi^2}\int_{-\infty}^{\infty}d\omega\frac{f_{\omega+\Omega}-f_{\omega}}{\Omega}\times\\
%		&\int dkk^3[g_+^L(\omega+\Omega)+g_-^L(\omega+\Omega)][g_+^M(\omega)+g_-^M(\omega)]
%\end{aligned}\end{equation}
When $\hbar\Omega\to0$, the dynamic shear viscosity $\eta_s(\Omega)$ reverts to the static shear viscosity described in Sec.~\ref{sec-dc-shear}. 
%At zero temperature$f_{\omega}$ tends to be a step function. According to the term $f_{\omega+\Omega}-f_{\omega}$ in Eq.~(\ref{eq:ac-shear-b0}), the states that contribute to $\eta_s(\Omega)$ need locate on the two sides of Fermi surface, which means that the frequencies $\omega+\Omega$ and $\omega$ in the integrand should satisfy $\omega<E<\omega+\Omega$ for $\Omega>0$, and $\omega-\Omega<E<\omega$ for $\Omega<0$. 
In this section, we will evaluate the $\eta_s(\Omega)$ under two other constraints at zero temperature: (1) $E\approx0\ll\Omega$ and (2) $0<\Omega\ll E$. Due to the electron-hole symmetry in graphene, the condition for negative $\Omega$ is symmetric to the case for positive $\Omega$ and is therefore ignored. In the regime of $\omega<0<\omega+\Omega$, the contribution from electron-hole (e-h) coherence plays a dominant role. In contrast, in the regime of $0<\omega\lesssim\omega+\Omega$, the electron-electron (e-e) coherent contribution due to collisions of thermally excited carriers is dominant. 

In the e-h dominant region $E\ll\Omega$, the $\eta_s(\Omega)$ is given by
\begin{equation}\label{eq:ac-shear-b0-1}
	\eta_s(\Omega)=\frac{\Omega^2}{16\hbar v_f^2}(\frac{1}{2}+\frac{16}{15}\frac{1}{A}),
\end{equation}
In the e-e dominant region $0<\Omega\ll E$, the $\eta_s(\Omega)$ arrives at
\begin{equation}\label{eq:ac-shear-b0-2}
	\eta_s(\Omega)=\frac{E^2}{2\pi^2\hbar v_f^2}
	\left(\frac{\pi^2}{A}+\frac{AE^2}{\frac{A^2}{\pi^2}\Omega^2+4E^2}\right)
\end{equation}
We compare and analyze these two results for dynamic shear viscosity in the e-h dominant and e-e dominant regions from both frequency-dependent and disorder effects perspectives. We find that the dynamic shear viscosity $\eta_s$ is positively correlated with frequency $\Omega$ in the e-h dominant region, but negatively correlated with frequency $\Omega$ in the e-e dominant region. The dependence of $\eta_s(\Omega)$ on disorder scattering is more complicated. In the e-h dominant region, $\eta_s$ is monotonically increasing with the disorder scattering. When the e-h coherence is dominant, however, $\eta_s$ also relies on the Fermi energy. As disorder scattering increases, $\eta_s$ declines at $4E^2\gg\frac{A^2}{\pi^2}\Omega^2$ and strengthens at $4E^2\ll\frac{A^2}{\pi^2}\Omega^2$. This opposite behavior of dynamic $\eta_s$ versus disorder scattering over different Fermi energy region is reminiscent of the static shear viscosity shown in Fig.~\ref{fig:dcshearb0}, where one finds similar behaviors in two different regions: ``normal region" ($|E|>1.1eV$) and ``anomalous region" ($|E|\lesssim1.1eV$). 

To clearly show the above properties of dynamic shear viscosity, we also illustrate the numerical results of $\eta_s$ vs $\Omega$ in the disordered graphene at zero temperature in Fig.~\ref{fig:acshearb0}(a-c), where the Fermi energies are set to: $E=0eV$, $0.5eV$, and $1.5eV$. In the results of $E=0.0eV$ which belongs to e-h dominant region, $\eta_s$ increases superlinearly with $\Omega$ and decreases with $A$, both behaviors are well agree with the analytic prediction Eq.~(\ref{eq:ac-shear-b0-1}). In the case of $E=1.5eV$ and low frequency , which belongs to e-e dominant region, $\eta_s$ drops sharply with $\Omega$, and the $\eta_s(\Omega)$ curves of different disorder strengths have a cross. These behaviors also meet expectation of Eq.~(\ref{eq:ac-shear-b0-2}). In the case of $E=0.5eV$, the transition between the e-h dominant region and e-e dominant region can be seen. The different transport mechanisms for these Fermi energies are also demonstrated by comparing of the contributions of electron-electron coherence and electron-hole coherence in Fig.~\ref{fig:acshearb0}(d-e).

Another remarkable piece of information we gain from the numerical results is a link between dynamic and static shear viscosities. We introduced ``normal region" ($|E|>1.1eV$) and ``anomalous region" ($|E|\lesssim1.1eV$) when analyzing the static shear viscosity. In the case of $E=1.5eV$, the increase of $\Omega$ can let available $\omega$ cross the ``normal region" into ``anomalous region". The critical frequency is $\hbar\Omega\approx0.4eV$, which quite matches the cross point shown in Fig.~\ref{fig:acshearb0}(c).

\begin{figure}
	%\centering
	\flushleft
	\includegraphics[width=0.48\textwidth]{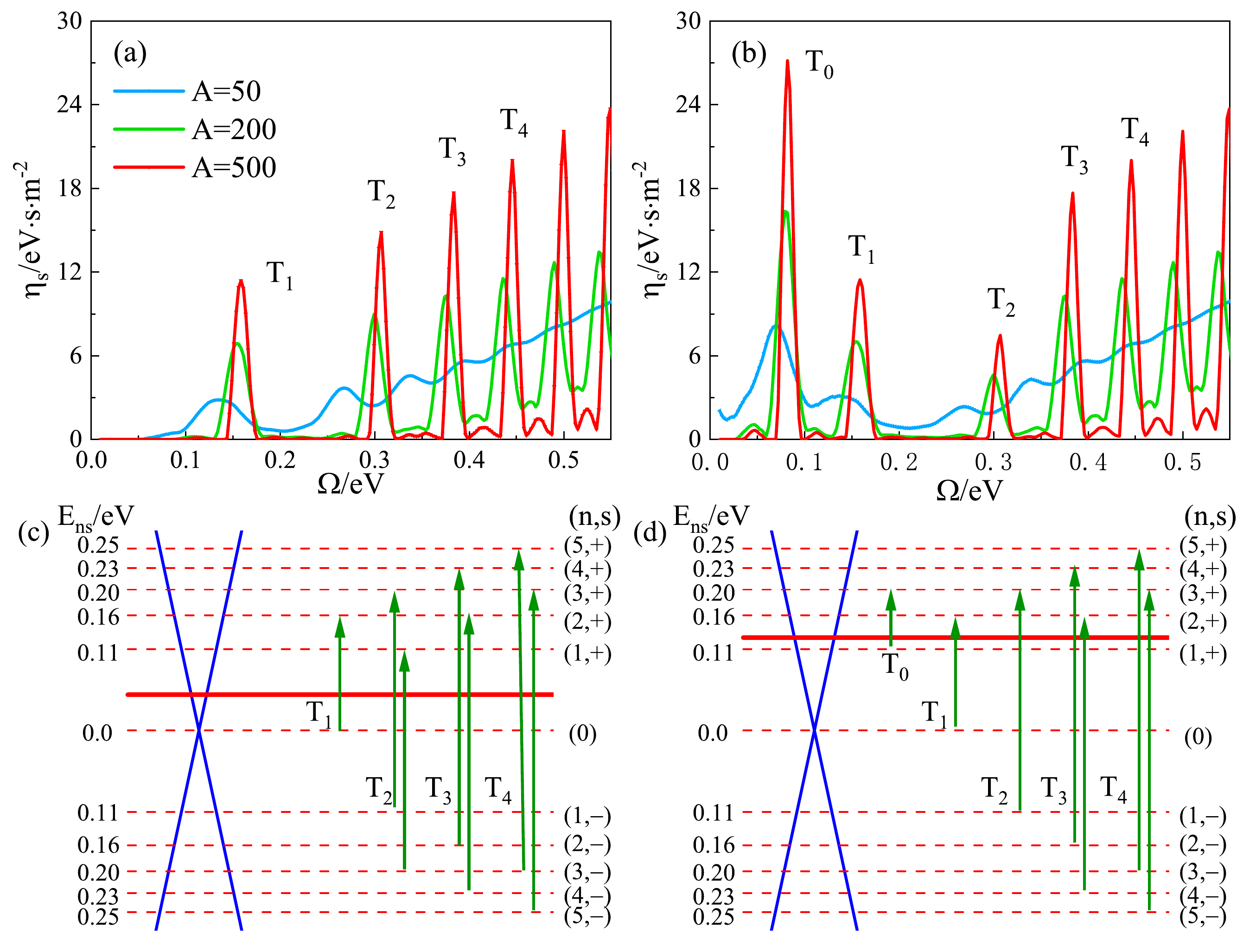}
	\caption{(color online). (a)(b) The numerical results of $\eta_s$ vs $\Omega$ for the magnetic field $B=10T$ and Fermi energy: (a) $E=0.05eV$ and (b) $E=0.13eV$. (c)(d) The schematic representation of the transitions correspond to the peaks $T_i$ in (a)(b).}
	\label{fig:acshearb10}
\end{figure}

\subsubsection{$B\neq0$}
In the presence of magnetic field $\bm{B}=B\hat{z}$, the total dynamic shear viscosity can be written as
\begin{equation}\begin{aligned}\label{eq:ac-shear-bn0-2}
		\eta_s(\Omega)=&-\frac{\hbar^3\omega_c^2}{8\pi^2l_B^2}\int_{-\infty}^{\infty}d\omega\frac{f_{\omega+\Omega}-f_{\omega}}{\Omega}\sum_{n,s,s'}(n+1)\\
		&[{\rm Im}G^R_{ns}(\omega+\Omega){\rm Im}G^R_{n+2,s'}(\omega)+(\omega+\Omega\leftrightarrow\omega)]
\end{aligned}\end{equation}
At the well separated Landau levels region and zero temperature, it is evaluated as
\begin{equation}\begin{aligned}\label{eq:ac-shear-bn0-3}
		\eta_s(\Omega)=&\frac{\hbar^3\omega_c^2}{8l_B^2}\sum_{n,s,s'}\frac{n+1}{\Omega}\\
		&\left[(f_{E_{ns}}-f_{E_{n+2,s'}})\delta(\Omega+E_{n+2,s'}-E_{ns})+(E_{ns}\leftrightarrow E_{n+2,s'})\right]
\end{aligned}\end{equation}
One can easily find that in this case the dynamic shear viscosity is determined by the state transitions between Landau levels, $E_{ns}\leftrightarrow E_{ms'}$, where the level indices satisfy $|m-n|=2$ and there is no restriction on $s$ and $s'$ (see Appendix II). Similar selection rule exists in the calculation of magneto-optical conductivity in graphene, but in which level index of the allowed transitions satisfies $|m-n|=1$ \cite{Gusynin2006jpcm,Gusynin2006prb,Gusynin2006prl,Gusynin2007prl}.

In Fig~\ref{fig:acshearb10}(a-b), we show the results of $\eta_s(\Omega)$ as a function of frequency $\Omega$. When the disorder scattering is extremely small, $\eta_s(\Omega)$ shows a series of resonant peaks. These peaks correspond to the transitions between the separated Landau levels. Meanwhile, the schematic diagrams which can help us understand the transitions in Fig~\ref{fig:acshearb10}(a) and (b) are given in Fig~\ref{fig:acshearb10}(c) and (d). It can be seen from the numerical results combining with the schematic diagrams that the transitions between the same two Landau levels contributes the same intensity to the resonance peak of $\eta_s$ despite the location of Fermi energy. The peaks $T_1$, $T_3$, and $T_4$ in Fig~\ref{fig:acshearb10}(a) and (b) have same intensities since they correspond to the same transitions in two cases, as shown in Fig~\ref{fig:acshearb10}(c) and (d). The peak $T_2$ in Fig.~\ref{fig:acshearb10}(b) is half of that in Fig.~\ref{fig:acshearb10}(a) because the peak $T_2$ in Fig.~\ref{fig:acshearb10}(a) contains the transitions $E_{3-}$ to $E_{1+}$ and $E_{1-}$ to $E_{3+}$ but the peak $T_2$ in Fig.~\ref{fig:acshearb10}(a) has only contributions from $E_{1-}$ to $E_{3+}$.

Another notable feature  is the dependence on frequency $\Omega$. 
For $\Omega>0$, in the electron-hole (e-h) transitions, $\frac{n+1}{\Omega}\propto\Omega$ due to $\hbar\Omega=|E_{n+2,s'}-E_{ns}|\approx2\sqrt{n}\hbar\omega_c$, but in the electron-electron (e-e) transitions, $\frac{n+1}{\Omega}\propto\Omega^{-3}$ since $\hbar\Omega=|E_{n+2,s'}-E_{ns}|\approx\frac{\hbar\omega_c}{\sqrt{n}}$. This is also confirmed by the numerical results in Fig.~\ref{fig:acshearb10}(b), where the $\eta_s$ contributed by electron-electron transition tends to diverge as $\Omega$ approaches zero. As the Landau levels gradually overlap, the peak of $\eta_s(\Omega)$ will drop and merge with others, but as shown by the numerical results, the magnitude of $\eta_s$ still maintains a general trend of decreasing with $\Omega$ for the e-e transitions and increasing with $\Omega$ for the e-h transitions.

% The Fermi energy for the results of Fig~\ref{fig:acshearb10}(a) is located at the Dirac point, so that transitions can only occur between hole and electron states.

\begin{figure*}
	\centering
	%\flushleft
	\includegraphics[width=1\textwidth]{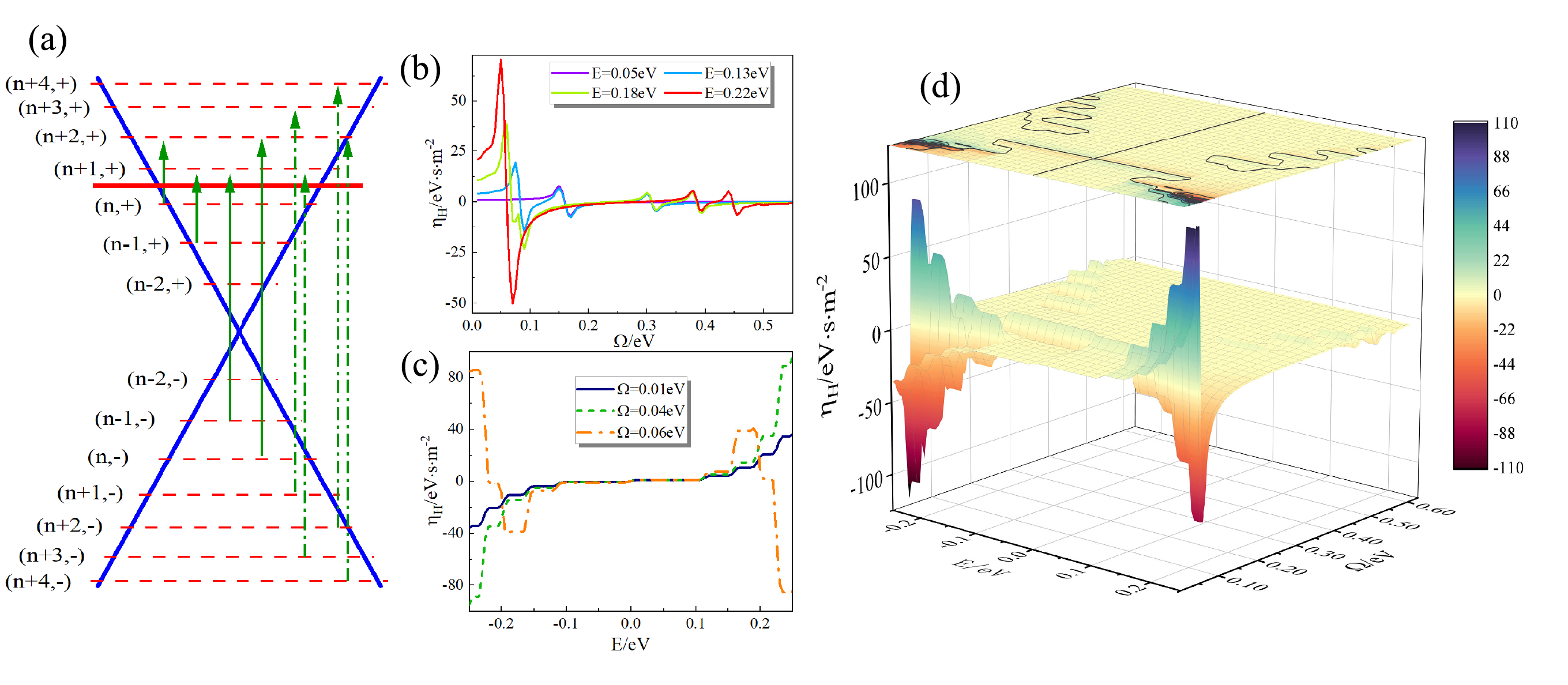}
	\caption{(color online) (a) The schematic diagram of the allowed transitions in calculating dynamic Hall viscosity. (b) dynamic Hall viscosity $\eta_H$ vs frequency $\Omega$ for magnetic field $B=10T$ and several Fermi energies. (c) dynamic Hall viscosity $\eta_H$ vs Fermi energy $E$ for same magnetic field and several frequencies. (d) A global dependence of $\eta_H$ on $\Omega$ and $E$.  }
	\label{fig:achall}
\end{figure*}

\subsection{Dynamic Hall Viscosity $\eta_H(\Omega)$}
\label{sec-ac-Hall}
The Kubo formula of dynamic Hall viscosity is similar as the one of the static shear viscosity except the finite frequency $\Omega$. It is also divided into three parts as
\begin{equation}\label{eq:ac-hall-1}
	\eta_H(\Omega)={\rm Re}\left[\eta_H^{I,RA}(\Omega)-\eta_H^{I,RR}(\Omega)+\eta_H^{II}(\Omega)\right]
\end{equation}
%where
%\begin{equation}\begin{aligned}\label{eq:ac-hall-2}
%	\eta_H^{I,LM}(\Omega)=&-\frac{\hbar}{\pi\mathcal{V}}\int d\omega\frac{f_{\omega+\Omega}-f_{\omega}}{\Omega}\times\\
%	&{\rm Tr}\left[G^L(\omega+\Omega)(T_{xy}-T_{yy})G^M(\omega)T_{xy}\right]
%\end{aligned}\end{equation}
%\begin{equation}\begin{aligned}\label{eq:ac-hall-3}
%		\eta_H^{II}(\Omega)=&-\frac{\hbar}{\pi\mathcal{V}}\int d\omega \frac{f_{\omega}}{\Omega}{\rm Tr}\left[G^R(\omega+\Omega)(T_{xy}-T_{yy})G^R(\omega)T_{xy}\right.\\
%		&\left.-G^R(\omega)(T_{xy}-T_{yy})G^R(\omega+\Omega)T_{xy}\right]
%\end{aligned}\end{equation}
%Plugging the expressions of stress tensor $T_{xy}$ and $T_{xx}-T_{yy}$ in $(n,k_y,s)$-basis, i.e. Eq.~(\ref{eq:Txy-Bn0}) and Eq.~(\ref{eq:Txx-Tyy-Bn0}), one can rewrite $\eta_H(\Omega)$ as
%\begin{equation}\begin{aligned}\label{eq:ac-hall-4}
%	\eta_H(\Omega)&=\frac{\hbar^3\omega_c^2}{8\pi^2l_B^2}\int d\omega\sum_{n,s,s'}(n+1)\\
%	&\left\{\frac{f_{\omega+\Omega}-f_{\omega}}{\Omega}\left[{\rm Re}G_{ns}^R(\omega+\Omega){\rm Im}G_{n+2,s'}^R(\omega)-(\omega+\Omega\leftrightarrow\omega)\right]\right.\\
%	&-\frac{f_{\omega}}{\Omega}\left[{\rm Re}G_{ns}^R(\omega+\Omega){\rm Im}G_{n+2,s'}^R(\omega)-(\omega+\Omega\leftrightarrow\omega)\right]\\
%	&\left.-\frac{f_{\omega}}{\Omega}\left[{\rm Im}G_{ns}^R(\omega+\Omega){\rm Re}G_{n+2,s'}^R(\omega)-(\omega+\Omega\leftrightarrow\omega)\right]\right\}
%\end{aligned}\end{equation}
Then, we evaluate dynamic Hall viscosity at the well separated Landau levels region and zero temperature, 
\begin{equation}\begin{aligned}\label{eq:ac-Hall-5}
	&\eta_H(\Omega)
	=\frac{\hbar^3\omega_c^2}{8\pi l_B^2}\sum_{n,s,s'}\frac{n+1}{\Omega} \\
	&\left\{2 (f_{E_{ns}+\Omega}-f_{E_{ns}})\frac{\Omega-E_{n+2,s'}+E_{ns}}{(\Omega-E_{n+2,s'}+E_{ns})^2+\Gamma^2}\right.\\
	%&\left.-(E_{ns}\leftrightarrow E_{n+2,s'})\right]\\
	&+(f_{E_{n+2,s'}+\Omega}-f_{E_{ns}-\Omega})\frac{\Omega+E_{n+2,s'}-E_{ns}}{(\Omega+E_{n+2,s'}-E_{ns})^2+\Gamma^2} \\
	&\left.-(E_{ns}\leftrightarrow E_{n+2,s'})
	\right\}
\end{aligned}\end{equation}
where the self-energy is assumed to be a small pure imaginary number $-i\Gamma$. It is obvious that, at the limit $\Omega\to0$, the first and second terms in the curly brace represent the Fermi surface and Fermi sea contributions, which have been analyzed in the section of static Hall viscosity. Here, we focus on the dependence of $\eta_H$ on $\Omega$. Since the term $\frac{\Omega\mp E_{n+2,s'}\pm E_{ns}}{(\Omega\mp E_{n+2,s'}\pm E_{ns})^2+\Gamma^2}$ in Eq.~(\ref{eq:ac-Hall-5}) describes a kink with the center $\Omega_c=\pm(E_{n+2,s'}-E_{ns})$, we assume the $\Omega$ in the distribution function multiplied by the kind function tends to $\Omega_c$. Thus, the dynamic Hall viscosity can be further simplified as
\begin{equation}\begin{aligned}\label{eq:ac-Hall-6}
		&\eta_H(\Omega)\approx\frac{\hbar^3\omega_c^2}{8\pi l_B^2}\sum_{n,s,s'}\frac{n+1}{\Omega}\\
		&\left[(f_{E_{n+2,s'}}-f_{E_{ns}})\frac{\Omega-E_{n+2,s'}+E_{ns}}{(\Omega-E_{n+2,s'}+E_{ns})^2+\Gamma^2}-(E_{ns}\leftrightarrow E_{n+2,s'})\right]\\
\end{aligned}\end{equation}

This expression is somewhat analogous to the evaluation of dynamic shear viscosity in the well separated Landau levels region. They have the same transition rule: $n\leftrightarrow n+2$, but different types of transition function: delta structure for $\eta_s(\Omega)$ and kink structure for $\eta_H(\Omega)$. Another essential different between these two is the sign before the exchange $(E_{ns}\leftrightarrow E_{n+2,s'})$. In the expression of $\eta_s(\Omega)$, it is a plus sign, which means the addition of the counterparts. In the expression of $\eta_H(\Omega)$, however, it is a minus sign, so the counterparts cancel each other out. Figure~\ref{fig:achall}(a) shows the schematic diagram of the transitions in a general case where the Fermi level is assumed to fall in the gap between the Landau levels with indices $(n,+)$ and $(n+1,+)$. It is noticeable that there are only four transitions left, as the others cancel out with their counterparts under the exchange $(ns)\leftrightarrow (n+2,s')$.

The numerical results of dynamic Hall viscosity are shown in Fig.~\ref{fig:achall}(b-d). In Fig.~\ref{fig:achall}(b), we plot dynamic Hall viscosity $\eta_H$ vs frequency $\Omega$ in the presence of magnetic field $B=10T$. The energies of the low-order Landau levels in this condition are shown in Fig.~\ref{fig:acshearb10}(c)(d). For Fermi energy $E=0.05eV$, there is only one resonance kink structure around $\Omega=0.16eV$, which is consistent with the prediction of Eq.~(\ref{eq:ac-Hall-6}), since there is only one single transition with frequency $\Omega=0.16eV$ in this case. For Fermi energy $E=0.13eV$, there are three resonance kink structures corresponding to the three single transitions at: $\Omega=0.09eV$, $0.16eV$, and $0.31eV$. For Fermi energy $E=0.18eV$ and $0.22eV$, there should be four resonance kink structures according to the Eq.~(\ref{eq:ac-Hall-6}). However, in the numerical results, the lowest two resonance structure are partially and fully overlap for $E=0.18eV$ and $E=0.22eV$, since their transition frequencies are too close together. 

In Fig.~\ref{fig:achall}(b), we plot $\eta_H$ vs $E$ for different frequencies. In Fig.~\ref{fig:achall}(b), we plot a global dependence of $\eta_H$ on $\Omega$ and $E$. Both of these figures show how the static quantized Hall viscosity should evolve into the dynamic Hall viscosity. The plateau structure is retained, but the heights of the plateaus vary with frequency and resonate around the frequencies that lead to single transitions.

%%%%%%%%%%%%%%%%%%%%%%%%%%%%%%%%%%%
\section{Discussion}
\label{sec:discussion}
We have presented a microscopic theory on hydrodynamic  electronics in disordered graphene. We provide a unified description for both undoped and doped graphene, with or without external magnetic fields. Surprisingly, we identify the shear viscosity exhibits an anomalous enhancement around the Dirac point. By considering the external magnetic field, the shear viscosity can be further enhanced. These findings are in remarkable contrast to the normal Fermi liquids, where hydrodynamic properties are usually suppressed by disorder-scattering or external magnetic field.
%Additionally, it is found the anomalous change of viscosity can be also revealed in the dynamic and magneto- hydrodynamic responses.  
Crucially, we present what we believe to be compelling evidences that this anomalous phenomena is from electron-hole collisions in the vicinity of Dirac point.

%Our findings reveal that the impurities offer a counterintuitive effect on the collective motion of electrons in graphene near the Dirac point, due to electron-hole collisions that do not exist in conventional 2DEG. It implies that undoped graphene is \oim{nice} to realize hydrodynamic electronics. 
%Similar to the minimal conductivity of graphene due to impurity assisted resonant tunneling through the electron-hole coherence and retarded-retarded ($RR$) channel \cite{Chen2021}, this anomalous term in static shear viscosity also mainly comes from the electron-hole coherence and $RR$ term.

The reason for selecting graphene as the platform to realize hydrodynamics in the existing experiments mainly relys on two facts. First, graphene is stiff so the electron-phonon scattering is greatly suppressed. Second, typical mean-free path due to
electron-electron scattering could become the smallest length scale above certain temperature in graphene, which fits the Gurzhi condition for viscous electronic flow \cite{Gurzhi1963}. In this context, the current work promotes a third viewpoint, i.e. hydrodynamic phenomena can be further amplified due to disorder-assisted electron-hole collisions  in the vicinity of Dirac point.  
So our study implies that undoped graphene is ethereal for  hydrodynamic electronics.
Moreover, momentum currents contains single-particle contributions compared to charge currents, which makes the disorder-assisted effect significantly larger in viscosity than that in conductivity.
Uncovering the critical relevance of hydrodynamic electronics with electron-hole coherent collisions affords a unique link between quantum-critical electron transport and the wealth of fluid dynamics phenomena.

%The current work is complementary to the existing studies on electron viscosity. Generally speaking,
%The effect of impurities is believed to suppress the scattering events with small momentum transfer \cite{Principi2016} and collective motion  of electrons  \cite{Polini2020}. Here, our findings demonstrate that the effect of impurities on the viscosity in graphene needs a careful study. 

Furthermore, electrons in graphene behave as quasi-relativistic gas of quasiparticles satisfying the relativistic equation of motion, providing a playground to study relativistic effects in fluid dynamics. In this work, we have shown that the electron-hole coherence related to quasi-relativistic nature of graphene results in numerous peculiar behaviors in both static and dynamic viscosities. It calls for more careful studies on the Navier-Stokes equation due to relativistic effects in describing the flow of Dirac electrons.

\begin{acknowledgments}
	We thank Igor S. Burmistrov, S. G. Xu and X. Lin for discussion. 
	W.W.C. and W.Z. are supported by
	National Science Foundation of China (92165102,11974288), and the foundation from Westlake University. This work was supported by ``Pioneer" and ''Leading Goose" R\&D Program of Zhejiang (2022SDXHDX0005), the Key R\&D Program of Zhejiang Province (2021C01002).
\end{acknowledgments}

%\textbf{Author contributions.} W.Z. designed this project. W.W.C. carried out the calculations and drafted the article. All authors discussed the results and revised the contents of the manuscript.

%\textbf{Data availability.} The datasets generated during the current study are available from the corresponding author on reasonable request. 

%\textbf{Additional information.} The authors declare no competing interests. 

%%%%%%%%%%%%%%%%%%%%%%%%%%%%%%%%%%%%%%%%%%%%%


\begin{thebibliography}{99}


\bibitem{Gurzhi1963}R. N. Gurzhi, J. Exp. Theor. Phys. \textbf{17}, 521 (1963)

	
\bibitem{Bandurin2016} D. A. Bandurin, I. Torre, R. Krishna Kumar, M. Ben Shalom, A. Tomadin, A. Principi, G. H. Auton, E. Khestanova, K. S. Novoselov, I. V. Grigorieva, L. A. Ponomarenko, A. K. Geim, and M. Polini, 
Negative local resistance caused by viscous electron backflow in graphene, 
\href{https://www.science.org/doi/10.1126/science.aad0201}
{Science \textbf{351}, 1055 (2016)}.

\bibitem{Bandurin2018} Denis A. Bandurin, Andrey V. Shytov, Leonid S. Levitov, Roshan Krishna Kumar, Alexey I. Berdyugin, Moshe Ben Shalom, Irina V. Grigorieva, Andre K. Geim, and Gregory Falkovich,
Fluidity onset in graphene,
\href{https://www.nature.com/articles/s41467-018-07004-4}
{Nat. Comm. \textbf{9}, 4533 (2018).}

\bibitem{Kumar2017} R. Krishna Kumar, D. A. Bandurin, F. M. D. Pellegrino, Y. Cao, A. Principi, H. Guo, G. H. Auton, M. Ben Shalom, L. A. Ponomarenko, G. Falkovich, K. Watanabe, T. Taniguchi, I. V. Grigorieva, L. S. Levitov, M. Polini and A. K. Geim,
Superballistic flow of viscous electron fluid through graphene constrictions,
\href{https://www.nature.com/articles/nphys4240}
{Nat. Phys. \textbf{13}, 1182 (2017)}.

\bibitem{Guo2017} Haoyu Guo, Ekin Ilseven, Gregory Falkovich, and Leonid S. Levitov,
Higher-than-ballistic conduction of viscous electron flows,
\href{https://www.pnas.org/content/114/12/3068}
{Proc. Natl. Acad. Sci, U.S.A \textbf{114} 3068 (2017)}.

\bibitem{Keser2021} Aydın Cem Keser, Daisy Q. Wang, Oleh Klochan, Derek Y. H. Ho, Olga A. Tkachenko, Vitaly A. Tkachenko, Dimitrie Culcer, Shaffique Adam, Ian Farrer, David A. Ritchie, Oleg P. Sushkov, and Alexander R. Hamilton,
Geometric Control of Universal Hydrodynamic Flow in a Two-Dimensional Electron Fluid
\href{https://journals.aps.org/prx/abstract/10.1103/PhysRevX.11.031030}
{Phys. Rev. X \textbf{11}, 031030 (2021)}.

\bibitem{Levitov2016} Leonid Levitov and Gregory Falkovich,
Electron viscosity, current vortices and negative nonlocal resistance in graphene,
\href{https://www.nature.com/articles/nphys3667}
{Nat. Phys. \textbf{12}, 672 (2016)}.

\bibitem{Mayzel2019} Jonathan Mayzel, Victor Steinberg and Atul Varshney,
Stokes flow analogous to viscous electron current in graphene,
\href{https://www.nature.com/articles/s41467-019-08916-5}
{Nat. Comm. \textbf{10}, 937 (2019)}.

\bibitem{Falkovich2017} Gregory Falkovich and Leonid Levitov,
Linking Spatial Distributions of Potential and Current in Viscous Electronics,
\href{https://journals.aps.org/prl/abstract/10.1103/PhysRevLett.119.066601}
{Phys. Rev. Lett. \textbf{119}, 066601 (2017)}.

\bibitem{Sulpizio2019} Joseph A. Sulpizio, Lior Ella, Asaf Rozen, John Birkbeck, David J. Perello, Debarghya Dutta, Moshe Ben-Shalom, Takashi Taniguchi, Kenji Watanabe, Tobias Holder, Raquel Queiroz, Alessandro Principi, Ady Stern, Thomas Scaffidi, Andre K. Geim and Shahal Ilani,
Visualizing Poiseuille flow of hydrodynamic electrons,
\href{https://www.nature.com/articles/s41586-019-1788-9}
{Nature \textbf{576}, 75 (2019)}.

\bibitem{Moll2016} Philip J. W. Moll, Pallavi Kushwaha, Nabhanila Nandi, Burkhard Schmidt, Andrew P. Mackenzie,
Evidence for hydrodynamic electron flow in PdCoO$_2$,
\href{https://www.science.org/doi/10.1126/science.aac8385}
{Science \textbf{351}, 1061 (2016)}.

\bibitem{Berdyugin2019} A. I. Berdyugin, S. G. Xu, F. M. D. Pellegrino, R. Krishna Kumar, A. Principi, I. Torre, M. Ben Shalom, T. Taniguchi, K. Watanabe, I. V. Grigorieva, M. Polini,
A. K. Geim, D. A. Bandurin,
Measuring Hall viscosity of graphene’s electron fluid,
\href{https://www.science.org/doi/10.1126/science.aau0685}
{Science \textbf{436}, 162 (2019)}.

\bibitem{Crossno2016} Jesse Crossno, Jing K. Shi, Ke Wang, Xiaomeng Liu, Achim Harzheim,
Andrew Lucas, Subir Sachdev, Philip Kim, Takashi Taniguchi, Kenji Watanabe,
Thomas A. Ohki, Kin Chung Fong,
Observation of the Dirac fluid and the breakdown of the Wiedemann-Franz law in graphene,
\href{https://www.science.org/doi/10.1126/science.aad0343}
{Science \textbf{351}, 1058 (2016)}.

\bibitem{Gooth2018} J. Gooth, F. Menges, N. Kumar, V. S\"u$\beta$, C. Shekhar, Y. Sun, U. Drechsler, R. Zierold, C. Felser and B. Gotsmann,
Thermal and electrical signatures of a hydrodynamic electron fluid in tungsten diphosphide,
\href{https://www.nature.com/articles/s41467-018-06688-y}
{Nat. Comm. \textbf{90}, 4093 (2018)}.


\bibitem{Briskot2015} U. Briskot, M. Sch\"utt, I. V. Gornyi, M. Titov, B. N. Narozhny, and A. D. Mirlin,
Collision-dominated nonlinear hydrodynamics in graphene,
\href{https://journals.aps.org/prb/abstract/10.1103/PhysRevB.92.115426}
{Phys. Rev. B \textbf{92}, 115426 (2015)}.

\bibitem{Muller2009} Markus M\"uller, J\"org Schmalian, and Lars Fritz,
Graphene: A Nearly Perfect Fluid,
\href{https://journals.aps.org/prl/abstract/10.1103/PhysRevLett.103.025301}
{Phys. Rev. Lett. \textbf{103}, 025301 (2009)}.

\bibitem{Narozhny2019} B. N. Narozhny and M. Sch\"utt,
Magnetohydrodynamics in graphene: Shear and Hall viscosities
\href{https://journals.aps.org/prb/abstract/10.1103/PhysRevB.100.035125}
{Phys. Rev. B \textbf{100}, 035125 (2019)}.

\bibitem{Torre2015} Iacopo Torre, Andrea Tomadin, Andre K. Geim, and Marco Polini,
Nonlocal transport and the hydrodynamic shear viscosity in graphene,
\href{https://journals.aps.org/prb/abstract/10.1103/PhysRevB.92.165433}
{Phys. Rev. B \textbf{92}, 165433 (2015)}.

\bibitem{Pellegrino2016prb} Francesco M. D. Pellegrino, Iacopo Torre, Andre K. Geim, and Marco Polini,
Electron hydrodynamics dilemma: Whirlpools or no whirlpools,
\href{https://journals.aps.org/prb/abstract/10.1103/PhysRevB.94.155414}
{Phys. Rev. B \textbf{94}, 155414 (2016)}

%\bibitem{Narozhny2015} B. N. Narozhny, I. V. Gornyi, M. Titov, M. Schütt, and A. D. Mirlin,
%Hydrodynamics in graphene: Linear-response transport
%\href{https://journals.aps.org/prb/abstract/10.1103/PhysRevB.91.035414}
%{Phys. Rev. B \textbf{91}, 035414 (2015)}.

\bibitem{Lucas2015} Andrew Lucas, Hydrodynamic transport in strongly coupled disordered quantum field theories,
\href{https://iopscience.iop.org/article/10.1088/1367-2630/17/11/113007}
{J. Phys.: Condens. Matter \textbf{17}, 113007 (2015).}

\bibitem{Lucas2018} Andrew Lucas and Kin Chung Fong, Hydrodynamics of electrons in graphene,
\href{https://iopscience.iop.org/article/10.1088/1361-648X/aaa274}
{J. Phys.: Condens. Matter \textbf{30}, 053001 (2018).}

\bibitem{Novoselov2005}Novoselov, K. S. et al. Two-dimensional gas of massless Dirac fermions in graphene. Nature 438, 197–200 (2005).


\bibitem{Katsnelson2006} M. I. Katsnelson, K. S. Novoselov, and A. K. Geim,
Chiral tunnelling and the Klein paradox in graphene,
\href{https://www.nature.com/articles/nphys384}
{Nat. Phys. \textbf{2}, 620–625 (2006).}

\bibitem{Titov2007} M. Titov, EPL 79, 17004 (2007).

\bibitem{Nomura2007} K. Nomura, A. H. MacDonald, 
  Phys. Rev. Lett. 98, 076602 (2007).

\bibitem{Adam2007}S. Adam, E. H. Hwang, V. M. Galitski, and S. Das Sarma
 PNAS 104, 18392-18397 (2007).


\bibitem{Burmistrov2019} Igor S. Burmistrov, Moshe Goldstein, Mordecai Kot, Vladislav D. Kurilovich, and Pavel D. Kurilovich,
Dissipative and Hall Viscosity of a Disordered 2D Electron Gas,
\href{https://journals.aps.org/prl/abstract/10.1103/PhysRevLett.123.026804}
{Phys. Rev. Lett. \text{123}, 026804 (2019)}.

\bibitem{Alekseev2016} P. S. Alekseev,
Negative Magnetoresistance in Viscous Flow of Two-Dimensional Electrons,
\href{https://journals.aps.org/prl/abstract/10.1103/PhysRevLett.117.166601}
{Phys. Rev. Lett. \textbf{117}, 166601 (2016).}

\bibitem{Pellegrino2017} Francesco M. D. Pellegrino, Iacopo Torre, and Marco Polini,
Nonlocal transport and the Hall viscosity of two-dimensional hydrodynamic electron liquids,
\href{https://journals.aps.org/prb/abstract/10.1103/PhysRevB.96.195401}
{Phys. Rev. B \textbf{96}, 195401 (2017)}.

\bibitem{Chen2021} Weiwei Chen, Yedi Shen, Bo Fu, Qinwei Shi, W. Zhu,
On the sample-dependent minimal conductivity in weakly disordered graphene,
\href{https://arxiv.org/abs/2108.13787}
{arXiv:2108.13787}

\bibitem{Sherafati2016} Mohammad Sherafati, Alessandro Principi, and Giovanni Vignale,
Hall viscosity and electromagnetic response of electrons in graphene,
\href{https://journals.aps.org/prb/abstract/10.1103/PhysRevB.94.125427}
{Phys. Rev. B \textbf{94}, 125427 (2016)}.

\bibitem{Hu2021} Liangdong Hu, Zhao Liu, D. N. Sheng, F. D. M. Haldane, and W. Zhu,
Microscopic diagnosis of universal geometric responses in fractional quantum Hall liquids
\href{https://journals.aps.org/prb/abstract/10.1103/PhysRevB.103.085103}
{Phys. Rev. B \textbf{103}, 085103 (2021)}.

\bibitem{Hoyos2012} Carlos Hoyos and Dam Thanh Son,
Hall Viscosity and Electromagnetic Response
\href{https://journals.aps.org/prl/pdf/10.1103/PhysRevLett.108.066805}
{Phys. Rev. Lett. \textbf{108}, 066805 (2012)}.

\bibitem{Sarma2011} S. Das Sarma, Shaffique Adam, E. H. Hwang, and Enrico Rossi,
Electronic transport in two-dimensional graphene,
\href{https://journals.aps.org/rmp/abstract/10.1103/RevModPhys.83.407}
{Rev. Mod. Phys. \textbf{83}, 407 (2011)}.

\bibitem{Gusynin2006prl} V. P. Gusynin, S. G. Sharapov, and J. P. Carbotte,
Unusual Microwave Response of Dirac Quasiparticles in Graphene,
\href{https://journals.aps.org/prl/abstract/10.1103/PhysRevLett.96.256802}
{Phys. Rev. Lett. \textbf{96}, 256802 (2006)}.

\bibitem{Gusynin2006prb} V. P. Gusynin and S. G. Sharapov,
Transport of Dirac quasiparticles in graphene: Hall and optical conductivities,
\href{https://journals.aps.org/prb/abstract/10.1103/PhysRevB.73.245411}
{Phys. Rev. B \textbf{73}, 245411 (2006)}.

\bibitem{Gusynin2006jpcm} V. P. Gusynin, S. G. Sharapov and J. P. Carbotte,
Magneto-optical conductivity in graphene,
\href{https://iopscience.iop.org/article/10.1088/0953-8984/19/2/026222}
{J. Phys.: Condens. Matter. \textbf{19}, 026222 (2006)}.

\bibitem{Gusynin2007prl} V. P. Gusynin, S. G. Sharapov, and J. P. Carbotte,
Anomalous Absorption Line in the Magneto-Optical Response of Graphene
\href{https://journals.aps.org/prl/abstract/10.1103/PhysRevLett.98.157402}
{Phys. Rev. Lett. \textbf{98}, 157402 (2007)}.

\bibitem{Principi2016} Alessandro Principi, Giovanni Vignale, Matteo Carrega, and Marco Polini,
Bulk and shear viscosities of the two-dimensional electron liquid in a doped graphene sheet,
\href{https://journals.aps.org/prb/abstract/10.1103/PhysRevB.93.125410}
{Phys. Rev. B \textbf{93}, 125410 (2016)}.

\bibitem{Polini2020} Marco Polini and Andre Geim,
Viscous electron fluids,
\href{https://physicstoday.scitation.org/doi/10.1063/PT.3.4497}
{Physics Today \textbf{73}, 6, 28 (2020)}.






%\bibitem{Avron1998} J. E. Avron,
%Odd viscosity,
%\href{https://link.springer.com/article/10.1023/A:1023084404080}
%{J. Stat. Phys. \textbf{92}, 543 (1998)}.	
%
%\bibitem{Bradlyn2012} Barry Bradlyn, Moshe Goldstein, and N. Read,
%Kubo formulas for viscosity: Hall viscosity, Ward identities, and the relation with conductivity,
%\href{https://journals.aps.org/prb/abstract/10.1103/PhysRevB.86.245309}
%{Phys. Rev. B \textbf{86}, 245309 (2012)}.




%\bibitem{Nielsen1985} O. H. Nielsen and Richard M. Martin,
%Quantum-mechanical theory of stress and force,
%\href{https://journals.aps.org/prb/abstract/10.1103/PhysRevB.32.3780}
%{Phys. Rev. B \textbf{32}, 3780 (1985)}.
%
%\bibitem{Kundu2008} Pijush K. Kundu and Ira M. Cohen,
%\emph{Fluid Mechanics}, Fourth Edition, Elsevier (2008).
%
%\bibitem{Band} Y. B. Band and Y. Avishai,
%\emph{Quantum mechanics with applications to nanotechnology and information science}, 
%Academic Press (2013).	

%\bibitem{Read2011} N. Read and E. H. Rezayi,
%Hall viscosity, orbital spin, and geometry: Paired superfluids and quantum Hall systems,
%\href{https://journals.aps.org/prb/abstract/10.1103/PhysRevB.84.085316}
%{Phys. Rev. B 84, 085316 (2011)}.
%
%\bibitem{Landau1970} L. D. Landau and E. M. Lifshitz,
%\emph{Theory of elasticity}, Second Edition, Elsevier (1970).
%
%\bibitem{Blaschke2016} D. N. Blaschke, R. Gieres. M. Reboud, and M. Schweda,
%The energy-momentum tensor(s) in classical gauge theories, 
%\href{https://www.sciencedirect.com/science/article/pii/S0550321316301845}
%{Nucl. Phys. \textbf{B912}, 192 (2016).}

%\bibitem{Palacios2006} Angel Fierros Palacios,
%\emph{The Hamilton-Type Principle in Fluid Dynamics},
%Springer, Vienna (2006).
%
%\bibitem{Link2018} Julia M. Link, Daniel E. Sheehy, Boris N. Narozhny, and J\"org Schmalian,
%Elastic response of the electron fluid in intrinsic graphene: The collisionless region,
%\href{https://journals.aps.org/prb/abstract/10.1103/PhysRevB.98.195103}
%{Phys. Rev. B \textbf{98}, 195103 (2018)}.
%
%\bibitem{Rao2020} Pranav Rao and Barry Bradlyn,
%Hall Viscosity in Quantum Systems with Discrete Symmetry: Point Group and Lattice Anisotropy,
%\href{https://journals.aps.org/prx/abstract/10.1103/PhysRevX.10.021005}
%{Phys. Rev. X \textbf{10}, 021005 (2020)}.

%\bibitem{note-1} Proof of symmetry of $\mathcal{J}'_{ij}$: $\mathcal{J}'_{ij}-\mathcal{J}'_{ji}=-(x_ip_j-x_jp_i)-\frac{i\hbar}{4}[\sigma_i,\sigma_j]=-\varepsilon^{ijk}L_k+\varepsilon^{ijk}\frac{\hbar}{2}\sigma_k=0$, since the pseudospin of graphene is commonly thought to be analogous to a spin 1/2 angular momentum.

%\bibitem{Ostrovsky2006} P. M. Ostrovsky, I. V. Gornyi, and A. D. Mirlin,
%Electron transport in disordered graphene,
%\href{https://journals.aps.org/prb/abstract/10.1103/PhysRevB.74.235443}
%{Phys. Rev. B \textbf{74}, 235443 (2006)}.










\end{thebibliography}
\end{document}

%\title{The static and dynamic shear and Hall viscosity in noninteracting graphene}
%\title{Viscous Dirac electron fluid}
\title{Supplementary Materials for ``Viscosity Enhancement by Electron-Hole Collisions in Dirac Electron Fluid''}

\author{Weiwei Chen}
\affiliation{School of Science, Westlake University, 18 Shilongshan Road, Hangzhou 310024,  China}
\affiliation{Institute of Natural Sciences, Westlake Institute for Advanced Study, 18 Shilongshan Road, Hangzhou 310024, China}

\author{W. Zhu}
%\thanks{zhuwei@westlake.edu.cn}
%\{zhuwei@westlake.edu.cn}
\affiliation{School of Science, Westlake University, 18 Shilongshan Road, Hangzhou 310024, China}
\affiliation{Institute of Natural Sciences, Westlake Institute for Advanced Study, 18 Shilongshan Road, Hangzhou 310024,  China}

\date{\today}

\maketitle

%\tableofcontents
{\let\clearpage\relax \tableofcontents} 
\thispagestyle{empty}

\clearpage
%\begin{appendix}

%%%%%%%%%%%%%%%%%%%%%%%%%%%%%%%%%%%%%%%%%%%%%%
\section{Methods}
\label{sec:methods}
\subsection{Definition of Shear and Hall Viscosity}
\label{sec:viscosity}
The viscosity, which relates the viscous stress in a fluid to the rate of change of a deformation (i.e., strain rate), is defined in a homogeneous Newtonian fluid by the following relation \cite{Avron1998},
\begin{equation}\label{eq-tau_ij}
	\tau_{ij}=\sum_{kl}\eta_{ij,kl}\frac{\partial\lambda_{kl}}{\partial t},
\end{equation}
where $\eta_{ij,kl}$ is the viscosity tensor, $\tau_{ij}$ is the stress tensor, $\lambda_{kl}=\frac{1}{2}(\frac{\partial u_k}{\partial x_l}+\frac{\partial u_l}{\partial x_k})$ is the symmetric strain tensor and $u_i$ is the deformation displacement along $i$-direction. In an isotropic system, $\tau_{ij}$ is also symmetric so that $\eta$ is symmetric under $i\leftrightarrow j$ and $k\leftrightarrow l$.
%so that one can separate the viscosity as $\eta=\eta^S+\eta^A$, where $\eta^S$ ($\eta^A$) is symmetric (anti-symmetric) under $\{ij\}\leftrightarrow\{kl\}$. 
%The strain rate can also be represented by gradients of the velocity field, $\frac{\partial\lambda_{kl}}{\partial t}=\frac{1}{2}(\frac{\partial v_k}{\partial x_l}+\frac{\partial v_l}{\partial x_k})$.
Thus, $\eta$ can be divided into symmetric and antisymmetric parts with respect to interchanging the first with the second pair of indices. Based on these features, the viscosity tensor of a two dimensional isotropic system is characterized in a natural basis $\eta=\sum\eta_{ab}\sigma_a\otimes\sigma_b$ ($a,b=0,x,z$) by three coefficients \cite{Avron1998,Pellegrino2017,Bradlyn2012,Sherafati2016}
\begin{widetext}
	\begin{equation}
		\eta_{ij,kl}=\zeta\sigma_0\otimes\sigma_0+\eta_s(\sigma_z\otimes\sigma_z+\sigma_x\otimes\sigma_x)+\eta_H(\sigma_z\otimes\sigma_x-\sigma_x\otimes\sigma_z)
		=\left(\begin{array}{cc}
			\left(\begin{array}{cc}\zeta+\eta_s&\eta_H\\\eta_H&\zeta-\eta_s\end{array}\right)_{kl}&
			\left(\begin{array}{cc}-\eta_H&\eta_s\\\eta_s&\eta_H\end{array}\right)_{kl}\\
			\left(\begin{array}{cc}-\eta_H&\eta_s\\\eta_s&\eta_H\end{array}\right)_{kl}&
			\left(\begin{array}{cc}\zeta-\eta_s&-\eta_H\\-\eta_H&\zeta+\eta_s\end{array}\right)_{kl}
		\end{array}\right)_{ij}
	\end{equation}
\end{widetext}
%Unlike the rotation generator $i\sigma_y$ in Ref.~\cite{Avron1998}, we choose $\sigma_z$ because the system is considered on the $x-y$ plane. 
where the symmetric components under the exchange ($ij\leftrightarrow kl$), $\zeta$ and $\eta_s$, denote bulk viscosity and shear viscosity, and the antisymmetric components under the exchange ($ij\leftrightarrow kl$), $\eta_H$, denotes Hall viscosity. In an incompressible fluid ($\frac{\partial \lambda_{ii}}{\partial t}=0$), the stress becomes independent of $\zeta$. Thus, the incompressible and isotropic fluids in two dimensions are characterized by two coefficients one for the even part $\eta_s$ and one for the odd part $\eta_H$. $\eta_s$ contributes to dissipation of energy, so it is also called dissipative viscosity. $\eta_H$ is dissipationless and only exist when time reversal symmetry is broken.  

\subsection{Strain Deformation, Stress Tensor}
\label{sec:strain-stress}
In the following, we derive the expression of the stress tensor in two ways. On the one hand, we associate the spatial strain transformation of the system with the unitary transformation of the Hamiltonian and derive the stress in quantum-mechanical theory. On the other hand, we simulate graphene by a  symmetric 2+1 dimensional Dirac field and obtain the stress tensor based on the Noether's theorem combining symmetrization procedure of Belinfante.

\paragraph{Quantum-mechanical theory of stress}
In quantum-mechanical theory, the stress is considered to be an intrinsic property of the quantum-mechanical ground state of matter response to deformation \cite{Nielsen1985}, so we start from using the unitary transformation in Hilbert space to describe the deformation . The Hamiltonian of the charge carriers of graphene near the half filling is described by a two-dimensional massless Dirac particles with the speed of light replaced by $v_f$,
\begin{equation}\label{Hamiltonian0}
	%H=\int d^2\bm{r}\hbar v_f\bm{\sigma}\cdot(-i\nabla)
	H=v_f\bm{\sigma}\cdot\bm{p}
\end{equation}
where $\bm{\sigma}=(\sigma_x,\sigma_y)$ are the Pauli matrices of pseudospin. In the presence of magnetic field, the Hamiltonian is changed by $\bm{p}\to\bm{\Pi}=\bm{p}+e\bm{A}$. %$\nabla\to\bm{D}=\nabla+i\frac{e}{\hbar}\bm{A}$.

the infinitesimal spatial deformation can be described as
\begin{equation}\label{eq-transformation-x}
	x_i\to x'_i=x_i+u_i(\bm{x})=u_{0,i}+\frac{\partial u_i}{\partial x_j}x_j+o(x^2)
\end{equation}
where $u_{0,i}$ corresponds to the translation so that is ignored here, and the repeated indices are summed in all cases. Then the matrix of deformation transformation ($\bm{x}'=\Lambda\bm{x}$) can be derived as \cite{Kundu2008}
\begin{equation}
	\Lambda=1+\frac{1}{2}\lambda_{ij}(x_j\partial_i+x_i\partial_j)-\frac{1}{4}r_{ij}(x_i\partial_j-x_j\partial_i)
\end{equation}
where we introduced strain tensor $\lambda_{ij}$ and rotation tensor $r_{ij}$:
\begin{equation}
	\lambda_{ij}=\frac{1}{2}(\frac{\partial u_i}{\partial x_j}+\frac{\partial u_j}{\partial x_i});\ \ \ \ r_{ij}=\frac{\partial u_i}{\partial x_j}-\frac{\partial u_j}{\partial x_i};
\end{equation}
It is noticed that the strain tensor is symmetric ($\lambda_{ij}=\lambda_{ji}$) and the rotation tensor is antisymmetric ($r_{ij}=-r_{ji}$). Based on this infinitesimal transformation, we can define two generators, strain transformation generator $\mathcal{J}_{ij}$ and rotation transformation generator $L_{ij}$ which is the well-known angular momentum:
\begin{equation}
	\mathcal{J}_{ij}=-\frac{1}{2}(x_ip_j+x_jp_i);\ \ \ \ \ \ L_{ij}=x_ip_j-x_jp_i
\end{equation}
Respectively, the strain transformation generator is symmetric ($\mathcal{J}_{ij}=\mathcal{J}_{ji}$) and the rotation transformation generator is antisymmetric ($L_{ij}=-L_{ji}$). Thus, one can parametrize the representation of spatial strain transformation $\mathcal{S}(\lambda)$ in terms of the strain transformation generator as
\begin{equation}
	\mathcal{S}(\lambda)=e^{-i\lambda_{ij}\mathcal{J}_{ij}/\hbar}
\end{equation}
Then, the deformed Hamiltonian is obtained by using time-dependent unitary operator $\mathcal{S}[\lambda(t)]$ \cite{Band}
\begin{equation}
	H_{\lambda}(t)=\mathcal{S}H\mathcal{S}^{-1}+i\hbar\frac{\partial\mathcal{S}}{\partial t}\mathcal{S}^{-1}=H-\frac{i\lambda_{ij}}{\hbar}[\mathcal{J}_{ij},H]+\frac{\partial\lambda_{ij}}{\partial t}\mathcal{J}_{ij}
\end{equation}
Thus one can obtain the integral stress tensor by the fundamental thermodynamic relation for deformed bodies \cite{Read2011,Bradlyn2012,Landau1970}
\begin{equation}\label{eq-Tij}
	T_{ij}=\int d\bm{r}\tau_{ij}=-\frac{\partial H_{\lambda}}{\partial \lambda_{ij}}=\frac{i}{\hbar}[\mathcal{J}_{ij},H]
\end{equation}
Substituting the Hamiltonian Eq.(\ref{Hamiltonian0}) into the above equation and performing Fourier transformation, the symmetric stress tensor is obtained as
\begin{equation}\label{eq-Tij-2}
	T_{ij}=\frac{v_f}{2}(\sigma_ip_j+\sigma_jp_i)
\end{equation}
which does not contain the effect from disorder potential since that one strains the electron liquid rather than the host materials \cite{Burmistrov2019}.
%However, this is not the final expression for the spin system since the generic coordinate transformation also include spatial rotation, which implies the stress generators $J_{ij}$ are related to the angular momentum. In the case of graphene, the total angular momentum also includes the pseudospin \cite{Link2018,Rao2020}. By establishing the relation between $J_{ij}$ and the angular momentum $L_{ij}$: $J_{ij}-J_{ji}=-(x_ip_j-x_jp_i)=-L_{ij}$ and considering the opposite symmetries of the orbital angular momentum ($L_{ij}=-L_{ji}$) and spin angular momentum ($S_{ij}=S_{ji}=\frac{\hbar}{4i}\epsilon^{ijk}\sigma_i\sigma_j$), the modified stress generators $\mathcal{J}_{ij}$ is derived as
%\begin{equation}
%	\mathcal{J}_{ij}=-\frac{1}{2}\{x_i,p_j\}+\frac{i\hbar}{8}[\sigma_i,\sigma_j]
%\end{equation} 
%Then, we substitute $\mathcal{J}_{ij}$ for $J_{ij}$ and covariant derivation $D_i$ for canonical derivation $\partial_i$. Finally, after integrating over the space, we get the corrected integral stress tensor 
%\begin{equation}\label{eq-Tij-2}
%	T_{ij}=\frac{v_f}{2}(\Pi_i\sigma_j+\Pi_j\sigma_i). 
%\end{equation}

In the presence of magnetic field, the spatial strain transformation should couple a gauge transformation $\mathcal{S}(\lambda)\to\mathcal{S}(\lambda)e^{-\frac{ie}{\hbar}\xi}$. Respectively, the Eq.~(\ref{eq-Tij}) becomes $T_{ij}\to\frac{i}{\hbar}e^{-\frac{ie}{\hbar}\xi}[\mathcal{J}_{ij},H]e^{\frac{ie}{\hbar}\xi}$. It is obvious that the stress tensor under magnetic field is obtained by $\bm{p}\to\bm{\Pi}=\bm{p}+e\bm{A}$.

Above, we obtained the stress tensor based on the thermodynamics of deformation, in fact, the Eq.~(\ref{eq-Tij}) is equal to the definition in a metric compatible Riemannian manifold $T_{\mu\nu}=2\frac{1}{\sqrt{|g|}}\frac{\delta S}{\delta g^{\mu\nu}}$ \cite{Blaschke2016}, where the small deformations strain tensor is replaced by the fundamental metric tensor, which have the relation $\lambda_{ij}=\frac{1}{2}(g_{ij}-\delta_{ij})$. The $\delta_{ij}$ plays the role of the fundamental metric tensor for undeformed region whose geometry corresponds to a flat Euclidean Space \cite{Palacios2006}.

Another thing worth noting is that for the same stress tensor $T_{ij}$, the strain transformation generator $\mathcal{J}_{ij}$ that satisfies the relation, $T_{ij}=\frac{i}{\hbar}[\mathcal{J}_{ij},H]$, is not unique. In some previous work \cite{Link2018,Rao2020}, scientists established a strain transformation generator including strain transformations in pseudospin of graphene as $\mathcal{J}'_{ij}=-\frac{1}{2}\{x_i,p_j\}+\frac{i\hbar}{8}[\sigma_i,\sigma_j]$, and they acquired the same stress tensor as Eq.(\ref{eq-Tij-2}). The similarity between $\mathcal{J}_{ij}$ and $\mathcal{J}'_{ij}$ is that they are both symmetric \cite{note-1}. The latter is closer to the symmetrization procedure of Belinfante in field theory, which we will simply derive in 2+1 dimensional Dirac field.

%where $S$ is action and $g^{\mu\nu}$ is metric, due to the relation between the small deformations strain tensor and the fundamental metric tensor: $\lambda_{ij}=\frac{1}{2}(g_{ij}-\delta_{ij})$, where $\delta_{ij}$ plays the role of the fundamental metric tensor for undeformed region $R$ whose geometry corresponds to a flat Euclidean Space \cite{Palacios2006}. 

\paragraph{Belinfante stress-energy-momentum tensor in field theory}
The symmetric stress tensor can also be arrived from Belinfante stress-energy-momentum tensor in field theory. We start with a symmetric Lagrangian density of a 2+1 dimensional Dirac field
\begin{equation}
	\mathcal{L}=\frac{i}{2}\hbar v_f\bar{\psi}\bar{\sigma}^{\mu}(\overrightarrow{\partial}_{\mu}-\overleftarrow{\partial}_{\mu})\psi,
\end{equation}
where $\bar{\psi}=\psi^{\dagger}\sigma_z$ is adjoint of the field $\psi$, $D_{\nu}\equiv(\partial_{v_ft},\nabla)$ is the covariant derivation in time-space coordinate, the Dirac matrices are chosen as 
\begin{equation}
	\bar{\sigma}^0=\sigma_z;\ \bar{\sigma}^1=i\sigma_y;\ \bar{\sigma}^2=-i\sigma_x;
\end{equation}
which satisfy Clifford algebra $\{\bar{\sigma}^{\mu},\bar{\sigma}^{\nu}\}=2g^{\mu\nu}$.

Based on the Noether's theorem, one can obtain the canonical stress-energy-momentum tensor \cite{Blaschke2016}:
\begin{equation}\begin{aligned}
	\tau^{\mu\nu}=&\frac{\partial\mathcal{L}}{\partial(\partial_{\mu}\psi)}\partial^{\nu}\psi+\partial^{\nu}\bar{\psi}\frac{\partial\mathcal{L}}{\partial(\partial_{\mu}\bar{\psi})}-g^{\mu\nu}\mathcal{L}\\
	=&\frac{i}{2}\hbar v_f\bar{\psi}\bar{\sigma}^{\mu}(\overrightarrow{\partial}^{\nu}-\overleftarrow{\partial}^{\nu})\psi
\end{aligned}\end{equation}
By adding the divergence of a Belinfante tensor $B^{\alpha\mu\nu}$ antisymmetric in the first two indices ($B^{\alpha\mu\nu}=-B^{\mu\alpha\nu}$):
\begin{equation}
	B^{\alpha\mu\nu}=\frac{1}{8}\hbar v_f\bar{\psi}\{\bar{\sigma}^{\alpha},\frac{i}{2}[\bar{\sigma}^{\mu},\bar{\sigma}^{\nu}]\}\psi
\end{equation}
one can obtain the Belinfante stress-energy-momentum tensor as
\begin{equation}\begin{aligned}
	\tau^{\mu\nu}_B=&T^{\mu\nu}+\partial_{\alpha}B^{\alpha\mu\nu}\\
	=&\frac{i}{4}\hbar v_f\bar{\psi}\left[\bar{\sigma}^{\mu}(\overrightarrow{\partial}^{\nu}-\overleftarrow{\partial}^{\nu})+\bar{\sigma}^{\nu}(\overrightarrow{\partial}^{\mu}-\overleftarrow{\partial}^{\mu})\right]\psi
\end{aligned}\end{equation}
Here, $\tau^{00}_B$ represents the energy density, $\tau^{0i}_B$ the momentum density (or energy flux density) and $\tau^{ij}_B$ the stress tensor. Since this stress tensor is derived for the Lagrangian density, the integral form of $\tau^{ij}_B$ is consistent to the result Eq.~(\ref{eq-Tij-2}).

\subsection{Kubo Formula of Viscosity}
\label{sec:kubo}
\begin{figure}
	\centering
	\includegraphics[width=0.6\linewidth]{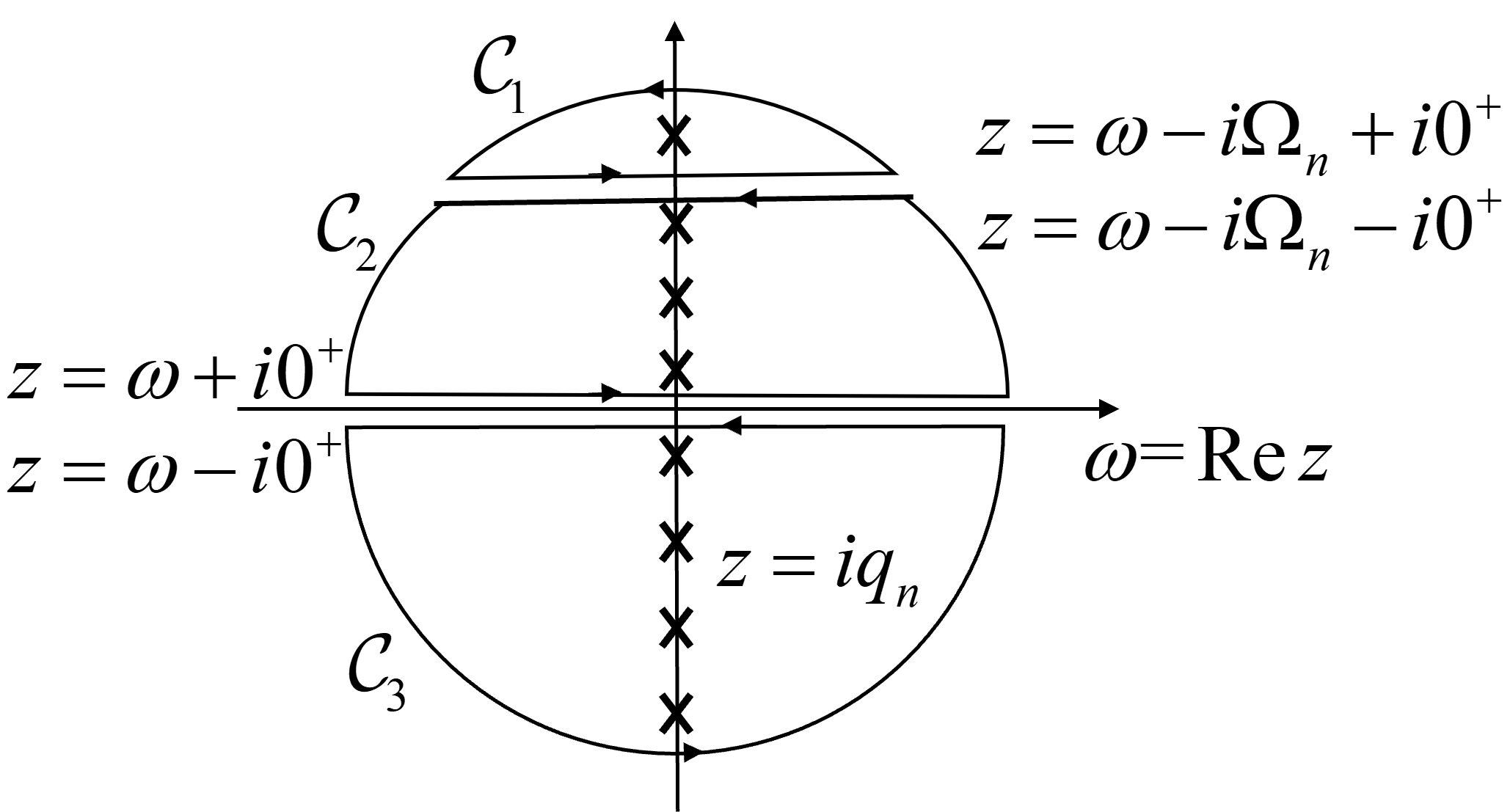}
	\caption{Contour integral with two branch cuts: $z={\rm Re}z=\omega$ and $z=\omega-i\Omega_n$.}
	\label{fig:sketch-contour}
\end{figure}

Since we have got the expression of the deformed Hamiltonian and stress tensor, the viscosity defined in Eq.~(\ref{eq-tau_ij}) can be evaluated by the linear response theory \cite{Bradlyn2012},
\begin{equation}\begin{aligned}
		\langle T_{ij}\rangle(t)
		=&-i\int^{\infty}_{-\infty}dt'\theta(t-t')\langle[T_{ij}(t),\mathcal{J}_{kl}(t')]\frac{\partial \lambda_{\alpha\beta}(t')}{\partial t'}\rangle\\
		=&\int_{-\infty}^{\infty}dt'X_{ijkl}(t-t')\frac{\partial \lambda_{\alpha\beta}(t')}{\partial t'}
\end{aligned}\end{equation}
where
\begin{equation}
	X_{ijkl}(t-t')=-i\theta(t-t')\langle[T_{ij}(t),\mathcal{J}_{kl}(t')]\rangle
\end{equation}
is stress-strain correlation function and $\langle\cdots\rangle$ means average over disorder. The Fourier transformation of it is
\begin{equation}\label{linear-viscosity-1}
	X_{ijkl}(\Omega)=i\int^{\infty}_0dt\langle[T_{ij}(t),\mathcal{J}_{kl}(0)]\rangle e^{i\Omega^+t}
\end{equation}
where $\Omega^+=\Omega+i0^+$. We can express Eq.(\ref{linear-viscosity-1}) in an additional equivalent form as stress-stress form by using time-translation invariance and the relation
\begin{equation}\begin{aligned}
		\int^{\infty}_0\mathcal{J}_{kl}(-t)e^{i\omega^+t}dt=&\int^0_{-\infty}\mathcal{J}_{kl}(t)e^{-i\Omega^+t}dt\\
		=&\frac{1}{-i\Omega^+}\left[\mathcal{J}_{kl}(0)+\int^{\infty}_{0}T_{kl}(-t)e^{i\Omega^+t}dt\right]
\end{aligned}\end{equation}
based on
\begin{equation}
	T_{ij}=\frac{i}{\hbar}[\mathcal{J}_{ij},H]=-\frac{\partial \mathcal{J}_{ij}}{\partial t}
\end{equation}
Thus the response function can rewritten as
\begin{equation}\begin{aligned}
	X_{ijkl}(\Omega)=&\frac{1}{\Omega^+}\left\{\langle[T_{ij}(0),\mathcal{J}_{kl}(0)]\rangle\right.\\
	&\left.+\int^{\infty}_0dt\langle[T_{ij}(t),T_{kl}(0)]\rangle e^{i\Omega^+t}\right\}
\end{aligned}\end{equation}
where the first term, which is called a contact term analogous to the diamagnetic conductivity, contributes to the bulk viscosity \cite{Bradlyn2012,Zakharov2021}. In this work we focus on the shear and Hall viscosity, we then obtain the viscosity to be
\begin{equation}\label{eq:kubo-1}
	\eta_{ijkl}(\Omega)=\frac{1}{\Omega^+}\int^{\infty}_0dt\langle[T_{ij}(t),T_{kl}(0)]\rangle e^{i\Omega^+t}
\end{equation}
which is called Kubo formula of viscosity. The calculation of Eq.~(\ref{eq:kubo-1}) is similar to the current-current correlation function, which have been detailed discussed in some books of many-body quantum theory \cite{Mahan,Bruus}, so here we only briefly outline some significant steps. To calculate this stress-stress correlation function, one can transform it into Matsubara function by analytical continuations: $\Omega^+\to i\Omega_n$ and $it\to\tau$,
\begin{equation}\label{eq-vis-Mat}
	\eta_{ijkl}(i\Omega_n)=\frac{1}{i}\frac{1}{i\Omega_n}\int^{\beta}_0d\tau\langle {\rm T}_{\tau}T_{ij}(\tau)T_{kl}(0)\rangle e^{i\Omega_n\tau}
\end{equation}
where the factor $\frac{1}{i}$ origins from $dt\to\frac{1}{i}d\tau$, and $\beta=\frac{1}{k_BT}$. For the disordered system, the integrand can be evaluated by the perturbation expansion as
\begin{equation}
	\langle {\rm T}_{\tau}T_{ij}(\tau)T_{kl}(0)\rangle=\frac{\langle{\rm T}_{\tau}\psi^{\dagger}(\tau)T_{ij}\psi(\tau)\psi(0)^{\dagger}T_{kl}\psi(0)U_{\beta}\rangle_0}{\langle U_{\beta}\rangle_0}
\end{equation}
where the factor $\frac{1}{i}$ origins from $dt\to\frac{1}{i}d\tau$, and $\beta=\frac{1}{k_BT}$. For the disordered system, the integrand can be evaluated by the perturbation expansion as
\begin{equation}\label{eq-correlation-tau}
	\langle {\rm T}_{\tau}T_{ij}(\tau)T_{kl}(0)\rangle=\frac{\langle{\rm T}_{\tau}\psi^{\dagger}(\tau)T_{ij}\psi(\tau)\psi(0)^{\dagger}T_{kl}\psi(0)U_{\beta}\rangle_0}{\langle U_{\beta}\rangle_0}
	=-{\rm Tr}\langle T_{ij}\mathcal{G}(\tau)T_{kl}\mathcal{G}(-\tau)\rangle
\end{equation}
where a closed fermion loop always gives a factor of -1 and the trace of a product of Dirac matrices, $\langle\cdots\rangle$ means the disorder average, 
%\begin{equation}
$U_{\beta}={\rm T}_{\tau}\left\{\exp[-\int_0^{\beta}d\tau'V(\tau')]\right\}$
%\end{equation}
is the time evolution operator in Matsubara formalism, and $\mathcal{G}$ is Matsubara Green's function. Plugging Eq.~(\ref{eq-correlation-tau}) into Eq.~(\ref{eq-vis-Mat}), we get
\begin{equation}\begin{aligned}\label{eq-vis-Mat-2}
		\eta_{ijkl}(i\Omega_n)=&-\frac{1}{i}\frac{1}{i\Omega_n\mathcal{V}}\int^{\beta}_0d\tau e^{i\Omega_n\tau}{\rm Tr}\langle T_{ij}\mathcal{G}(\tau)T_{kl}\mathcal{G}(-\tau)\rangle\\
		=&-\frac{1}{i}\frac{1}{i\Omega_n\mathcal{V}}\int^{\beta}_0d\tau\int^{\beta}_0d\tau_1\frac{1}{\beta}\sum_{iq_n}e^{-iq_n(\tau_1-\tau)} e^{i\Omega_n\tau}{\rm Tr}\langle T_{ij}\mathcal{G}(\tau)T_{kl}\mathcal{G}(-\tau_1)\rangle\\
		=&-\frac{1}{i}\frac{1}{i\Omega_n\mathcal{V}}\frac{1}{\beta}\sum_{iq_n}\int^{\beta}_0d\tau\int^{\beta}_0d\tau_1e^{-i(q_n+\Omega_n)\tau}{\rm Tr}\langle T_{ij}\mathcal{G}(\tau)T_{kl}\mathcal{G}(-\tau_1)\rangle_0e^{-iq_n\tau_1}\\
		=&-\frac{1}{i}\frac{1}{i\Omega_n\mathcal{V}}\frac{1}{\beta}\sum_{iq_n}{\rm Tr}\langle T_{ij}\mathcal{G}(iq_n+i\Omega_n)T_{kl}\mathcal{G}(iq_n)\rangle
\end{aligned}\end{equation}
where we have inserted a imaginary time integral $\int^{\beta}_0\delta(\tau_1-\tau)d\tau_1$ and changed $\delta(\tau_1-\tau)$ by a frequency summation $\frac{1}{\beta}\sum_{iq_n}e^{-iq_n(\tau_1-\tau)}$. Here, $iq_n=i(2n+1)\pi/\beta$ denotes Fermi frequency and $i\Omega_n=i2n\pi/\beta$ denotes Boson frequency. Then, we use the techniques for summations of Matsubara Green’s functions with known branch cuts. According to the summation of $iq_n$, one can introduce a contour integral 
\begin{equation}
	I=\oint_{\mathcal{C}_1+\mathcal{C}_2+\mathcal{C}_3}\frac{dz}{2\pi i}f(z)\mathcal{G}(z+i\Omega_n)\mathcal{G}(z)
\end{equation}
where $f_{\omega}=\frac{1}{\exp[\beta(\omega-E)]+1}$ is the Fermi-Dirac distribution function and the contours are shown in Fig.~\ref{fig:sketch-contour}, since for variable $z$ the branch cut (also singularity line) of $\mathcal{G}(z)$ is real axis, $z={\rm Re}z=\omega$, and the branch cut of $\mathcal{G}(z+i\Omega_n)$ is $z=\omega-i\Omega_n$. On the one hand, we calculate $I$ by residue theorem
\begin{equation}\label{eq-I-1}
	I=\sum_{iq_n}\lim_{z\to iq_n=i(2n+1)\pi/\beta}\frac{z-iq_n}{e^{iq_n\beta}+1}\mathcal{G}(iq_n+i\Omega_n)\mathcal{G}(iq_n)=-\frac{1}{\beta}\sum_{iq_n}\mathcal{G}(iq_n+i\Omega_n)\mathcal{G}(iq_n)
\end{equation}
On the other hand, we calculate $I$ by separating contour integrals
\begin{equation}\begin{aligned}\label{eq-I-2}
		I=&\int_{-\infty}^{\infty}\frac{d\omega}{2\pi i}f(z)\mathcal{G}(z+i\Omega_n)\mathcal{G}(z)\bigg|_{z=\omega-i\Omega_n+i0^+}+\int_{\infty}^{-\infty}\frac{d\omega}{2\pi i}f(z)\mathcal{G}(z+i\Omega_n)\mathcal{G}(z)\bigg|_{z=\omega-i\Omega_n-i0^+}\\
		&+\int_{-\infty}^{\infty}\frac{d\omega}{2\pi i}f(z)\mathcal{G}(z+i\Omega_n)\mathcal{G}(z)\bigg|_{z=\omega+i0^+}+\int_{\infty}^{-\infty}\frac{d\omega}{2\pi i}f(z)\mathcal{G}(z+i\Omega_n)\mathcal{G}(z)\bigg|_{z=\omega-i0^+}\\
		=&\int_{-\infty}^{\infty}\frac{d\omega}{2\pi i}f(\omega)[\mathcal{G}(\omega+i0^+)\mathcal{G}(\omega-i\Omega_n)-\mathcal{G}(\omega-i0^+)\mathcal{G}(\omega-i\Omega_n)+\mathcal{G}(\omega+i\Omega_n)\mathcal{G}(\omega+i0^+)-\mathcal{G}(\omega+i\Omega_n)\mathcal{G}(\omega-i0^+)]
\end{aligned}\end{equation}

Based on the comparison of the Eq.~(\ref{eq-I-1}) and Eq.~(\ref{eq-I-2}), we can rewrite Eq.~(\ref{eq-vis-Mat-2}) as
\begin{equation}\begin{aligned}
		\eta_{ijkl}(i\Omega_n)=&\frac{1}{i}\frac{1}{i\Omega_n\mathcal{V}}\int_{-\infty}^{\infty}\frac{d\omega}{2\pi i}f(\omega){\rm Tr}\langle T_{ij}\mathcal{G}(\omega+i0^+)T_{kl}\mathcal{G}(\omega-i\Omega_n)-T_{ij}\mathcal{G}(\omega-i0^+)T_{kl}\mathcal{G}(\omega-i\Omega_n)\\
		&+T_{ij}\mathcal{G}(\omega+i\Omega_n)T_{kl}\mathcal{G}(\omega+i0^+)-T_{ij}\mathcal{G}(\omega+i\Omega_n)T_{kl}\mathcal{G}(\omega-i0^+)\rangle
\end{aligned}\end{equation}
Finally, we do the analytical continuity $i\Omega_n\to\Omega+i0^+$ and get
\begin{equation}\begin{aligned}
		\eta_{ijkl}(\Omega)=&\frac{1}{i}\frac{1}{\Omega\mathcal{V}}\int_{-\infty}^{\infty}\frac{d\omega}{2\pi i}f_{\omega}{\rm Tr}\langle T_{ij}\mathcal{G}(\omega+i0^+)T_{kl}\mathcal{G}(\omega-\Omega-i0^+)-T_{ij}\mathcal{G}(\omega-i0^+)T_{kl}\mathcal{G}(\omega-\Omega-i0^+)\\
		&+T_{ij}\mathcal{G}(\omega+\Omega+i0^+)T_{kl}\mathcal{G}(\omega+i0^+)-T_{ij}\mathcal{G}(\omega+\Omega+i0^+)T_{kl}\mathcal{G}(\omega-i0^+)\rangle\\
		=&-\frac{1}{\Omega\mathcal{V}}\int_{-\infty}^{\infty}\frac{d\omega}{2\pi}f_{\omega}{\rm Tr}\langle T_{ij}G^R(\omega)T_{kl}G^A(\omega-\Omega)-T_{ij}G^A(\omega)T_{kl}G^A(\omega-\Omega)\\
		&+T_{ij}G^R(\omega+\Omega)T_{kl}G^R(\omega)-T_{ij}G^R(\omega+\Omega)T_{kl}G^A(\omega)\rangle\\
		=&-\frac{1}{\Omega\mathcal{V}}{\rm Tr}\int_{-\infty}^{\infty}\frac{d\omega}{2\pi}f_{\omega+\Omega}\langle T_{ij}G^R(\omega+\Omega)T_{kl}G^A(\omega)-T_{ij}G^A(\omega+\Omega)T_{kl}G^A(\omega)\rangle\\
		&+f_{\omega}\langle T_{ij}G^R(\omega+\Omega)T_{kl}G^R(\omega)-T_{ij}G^R(\omega+\Omega)T_{kl}G^A(\omega)\rangle\\
		=&-\frac{1}{\Omega\mathcal{V}}{\rm Tr}\int_{-\infty}^{\infty}\frac{d\omega}{2\pi}(f_{\omega+\Omega}-f_{\omega})\langle T_{ij}G^R_{\omega+\Omega}T_{kl}G^A_{\omega}-T_{ij}G^A_{\omega+\Omega}T_{kl}G^A_{\omega}\rangle+f_{\omega}\langle T_{ij}G^R_{\omega+\Omega}T_{kl}G^R_{\omega}-T_{ij}G^A_{\omega+\Omega}T_{kl}G^A_{\omega}\rangle\\
\end{aligned}\end{equation}

In the limit $\Omega\to0$, $\eta_{ijkl}(\Omega)$ can be reduced into two different forms under shear and Hall conditions. In the case of shear viscosity, we reserve the real part of $\eta_{ijkl}(\Omega)$ and choose $T_{ij}=T_{kl}=T_{xy}$ so that
\begin{equation}\begin{aligned}\label{shear-1}
		\eta_s(\Omega\to0)=&-\lim_{\Omega\to0}\frac{\hbar}{\Omega\mathcal{V}}{\rm ReTr}\int^{\infty}_{-\infty}\frac{d\omega}{2\pi }[(f_{\omega+\Omega}-f_{\omega})(T_{xy}G^R_{\omega+\Omega}T_{xy}G^A_{\omega}-T_{xy}G^A_{\omega+\Omega}T_{xy}G^A_{\omega})\\
		&+f_{\omega}(T_{xy}G^R_{\omega+\Omega}T_{xy}G^R_{\omega}-T_{xy}G^A_{\omega+\Omega}T_{xy}G^A_{\omega})]\\
		%	=&-\lim_{\Omega\to0}\frac{\hbar}{\Omega\mathcal{V}}{\rm ReTr}\int^{\infty}_{-\infty}\frac{d\omega}{2\pi }[(f_{\omega+\Omega}-f_{\omega})(T_{xy}G^A_{\omega+\Omega}T_{xy}G^R_{\omega})-f_{\omega+\Omega}T_{xy}G^A_{\omega+\Omega}T_{xy}G^A_{\omega}+f_{\omega}T_{xy}G^R_{\omega+\Omega}T_{xy}G^R_{\omega}]
\end{aligned}\end{equation}
Based on the relation
\begin{equation}
	{\rm ReTr}[T_{xy}G^A_{\omega+\Omega}T_{xy}G^A_{\omega}]={\rm ReTr}[T_{xy}G^R_{\omega+\Omega}T_{xy}G^R_{\omega}]
\end{equation}
the term in the second line of Eq.~(\ref{shear-1}) is canceled, and Eq.~(\ref{shear-1}) can be rewritten as
\begin{equation}\begin{aligned}\label{shear-2}
		\eta_s(\Omega\to0)
		%=&-\lim_{\Omega\to0}\frac{\hbar}{\Omega\mathcal{V}}{\rm ReTr}\int^{\infty}_{-\infty}\frac{d\omega}{2\pi }[(f_{\omega+\Omega}-f_{\omega})(T_{xy}G^A_{\omega+\Omega}T_{xy}G^R_{\omega}-T_{xy}G^A_{\omega+\Omega}T_{xy}G^A_{\omega})\\
		%&+f_{\omega}(T_{xy}G^R_{\omega+\Omega}T_{xy}G^R_{\omega}-T_{xy}G^A_{\omega+\Omega}T_{xy}G^A_{\omega})]\\
		=&-\lim_{\Omega\to0}\frac{\hbar}{\mathcal{V}}{\rm ReTr}\int^{\infty}_{-\infty}\frac{d\omega}{2\pi }\frac{f_{\omega+\Omega}-f_{\omega}}{\Omega}(T_{xy}G^R_{\omega+\Omega}T_{xy}G^A_{\omega}-T_{xy}G^R_{\omega+\Omega}T_{xy}G^R_{\omega})
\end{aligned}\end{equation}
For the zero temperature, $\lim_{\Omega\to0}\frac{f_{\omega+\Omega}-f_{\omega}}{\Omega}=-\delta(\omega-E)$ where $E$ is Fermi energy, the above expression can further reduced as
\begin{equation}\begin{aligned}\label{shear-3}
		\eta_s(\Omega\to0)=&\frac{\hbar}{2\pi\mathcal{V}}{\rm ReTr}[T_{xy}G^R(E)T_{xy}G^A(E)-T_{xy}G^R(E)T_{xy}G^R(E)]
\end{aligned}\end{equation}
This is the Eq.(1) in the main paper. 

On the other hand, in the case of Hall viscosity, we reserve the real part of $\eta_{ijkl}(\Omega)$ and choose $T_{ij}=T_{xy}$, $T_{kl}=\frac{1}{2}(T_{xx}-T_{yy})$, so that
\begin{equation}\begin{aligned}\label{Hall-1}
		\eta_H(\Omega\to0)=&-\lim_{\Omega\to0}\frac{\hbar}{\Omega\mathcal{V}}{\rm ReTr}\int^{\infty}_{-\infty}\frac{d\omega}{2\pi }\{[(f_{\omega+\Omega}-f_{\omega})(T_{xy}G^R_{\omega+\Omega}\frac{T_{xx}-T_{yy}}{2}G^A_{\omega}-T_{xy}G^A_{\omega+\Omega}\frac{T_{xx}-T_{yy}}{2}G^A_{\omega})\\
		&+f_{\omega}(T_{xy}G^R_{\omega+\Omega}\frac{T_{xx}-T_{yy}}{2}G^R_{\omega}-T_{xy}G^A_{\omega+\Omega}\frac{T_{xx}-T_{yy}}{2}G^A_{\omega})]\}\\
		=&-\lim_{\Omega\to0}\frac{\hbar}{\Omega\mathcal{V}}{\rm ReTr}\int^{\infty}_{-\infty}\frac{d\omega}{4\pi }\{(f_{\omega+\Omega}-f_{\omega})[G^R_{\omega+\Omega}(T_{xx}-T_{yy})G^A_{\omega}T_{xy}-G^R_{\omega}(T_{xx}-T_{yy})G^R_{\omega+\Omega}T_{xy}]\\
		&+f_{\omega}[G^R_{\omega+\Omega}(T_{xx}-T_{yy})G^R_{\omega}T_{xy}-G^R_{\omega}(T_{xx}-T_{yy})G^R_{\omega+\Omega}T_{xy}]\}\\
		=&-\lim_{\Omega\to0}\frac{\hbar}{\mathcal{V}}{\rm ReTr}\int^{\infty}_{-\infty}\frac{d\omega}{4\pi }\{\frac{f_{\omega+\Omega}-f_{\omega}}{\Omega}[G^R_{\omega+\Omega}(T_{xx}-T_{yy})G^A_{\omega}T_{xy}-G^R_{\omega}(T_{xx}-T_{yy})G^R_{\omega+\Omega}T_{xy}]\\
		&-f_{\omega}[G^R_{\omega}(T_{xx}-T_{yy})\frac{G^R_{\omega+\Omega}-G^R_{\omega}}{\Omega}T_{xy}-\frac{G^R_{\omega+\Omega}-G^R_{\omega}}{\Omega}(T_{xx}-T_{yy})G^R_{\omega}T_{xy}]\}\\
\end{aligned}\end{equation}
where we have used the relations
\begin{equation}
	{\rm ReTr}[T_{xy}G^A_{\omega+\Omega}(T_{xx}-T_{yy})G^A_{\omega}]={\rm ReTr}[G^R_{\omega}(T_{xx}-T_{yy})G^R_{\omega+\Omega}T_{xy}]
\end{equation}
For the zero temperature, $\lim_{\Omega\to0}\frac{f_{\omega+\Omega}-f_{\omega}}{\Omega}=-\delta(\omega-E)$ where $E$ is Fermi energy, the Eq.~(\ref{Hall-1}) can further reduced as
\begin{equation}\begin{aligned}\label{Hall-2}
		\eta_H(\Omega\to0)=&\frac{\hbar}{4\pi\mathcal{V}}{\rm Re}\left\{{\rm Tr}[G^R(E)(T_{xx}-T_{yy})G^A(E)T_{xy}-G^R(E)(T_{xx}-T_{yy})G^R_{\omega+\Omega}T_{xy}]\right.\\
		&+\left.\int d\omega f_{\omega}[G^R_{\omega}(T_{xx}-T_{yy})\frac{dG^R_{\omega}}{d\omega}T_{xy}-\frac{dG^R_{\omega}}{d\omega}(T_{xx}-T_{yy})G^R_{\omega}T_{xy}]\right\}\\
\end{aligned}\end{equation}
which is the Eq.(9) in the main paper.

\subsection{Eigenbasis of the Pure System and Disorder-induced Self-energy}
\label{sec:eigen-self}
In the absence of magnetic field, the eigenvalues and eigenstates of the pure graphene are
\begin{equation}
	E_{\bm{k}s}=s\hbar v_f k
\end{equation}
\begin{equation}
	\Psi_{\bm{k}s}(\bm{r})=\langle \bm{r}|\bm{k}s\rangle=
	\frac{e^{i\bm{k}\cdot\bm{r}}}{\sqrt{2A}}\left(\begin{array}{c}
		1\\se^{i\theta_{\bm{k}}}
	\end{array}
	\right)
\end{equation}
where $s=\pm$ denotes the chiral and $A$ is the area of sample. The basis formed by $\Psi_{\bm{k}s}$ is called ($\bm{k},s$)-basis. In this basis, the components of stress tensor $T_{xy}$ and $T_{xx}-T_{yy}$, which will be used in the calculation of shear and Hall viscosity, are written as
\begin{equation}
	T_{xy}(\bm{k})=\frac{\hbar v_fk}{2}(\sigma_z\sin2\theta_{\bm{k}}-\sigma_y\cos2\theta_{\bm{k}})
\end{equation}
\begin{equation}
	T_{xx}(\bm{k})-T_{yy}(\bm{k})=\hbar v_fk(\sigma_z\cos2\theta_{\bm{k}}+\sigma_y\sin2\theta_{\bm{k}})
\end{equation}

Then, we consider the graphene in the presence of a magnetic field perpendicular to the graphene, $\bm{B}=B\hat{z}$. The corresponding vector potential is given by $\bm{A}=Bx\hat{y}$ satisfying Landau gauge. The eigenenergy and eigenstates of the Hamiltonian in the magnetic field are 
\begin{equation}
	E_{n,k_y,s}=s\hbar\omega_c\sqrt{n};\ \ \ \ s=\pm1;\ \ \ \ n=0,1,2,\cdots
\end{equation}
and
\begin{equation}
	\Psi_{n,k_y,s}(x,y)=\frac{e^{ik_yy}}{\sqrt{L_y}}\left\{\begin{array}{ll}
		\left(\begin{array}{cc}
			0,&\phi_{0,k_y}
		\end{array}\right)^T;&n=0\\
		\frac{1}{\sqrt{2}}\left(\begin{array}{cc}
			-is\phi_{n-1,k_y},&\phi_{n,k_y}
		\end{array}\right)^T;&n\neq0
	\end{array}
	\right.
\end{equation}
with
\begin{equation}
		\phi_{n,k_y}(x)=\sqrt{\frac{1}{2^nn!\sqrt{\pi}l_B}}e^{-(\frac{x}{l_B}+l_Bk_y)^2/2}H_{n}(\frac{x}{l_B}+l_Bk_y)
\end{equation}
where $l_B=\sqrt{\hbar /eB}$ is magnetic length, $\omega_c=\sqrt{2}v_f/l_B$ is cyclotron frequency. The basis formed by $\Psi_{n,k_y,s}$ is called ($n,k_y,s$)-basis. In this basis, the components of stress tensor $T_{xy}$ and $T_{xx}-T_{yy}$ are written as
\begin{equation}\begin{aligned}\label{eq:Txy-Bn0}
		&\langle n,k_y,s|T_{xy}|n',k_y',s'\rangle\\
		=&\delta_{k_y,k_y'}\left\{\begin{array}{cc}
			0;&n=n'=0\\
			-is'\frac{\hbar\omega_c}{2\sqrt{2}}\delta_{0,n'-2};&n=0,n'\neq0\\
			is\frac{\hbar\omega_c}{2\sqrt{2}}\delta_{0,n-2};&n\neq0,n'=0\\
			\frac{\hbar\omega_c}{4}(is\sqrt{n-1}\delta_{n,n'+2}-is'\sqrt{n+1}\delta_{n,n'-2});&n,n'\neq0
		\end{array}\right.
\end{aligned}\end{equation}
\begin{equation}\begin{aligned}\label{eq:Txx-Tyy-Bn0}
		&\langle n,k_y,s|T_{xx}-T_{yy}|n',k_y',s'\rangle\\
		=&\delta_{k_y,k_y'}\left\{\begin{array}{cc}
			0;&n=n'=0\\
			-s'\frac{\hbar\omega_c}{\sqrt{2}}\delta_{0,n'-2};&n=0,n'\neq0\\
			-s\frac{\hbar\omega_c}{\sqrt{2}}\delta_{n-2,0};&n\neq0,n'=0\\
			-(s\sqrt{n-1}\delta_{n',n-2}+s'\sqrt{n+1}\delta_{n',n+2});&n,n'\neq0
		\end{array}\right.
\end{aligned}\end{equation}
which can be easily derived by the second quantization shown in Appendix.~\ref{appendix-eigen}.

The disorder $V(\bm{r})$ we considered is a short-range random potential having form in the $\bm{r}$-space as
\begin{equation}
	V(\bm{r})=\sum_{i}^{N_i}V_i\delta(\bm{r}-\bm{r}_i)
\end{equation}
with random strength distribution satisfying $\overline{V(\bm{r})}=0$ and $\overline{V(\bm{r})V(\bm{r}')}=n_iV_0^2\delta(\bm{r}-\bm{r}')$, where $\overline{\cdots}$ stands for averaging over disorder realizations, $V_0^2$ is the variance of impurity strength, and $n_i=N_i/\mathcal{V}$ is the concentration of impurity. In the following, we show the analytic expression of self-energy function based on self-consistent Born approximation (SCBA) (details in Appendix~\ref{appendix-self}), and express the density of state in term of self-energy.

\paragraph{In the absence of magnetic field $B=0$} At first, we evaluate the self-energy with SCBA in the absence of magnetic field 
\begin{equation}\label{eq:self-B0-1}
	\Sigma(E)=\frac{2(\hbar v_f)^2}{A}\int dkk\frac{E-\Sigma}{(E-\Sigma)^2-(\hbar v_fk)^2}
\end{equation}
where $A=\frac{4\pi(\hbar v_f)^2}{n_iV^2_0}$ is the dimensionless parameter characterizing the scattering strength. The solution of the above self-consistent equation in weak disorder limit is given by\cite{Ostrovsky2006}
\begin{equation}\label{eq:self-B0}
	{\rm Im}\Sigma(E)=-E_c e^{-A/2}-\frac{\pi}{A}|E|
\end{equation}
%\begin{equation}
%	\Sigma(E)=\left\{\begin{array}{ll}-i\Gamma_0-\frac{E}{\alpha},&|E|\ll\Gamma_0\\-\alpha E\ln\frac{E_c}{|E|}-i\frac{\pi\alpha |E|}{2}\left[1+2\alpha\ln\frac{E_c}{|E|}\right],&|E|\gg\Gamma_0\end{array}\right.
%\end{equation}
where $E_c\approx7.2eV$ is the energy cutoff. The density of states in terms of self-energy is given by
\begin{equation}\begin{aligned}\label{eq:density of states-B0}
		\rho(E)=-\frac{2A}{\pi^2(\hbar v_f)^2}{\rm Im}\Sigma(E)
\end{aligned}\end{equation}

\paragraph{In the presence of magnetic field $B\neq0$} In the presence of magnetic field, we write the SCBA equation of self-energy in Landau quantized ($n,k_y,s$)-basis,
\begin{equation}\label{eq:self-Bneq0-1}
	\Sigma(E)=\overline{\langle n,k_y,s|VG(E)V|n,k_y,s\rangle}=\frac{(\hbar \omega_c)^2}{2A}\sum_{ns}G_{ns}(E)
\end{equation}
where $G_{ns}(E)=(E-s\sqrt{n}\hbar\omega_c-\Sigma)^{-1}$ is the Green's function in the ($n,k_y,s$)-basis. $\omega_c=\sqrt{2}v_f/l_B$ is cyclotron frequency, $l_B=\sqrt{\hbar/eB}$ is magnetic length. Similar to the case of 2DEG, we consider the graphene in the presence of magnetic field in two classes: ``Well Separated" region, where the Landau levels are well separated with each other, and ``Overlapped" region, where the Landau subbands broaden and overlap with each other due to the disorder scattering. While the criterion distinguishing these two regions in 2DEG is the product of the cyclotron frequency, which is determined by the magnetic field, and the relaxation time, which is determined by the disorder strength, in graphene, due to the uneven distribution of Landau levels, it should also consider the location of the Fermi level. Thus, we introduce an effective cyclotron frequency $\tilde{\omega}_c=\frac{\hbar\omega_c^2}{2|E|}$, which can tend to the expression of cyclotron frequency in 2DEG, i.e. $\tilde{\omega}_c=\frac{eB}{m}$, by using effective mass $m=\frac{2|E|}{v_f^2}$. Then the ``Well Separated" region and ``Overlapped" region are divided by $\tilde{\Omega}_c\tau\gg1$ and $\tilde{\Omega}_c\tau\lesssim1$

%by the value of $\tilde{\omega}_c\tau$: separated region ($\tilde{\omega}_c\tau\gg1$) and overlapped region ($\tilde{\omega}_c\lesssim1$). Here $\tilde{\omega}_c=\frac{\hbar\omega_c^2}{2|E|}$ is an effective cyclotron frequency, which can be written in the form $\frac{eB}{m}$ by introducing 

%Scattering can destroy and modify these cyclotron motions, leading to Landau subband broadening.

In the region of well separated Landau levels, we get
\begin{equation}\label{eq:self-separated}
	{\rm Im}\Sigma(E)=-\hbar\omega_c\sqrt{\frac{1}{2A}-\varepsilon^2}
\end{equation}
where the Fermi energy $E$ is assumed locating close to the Landau level $E_{NS}$ and the distance is characterized by the $\varepsilon=(E-E_{NS})/2\hbar\omega_c$. 

In the region of overlapped Landau levels,
\begin{equation}\label{eq:self-overlap}
	{\rm Im}\Sigma(E)=-E_ce^{-A/2}-\frac{(\hbar\omega_c)^2}{2E_ce^{-A/2}}-\frac{\pi}{A}|E|\left[1+2\delta\cos\frac{\pi E}{\hbar\tilde{\omega}_c}\right]
\end{equation}
%\begin{equation}\label{eq:self-overlapped}
%	{\rm Im}\Sigma=\left\{\begin{array}{ll}-\left[E_ce^{-A/2}+\frac{(\hbar\omega_c)^2}{2E_c}e^{A/2}\right];&|E|\ll|{\rm Im}\Sigma|\\-\frac{\pi}{A}E\left[1+2\delta\cos\frac{\pi E}{\hbar\tilde{\omega}_c}\right];&|E|\gg|{\rm Im}\Sigma|\end{array}\right.
%\end{equation}
where $\delta=e^{-\frac{4\pi^2E^2}{A(\hbar\omega_c)^2}}$.

In this condition, the density of states is expressed as
\begin{equation}\label{eq:density of states}
		\rho(E)=-\frac{2}{\pi^2l_B^2}\frac{2A}{(\hbar\omega_c)^2}{\rm Im}\Sigma(E)
\end{equation}

\section{Second Quantization of Graphene in the Presence of Magnetic Field}\label{appendix-eigen}
One can second quantized the Hamiltonian of graphene in the presence of magnetic field as
\begin{equation}
	H=-i\int d\bm{r}\frac{\hbar\omega_c}{2}(b_-\sigma^+-b_+\sigma^-)
\end{equation}
where $b_{\pm}=\frac{1}{\sqrt{2}}(\mp\partial_{\xi}+\xi)$ is ladder operator, $l_B=\sqrt{\hbar c/eB}$ is magnetic length, $\omega_c=\sqrt{2}v_f/l_B$ is cyclotron frequency, $\xi=x/l_B+l_Bk_y$ dimensionless length scale, and $\sigma^{\pm}=\sigma_x\pm i\sigma_y$. The eigenenergy and eigenstates of the Hamiltonian in the magnetic field are 
\begin{equation}
	E_{n,k_y,s}=s\hbar\omega_c\sqrt{n};\ \ \ \ s=\pm1;\ \ \ \ n=0,1,2,\cdots
\end{equation}
and
\begin{equation}\label{eq:app-eigenstates-1}
	\Psi_{n,k_y,s}(x,y)=\frac{e^{ik_yy}}{\sqrt{L_y}}\left\{\begin{array}{ll}
		\left(\begin{array}{cc}
			0,&\phi_{0,k_y}
		\end{array}\right)^T;&n=0\\
		\frac{1}{\sqrt{2}}\left(\begin{array}{cc}
			-is\phi_{n-1,k_y},&\phi_{n,k_y}
		\end{array}\right)^T;&n\neq0
	\end{array}
	\right.
\end{equation}
where the $\phi_{n,k_y}$ satisfies
\begin{equation}\label{eq:app-eigenstates-2}
	b_+\phi_{n,k_y}=\sqrt{n+1}\phi_{n+1,k_y};\ \ b_-\phi_{n,k_y}=\sqrt{n}\phi_{n-1,k_y};\ \ b_-\phi_{0,k_y}=0
\end{equation}

In previous section, we have got the stress tensor of graphene with nonzero magnetic field
\begin{equation}
	T_{ij}=\frac{v_f}{2}(\Pi_i\sigma_j+\Pi_j\sigma_i);\ \ \ \ \bm{\Pi}=\bm{p}+\frac{e}{c}\bm{A}.
\end{equation}
Here, we shall rewrite them in the ($n,k_y,s$)-basis. At first, we express $T_{xy}$ in terms of ladder operators $b_{\pm}$ as
\begin{equation}\begin{aligned}
		T_{xy}=&\frac{v_f}{2}[p_x\sigma_y+(p_y+\frac{eB}{c}x)\sigma_x]=\frac{\hbar v_f}{2}[-i\partial_x\sigma_y+(k_y+\frac{x}{l_B^2})\sigma_x]\\
		=&\frac{\hbar\omega_c}{4}[-i(b_--b_+)\sigma_y+(b_-+b_+)\sigma_x]\\
		=&\frac{\hbar\omega_c}{2}\left(\begin{array}{cc}0&b_+\\b_-&0\end{array}\right)
\end{aligned}\end{equation}
The components of stress tensor $\langle n,k_y,s|T_{xy}|n',k_y',s'\rangle$ can be separated into three classes:

(i) $n=n'=0$;
\begin{equation}\begin{aligned}
		\langle 0,k_y|T_{xy}|0,k_y'\rangle=&\delta_{k_y,k_y'}\frac{\hbar\omega_c}{2}\left(\begin{array}{cc}0&\langle\phi_{0,k_y}|\end{array}\right)\left(\begin{array}{c}|\phi_{1,k_y}\rangle\\0\end{array}\right)\\
		=&0
\end{aligned}\end{equation}

(ii) $n=0,n'\neq0$ or $n\neq0,n'=0$;
\begin{equation}\begin{aligned}
		\langle nk_ys|T_{xy}|0k_y'\rangle=&\delta_{k_y,k_y'}\frac{\hbar\omega_c}{2\sqrt{2}}\left(\begin{array}{cc}is\langle\phi_{n-1,k_y}|&\langle\phi_{nk_y}|\end{array}\right)\left(\begin{array}{c}|\phi_{1k_y'}\rangle\\0\end{array}\right)\\
		=&is\frac{\hbar\omega_c}{2\sqrt{2}}\delta_{0,n-2}\delta_{k_y,k_y'}
\end{aligned}\end{equation}
Similarly,
\begin{equation}\begin{aligned}
		\langle 0k_y|T_{xy}|n'k_y's'\rangle=-is'\frac{\hbar\omega_c}{2\sqrt{2}}\delta_{0,n'-2}\delta_{k_y,k_y'}
\end{aligned}\end{equation}

(iii) $n\neq0$ and $n'\neq0$;
\begin{equation}\begin{aligned}
		&\langle n,k_y,s|T_{xy}|n',k_y',s'\rangle\\
		=&\delta_{k_y,k_y'}\frac{\hbar\omega_c}{4}\left(\begin{array}{cc}is\langle\phi_{n-1,k_y}|&\langle\phi_{n,k_y}|\end{array}\right)\left(\begin{array}{c}\sqrt{n'+1}|\phi_{n'+1,k_y'}\rangle\\-is'\sqrt{n'-1}|\phi_{n'-2,k_y'}\rangle\end{array}\right)\\
		=&\frac{\hbar\omega_c}{4}(is\sqrt{n-1}\delta_{n,n'+2}-is'\sqrt{n+1}\delta_{n,n'-2})\delta_{k_y,k_y'}
\end{aligned}\end{equation}

In a same way, we can express $T_{xx}$ and $T_{yy}$ in terms of $b_{\pm}$ as
\begin{equation}\begin{aligned}
		T_{xx}=-i\hbar v_f\partial_x\sigma_x=-i\frac{\hbar v_f}{\sqrt{2}l_B}(b_--b_+)\sigma_x
\end{aligned}\end{equation}
\begin{equation}\begin{aligned}
		T_{yy}=v_f(p_y+\frac{eB}{c}x)\sigma_y=\frac{\hbar v_f}{\sqrt{2}l_B}(b_-+b_+)\sigma_y
\end{aligned}\end{equation}
Thus
\begin{equation}
	T_{xx}-T_{yy}=i\hbar\omega_c\left(\begin{array}{cc}&b_+\\-b_-&\end{array}\right)
\end{equation}

Based the expression of eigenstates Eq.~(\ref{eq:app-eigenstates-1}) and the relations Eq.~(\ref{eq:app-eigenstates-2}), the $\langle n,k_y,s|T_{xx}-T_{yy}|n',k_y',s'\rangle$ can be given in three classes:

(i) $n=n'=0$; 
\begin{equation}
\langle 0,k_y|T_{xx}-T_{yy}|0,k_y'\rangle=0.
\end{equation}

(ii) $n=0,n'\neq0$ or $n\neq0,n'=0$;
\begin{equation}\begin{aligned}
		\langle0,k_y|T_{xx}-T_{yy}|n',k_y',s'\rangle=-s'\frac{\hbar\omega_c}{\sqrt{2}}\delta_{0,n'-2}\delta_{k_y,k_y'}
\end{aligned}\end{equation}

\begin{equation}\begin{aligned}
		\langle n,k_y,s|T_{xx}-T_{yy}|0,k_y'\rangle=-s\frac{\hbar\omega_c}{\sqrt{2}}\delta_{n-2,0}\delta_{k_y,k_y'}
\end{aligned}\end{equation}

(iii) $n\neq0$ and $n'\neq0$;
\begin{equation}\begin{aligned}
		&\langle n,k_y,s|T_{xx}-T_{yy}|n',k_y',s'\rangle\\
		=&-\frac{\hbar\omega_c}{2}(s\sqrt{n-1}\delta_{n',n-2}+s'\sqrt{n+1}\delta_{n',n+2})\delta_{k_y,k_y'}
\end{aligned}\end{equation}

\section{SCBA of Self-Energy}
\label{appendix-self}
\subsection{$B=0$}
In the absence of magnetic field, the self-energy function based on self-consistent Born approximation (SCBA) is evaluated as
\begin{equation}\begin{aligned}
		\Sigma(E)=&\frac{n_iV^2_0}{2}\sum_{s}\int\frac{d^2\bm{k}}{(2\pi)^2}G(\bm{k}s,E)
		=\frac{n_iV^2_0}{2}\sum_{s}\int\frac{d^2\bm{k}}{(2\pi)^2}\frac{1}{E-E_{\bm{k}s}-\Sigma}
		=\frac{n_iV^2_0}{2}\int\frac{d^2\bm{k}}{(2\pi)^2}\frac{1}{E-\hbar v_fk-\Sigma}+\frac{1}{E+\hbar v_fk-\Sigma}\\
		=&n_iV^2_0\int\frac{d^2\bm{k}}{(2\pi)^2}\frac{E-\Sigma}{(E-\Sigma)^2-(\hbar v_fk)^2}
		=\frac{n_iV^2_0}{2\pi}\int dkk\frac{E-\Sigma}{(E-\Sigma)^2-(\hbar v_fk)^2}\\
		=&\frac{2(\hbar v_f)^2}{A}\int dkk\frac{E-\Sigma}{(E-\Sigma)^2-(\hbar v_fk)^2}
\end{aligned}\end{equation}
Since self-energy independent with momentum, it can be integrated as
\begin{equation}
	\Sigma(E)=-\frac{(E-\Sigma)}{A}\ln\frac{-E_c^2}{(E-\Sigma)^2}
\end{equation} 
Then, we assume ${\rm Re}\Sigma(E)\to0$, and evaluate ${\rm Im}\Sigma$ in two limits. 

For $|E|\ll|{\rm Im}\Sigma|$,
\begin{equation}
	{\rm Im}\Sigma\approx\frac{{\rm Im}\Sigma}{A}\ln\frac{-E_c^2}{-{\rm Im}\Sigma^2}\ \ \Longrightarrow \ \ {\rm Im}\Sigma=-E_ce^{-A/2}
\end{equation}

For $|E|\gg|{\rm Im}\Sigma|$,
\begin{equation}\begin{aligned}
		\Sigma\approx&-\frac{E}{A}\ln\frac{-E_c^2}{E^2+i{\rm sgn}(E)0^+}=-\frac{E}{A}\ln\frac{E_c^2}{E^2e^{-i\pi{\rm sgn}(E)}}
		=-\frac{E}{A}\left[2\ln\frac{E_c}{|E|}+i\pi{\rm sgn}(E)\right]
\end{aligned}\end{equation}
\begin{equation}
	\Longrightarrow {\rm Im}\Sigma=-\frac{\pi}{A}|E|
\end{equation}
Combining the results in two limits, we can get
\begin{equation}
	{\rm Im}\Sigma=-E_ce^{-A/2}-\frac{\pi}{A}|E|
\end{equation}

\subsection{$\textbf{B}=B\hat{z}$}
Then, we evaluate the self-energy in the presence of magnetic with SCBA,
\begin{equation}\begin{aligned}
		\Sigma_{n,k_y,s}(E)=&\overline{\langle n,k_y,s|VG(E)V|n,k_y,s\rangle}=\sum_{n',k_y',s'}\overline{|\langle n,k_y,s|V|n',k_y',s'\rangle|^2}G_{n',k_y',s'}(E)\\
		=&n_iV_0^2\sum_{n',s'}G_{n',s'}(E)\left[\sum_{k_y'}\int d\bm{r}\Psi_{n,k_y,s}^{\dagger}(\bm{r})\Psi_{n',k_y',s'}(\bm{r})\Psi_{n',k_y',s'}^{\dagger}(\bm{r})\Psi_{n,k_y,s}(\bm{r})\right]\\
		%=&n_iV_0^2\sum_{n',s'}G_{n',s'}(E)\left[\frac{1}{L_y}\sum_{k_y'}\int dx\Psi_{n,k_y,s}^{\dagger}(x)\Psi_{n',k_y',s'}(x)\Psi_{n',k_y',s'}^{\dagger}(x)\Psi_{n,k_y,s}(x)\right]\\
		=&n_iV_0^2\sum_{n',s'}G_{n',s'}(E)\left[\int\frac{dk_y'}{2\pi}\int dx\Psi_{n,k_y,s}^{\dagger}(x)\Psi_{n',k_y',s'}(x)\Psi_{n',k_y',s'}^{\dagger}(x)\Psi_{n,k_y,s}(x)\right]\\
		=&n_iV_0^2\sum_{n',s'}G_{n',s'}(E)\left[\int\frac{dx'}{2\pi l_B^2}\int dx\Psi_{n,k_y,s}^{\dagger}(x)\Psi_{n',k_y',s'}(x')\Psi_{n',k_y',s'}^{\dagger}(x')\Psi_{n,k_y,s}(x)\right]\\
		=&\frac{n_iV_0^2}{2\pi l_B^2}\sum_{n',s'}G_{n',s'}(E)\int dxdx'
		\frac{1}{4}\left(\begin{array}{cc}is\phi_{n-1}(x)&\phi_n(x)\end{array}\right)\left(\begin{array}{c}-is'\phi_{n'-1}(x')\\\phi_{n'}(x')\end{array}\right)\left(\begin{array}{cc}is'\phi_{n'-1}(x')&\phi_{n'}(x')\end{array}\right)\left(\begin{array}{c}-is\phi_{n-1}(x')\\\phi_{n}(x')\end{array}\right)\\
		=&\frac{n_iV_0^2}{4\pi l_B^2}\sum_{n',s'}G_{n',s'}(E)=\frac{(\hbar\omega_c)^2}{2A}\sum_{n,s}G_{ns}(E)
\end{aligned}\end{equation}
It is clear to see that self-energy due to short-range scattering is Landau index ($n,k_y,s$) independent but only depends on Fermi energy. 

In the region of well separated Landau levels, the Fermi energy is close to the $N$th Landau level which is well separated from others. The self-consistent equation can be reduced as
\begin{eqnarray}
	{\rm Re}\Sigma&=&\frac{(\hbar\omega_c)^2}{2A}\frac{E-E_{NS}-{\rm Re}\Sigma}{(E-E_{NS}-{\rm Re}\Sigma)^2+({\rm Im}\Sigma)^2}\\
	{\rm Im}\Sigma&=&\frac{(\hbar\omega_c)^2}{2A}\frac{{\rm Im}\Sigma}{(E-E_{NS}-{\rm Re}\Sigma)^2+({\rm Im}\Sigma)^2}
\end{eqnarray}
and solved as
\begin{equation}
	{\rm Re}\Sigma(E)=\hbar\omega_c\varepsilon;\ \ \ \ {\rm Im}\Sigma(E)=-\hbar\omega_c\sqrt{\frac{1}{2A}-\varepsilon^2}
\end{equation}
which is shown in Eq.~(\ref{eq:self-separated}) in the main paper. 

In the region of overlapping Landau levels, we rewrite self-consistent equation as
\begin{equation}\label{self-SCBA-2}
	\Sigma=\frac{E-\Sigma}{A}\sum_n\frac{1}{(E-\Sigma)^2/(\hbar\omega_c)^2-n}
\end{equation}
Then, we introduce a integral
\begin{equation}
	I=\int_cdzf(z)\cot(\pi z)\ \ \ \ \text{with}\ \ f(z)=\frac{1}{(E-\Sigma)^2/(\hbar\omega_c)^2-z}
\end{equation}
and rewrite Eq.~(\ref{self-SCBA-2}) as
\begin{equation}
	\Sigma=\frac{E-\Sigma}{A}\sum_{n=0}^{\infty}f(n)
\end{equation}
If we do the integral over circle $|z|\to\infty$,
\begin{equation}\label{eq:residue theorem}
		0=I=2\pi i\left\{\frac{1}{\pi}\sum_{n=-\infty}^{\infty}f(n)-\cot\left[\pi\frac{(E-\Sigma)^2}{(\hbar\omega_c)^2}\right]\right\}
\end{equation}
so that
\begin{equation}\label{self-SCBA-3}
	\sum_{n=-\infty}^{\infty}f(n)=\pi\cot\left[\pi\frac{(E-\Sigma)^2}{(\hbar\omega_c)^2}\right]
\end{equation}

When $|E|/\hbar\omega_c\gg1$, the summation over $n$ in Eq.~(\ref{self-SCBA-3}) can be safely extended from $n=-\infty$ to $n=\infty$. By using Eq.~(\ref{eq:residue theorem}), the equation of self-energy becomes
\begin{equation}\begin{aligned}
		{\rm Re}\Sigma+i{\rm Im}\Sigma\approx\frac{E-\Sigma}{A}\pi\cot\left[\pi\frac{(E-\Sigma)^2}{(\hbar\omega_c)^2}\right]
		=-i\frac{\pi}{A}\left[(E-{\rm Re}\Sigma)-i{\rm Im}\Sigma\right]
		\left\{1+2\sum_{k=1}^{\infty}\exp\left[\frac{2ik\pi}{(\hbar\omega_c)^2}(E-{\rm Re}\Sigma-i{\rm Im}\Sigma)^2\right]\right\}\\
		%=&-i\frac{\pi}{A}\left[(E-{\rm Re}\Sigma)-i{\rm Im}\Sigma\right]\left\{1+2\sum_{k=1}^{\infty}e^{\frac{4k\pi(E-{\rm Re}\Sigma){\rm Im}\Sigma}{(\hbar\omega_c)^2}}e^{\frac{2ik\pi}{(\hbar\omega_c)^2}[(E-{\rm Re}\Sigma)^2-({\rm Im}\Sigma)^2]}\right\}
\end{aligned}\end{equation}
One can only keep $k=1$ since the exponential term is decaying with the increase with $k$. Meanwhile the real part of self-energy is ignored.
\begin{equation}
	{\rm Im}\Sigma\approx-\frac{\pi}{A}E\left[1+2\delta\cos\frac{2\pi E^2}{(\hbar\omega_c)^2}\right]
\end{equation}
When $|E|/\hbar\omega_c\lesssim1$, we evaluate the self-energy as
\begin{equation}\begin{aligned}
		i{\rm Im}\Sigma\approx&-i\frac{{\rm Im}\Sigma}{A}\left\{\pi\cot\left[-\pi\frac{({\rm Im}\Sigma)^2}{(\hbar\omega_c)^2}\right]-\sum_{n\geq1}\frac{(\hbar\omega_c)^2}{({\rm Im}\Sigma)^2+n(\hbar\omega_c)^2}\right\}\\
		A\approx&\pi\cot\frac{\pi({\rm Im}\Sigma)^2}{(\hbar\omega_c)^2}+\ln\frac{E_c^2}{({\rm Im}\Sigma)^2}\\
		A\approx&\frac{(\hbar\omega_c)^2}{({\rm Im}\Sigma)^2}+\ln\frac{E_c^2}{({\rm Im}\Sigma)^2}\\
		\frac{(\hbar\omega_c)^2}{E_c^2}e^A=&e^{\frac{(\hbar\omega_c)^2}{({\rm Im}\Sigma)^2}}\frac{(\hbar\omega_c)^2}{({\rm Im}\Sigma)^2}
\end{aligned}\end{equation}
The last equation can be rewritten as $z=we^w$ which is the definition of Lambert W function, thus 
\begin{equation}
	{\rm Im}\Sigma\approx-\frac{\hbar\omega_c}{\sqrt{W\left[\frac{(\hbar\omega_c)^2}{E_c^2}e^A\right]}}\approx -\left[E_ce^{-A/2}+\frac{(\hbar\omega_c)^2}{2E_ce^{-A/2}}\right]
\end{equation}
where $W(x)$ is Lambert W function. Combining the results in two limits, we can get
\begin{equation}
	{\rm Im}\Sigma=-E_ce^{-A/2}-\frac{(\hbar\omega_c)^2}{2E_ce^{-A/2}}-\frac{\pi}{A}|E|\left[1+2\delta\cos\frac{\pi |E|}{\hbar\tilde{\omega}_c}\right]
\end{equation}

\section{vertex correction}
\subsection{$B=0$}\label{app-vertex-b0}
The ``dressed" vertex function $\tilde{T}^{LM}_{xy}$ can be obtained by Bethe-Salpeter equation. In the absence of magnetic field, we write the Bethe-Salpeter equation in $(\bm{k},s)$-basis
\begin{equation}
	\tilde{T}^{LM}_{xy}(\bm{k},E)=T_{xy}(\bm{k})+\int\frac{d^2\bm{k}'}{(2\pi)^2}V^2(\bm{k}-\bm{k}')U^{\dagger}_{\bm{k}}U_{\bm{k}'}G^L(\bm{k}',E)\tilde{T}^{LM}_{xy}(\bm{k}',E)G^M(\bm{k}',E)U^{\dagger}_{\bm{k}'}U_{\bm{k}}
\end{equation}
where $U^{\dagger}_{\bm{k}}U_{\bm{k}'}$ denotes the spin rotation while momentum changing
\begin{equation}
	U_{\bm{k}}^{\dagger}U_{\bm{k}'}=
	\frac{1}{2}
	\left(\begin{array}{cc}
		1+e^{i\Delta\theta} & 1-e^{i\Delta\theta}\\1-e^{i\Delta\theta}&1+e^{i\Delta\theta}
	\end{array}\right)
\end{equation}
with $\Delta\theta=\theta_{\bm{k}'}-\theta_{\bm{k}}$, and $V^2(\bm{k}-\bm{k}')=n_iV^2_0$ with $n_i$ and $V_0$ denoting the concentration and strength of impurity for short-range disorder. In order to solve this self-consistent equation, we start with the first order approximation in the following
\begin{equation}\begin{aligned}
		\tilde{T}^{(1),LM}_{xy}(\bm{k},E)=&n_iV^2_0\int\frac{d^2\bm{k}'}{(2\pi)^2}U^{\dagger}_{\bm{k}}U_{\bm{k}'}G^L(\bm{k}',E)T_{xy}(\bm{k}')G^M(\bm{k}',E)U^{\dagger}_{\bm{k}'}U_{\bm{k}}\\
		=&v_f\frac{n_iV^2_0}{4}\int\frac{d^2\bm{k}'}{(2\pi)^2}\frac{\hbar v_fk}{2}
		\left(\begin{array}{cc}
			1+e^{i\Delta\theta} & 1-e^{i\Delta\theta}\\1-e^{i\Delta\theta}&1+e^{i\Delta\theta}
		\end{array}\right)
		\left(\begin{array}{cc}
			g^L_+&\\&g^L_-
		\end{array}\right)
		\left(\begin{array}{cc}
			\sin2\theta_{\bm{k}'}&i\cos2\theta_{\bm{k}'}\\-i\cos2\theta_{\bm{k}'}&-\sin2\theta_{\bm{k}'}
		\end{array}\right)
		\left(\begin{array}{cc}
			g^M_+&\\&g^M_-
		\end{array}\right)
		\left(\begin{array}{cc}
			1+e^{-i\Delta\theta} & 1-e^{-i\Delta\theta}\\1-e^{-i\Delta\theta}&1+e^{-i\Delta\theta}
		\end{array}\right)\\
		=&v_f\frac{n_iV^2_0}{4}\int\frac{d^2\bm{k}'}{(2\pi)^2}i\frac{\hbar v_fk}{4}\\
		&\left(\begin{array}{cc}
			(1+e^{i\Delta\theta})g^L_+ & (1-e^{i\Delta\theta})g^L_-\\(1-e^{i\Delta\theta})g^L_+&(1+e^{i\Delta\theta})g^L_-
		\end{array}\right)
		\left[e^{2i(\theta_{\bm{k}}+\Delta\theta)}\left(\begin{array}{cc}
			-1&1\\
			-1&1
		\end{array}\right)+e^{-2i(\theta_{\bm{k}}+\Delta\theta)}\left(\begin{array}{cc}
			1&1\\
			-1&-1
		\end{array}\right)\right]
		\left(\begin{array}{cc}
			(1+e^{-i\Delta\theta})g^M_+ & (1-e^{-i\Delta\theta})g^M_+\\(1-e^{-i\Delta\theta})g^M_-&(1+e^{-i\Delta\theta})g^M_-
		\end{array}\right)\\
		=&0
\end{aligned}\end{equation} 
Since the Bethe-Salpeter equation is a self-consistent iterative equation, we can get the vertex correction of shear viscosity to zero in the presence of short-range disorder. 

\newpage
\subsection{$\textbf{B}=B\hat{z}$}\label{app-vertex-bn0}
In the presence of magnetic field, we rewrite the Bethe-Salpeter equation in $(n,k_y,s)$-basis. Similar to the case of $B=0$, we start with the first order of corrected stress tensor 
\begin{equation}\begin{aligned}
		&\langle n,k_y,s|\tilde{T}^{(1)}_{xy}|n',k_y',s'\rangle\\
		=&\overline{\langle n,k_y,s|VGT_{xy}GV|n',k_y',s'\rangle}\\
		=&\sum_{n'',k_y'',s''}\sum_{n''',k_y''',s'''}\overline{\langle n,k_y,s|V|n'',k_y'',s''\rangle G_{n'',s''}\langle n'',k_y'',s''|T_{xy}|n''',k_y''',s'''\rangle G_{n''',s'''}\langle n''',k_y''',s'''|V|n',k_y',s'\rangle}\\
		=&n_iV_0^2\sum_{n'',k_y'',s''}\sum_{n''',k_y''',s'''}G_{n'',s''}G_{n''',s'''}\langle n'',s''|T_{xy}|n''',s'''\rangle\delta_{k_y'',k_y'''}
		\left[\int d\bm{r}\Psi_{n,k_y,s}^{\dagger}(\bm{r})\Psi_{n'',k_y'',s''}(\bm{r})\Psi_{n''',k_y''',s'''}^{\dagger}(\bm{r})\Psi_{n',k_y',s'}(\bm{r})\right]\\
		=&n_iV_0^2\sum_{n'',s''}\sum_{n''',s'''}G_{n'',s''}G_{n''',s'''}\langle n'',s''|T_{xy}|n''',s'''\rangle\\
		&\left[\sum_{k_y''}\int dxdy\Psi_{n,k_y,s}^{\dagger}(x)\frac{e^{-ik_yy}}{\sqrt{L_y}}\Psi_{n'',k_y'',s''}(x)\frac{e^{ik_y''y}}{\sqrt{L_y}}\Psi_{n''',k_y'',s'''}^{\dagger}(x)\frac{e^{-ik_y''y}}{\sqrt{L_y}}\Psi_{n',k_y',s'}(x)\frac{e^{-ik_y'y}}{\sqrt{L_y}}\right]\\
		=&n_iV_0^2\sum_{n'',s''}\sum_{n''',s'''}G_{n'',s''}G_{n''',s'''}\langle n'',s''|T_{xy}|n''',s'''\rangle\\
		&\left[\frac{1}{L_y^2}\sum_{k_y''}\int dxdye^{-i(k_y-k_y')y}\Psi_{n,k_y,s}^{\dagger}(x)\Psi_{n'',k_y'',s''}(x)\Psi_{n''',k_y'',s'''}^{\dagger}(x)\Psi_{n',k_y',s'}(x)\right]\\
		=&n_iV_0^2\delta_{k_y,k_y'}\sum_{n'',s''}\sum_{n''',s'''}G_{n'',s''}G_{n''',s'''}\langle n'',s''|T_{xy}|n''',s'''\rangle
		\left[\frac{1}{L_y}\sum_{k_y''}\int dx\Psi_{n,k_y,s}^{\dagger}(x)\Psi_{n'',k_y'',s''}(x)\Psi_{n''',k_y'',s'''}^{\dagger}(x)\Psi_{n',k_y,s'}(x)\right]\\
		=&n_iV_0^2\delta_{k_y,k_y'}\sum_{n'',s''}\sum_{n''',s'''}G_{n'',s''}G_{n''',s'''}\langle n'',s''|T_{xy}|n''',s'''\rangle
		\left[\frac{dx'}{2\pi l_B^2}\int dx\Psi_{n,k_y,s}^{\dagger}(x)\Psi_{n'',k_y'',s''}(x')\Psi_{n''',k_y'',s'''}^{\dagger}(x')\Psi_{n',k_y,s'}(x)\right]\\
		%	=&n_iV_0^2\sum_{n',s'}G_{n',s'}(E)\left[\frac{1}{L_y}\sum_{k_y'}\int dx\Psi_{n,k_y,s}^{\dagger}(x)\Psi_{n',k_y',s'}(x)\Psi_{n',k_y',s'}^{\dagger}(x)\Psi_{n,k_y,s}(x)\right]\\
		%	=&n_iV_0^2\sum_{n',s'}G_{n',s'}(E)\left[\int\frac{dk_y'}{2\pi}\int dx\Psi_{n,k_y,s}^{\dagger}(x)\Psi_{n',k_y',s'}(x)\Psi_{n',k_y',s'}^{\dagger}(x)\Psi_{n,k_y,s}(x)\right]\\
		%	=&n_iV_0^2\sum_{n',s'}G_{n',s'}(E)\left[\int\frac{dx'}{2\pi l_B^2}\int dx\Psi_{n,k_y,s}^{\dagger}(x)\Psi_{n',k_y',s'}(x')\Psi_{n',k_y',s'}^{\dagger}(x')\Psi_{n,k_y,s}(x)\right]\\
		=&\frac{n_iV_0^2}{2\pi l_B^2}\delta_{k_y,k_y'}\sum_{n'',s''}\sum_{n''',s'''}G_{n'',s''}G_{n''',s'''}\langle n'',s''|T_{xy}|n''',s'''\rangle\int dxdx'\\
		&\frac{1}{4}\left(\begin{array}{cc}is\phi_{n-1}(x)&\phi_n(x)\end{array}\right)\left(\begin{array}{c}-is''\phi_{n''-1}(x')\\\phi_{n''}(x')\end{array}\right)\left(\begin{array}{cc}is'''\phi_{n'''-1}(x')&\phi_{n'''}(x')\end{array}\right)\left(\begin{array}{c}-is'\phi_{n'-1}(x)\\\phi_{n'}(x)\end{array}\right)\\
		=&\frac{n_iV_0^2}{8\pi l_B^2}\delta_{k_y,k_y'}\sum_{n'',s''}\sum_{n''',s'''}G_{n'',s''}G_{n''',s'''}\langle n'',s''|T_{xy}|n''',s'''\rangle\int dxdx'\\
		&[ss''\phi_{n-1}(x)\phi_{n''-1}(x')+\phi_{n}(x)\phi_{n''}(x')][s'''s'\phi_{n'''-1}(x')\phi_{n'-1}(x)+\phi_{n'''}(x')\phi_{n'}(x)]\\
		=&\frac{n_iV_0^2}{8\pi l_B^2}\delta_{k_y,k_y'}\sum_{n'',s''}\sum_{n''',s'''}G_{n'',s''}G_{n''',s'''}\langle n'',s''|T_{xy}|n''',s'''\rangle\int dxdx'\\
		&[(ss's''s'''+1)\delta_{n,n'}\delta_{n'',n'''}+ss''\delta_{n,n'+1}\delta_{n'',n'''+1}+s's'''\delta_{n,n'-1}\delta_{n'',n'''-1}]\\
\end{aligned}\end{equation}
where the disorder average is reflected in the relation
\begin{equation}\begin{aligned}
		\overline{\langle n,k_y,s|V|n'',k_y'',s''\rangle\langle n''',k_y''',s'''|V|n',k_y',s'\rangle}
		=&\int d\bm{r}d\bm{r}'\overline{\Psi_{n,k_y,s}^{\dagger}(\bm{r})V(\bm{r})\Psi_{n'',k_y'',s''}(\bm{r})\Psi_{n''',k_y''',s'''}^{\dagger}(\bm{r}')V(\bm{r}')\Psi_{n',k_y',s'}(\bm{r}')}\\
		=&\int d\bm{r}d\bm{r}'\overline{V(\bm{r})V(\bm{r}')}\Psi_{n,k_y,s}^{\dagger}(\bm{r})\Psi_{n'',k_y'',s''}(\bm{r})\Psi_{n''',k_y''',s'''}^{\dagger}(\bm{r}')\Psi_{n',k_y',s'}(\bm{r}')\\
		=&n_iV_0^2\int d\bm{r}\Psi_{n,k_y,s}^{\dagger}(\bm{r})\Psi_{n'',k_y'',s''}(\bm{r})\Psi_{n''',k_y''',s'''}^{\dagger}(\bm{r})\Psi_{n',k_y',s'}(\bm{r})
\end{aligned}\end{equation}
The corrected stress tensor is zero $\langle n'',s''|\tilde{T}_{xy}|n''',s'''\rangle=0$, since $\langle n'',s''|T_{xy}|n''',s'''\rangle$ is finite only when $n''=n'''\pm2$.

\newpage
\section{Derivation of static shear viscosity $\eta_s$ in the presence of $B=0$}\label{appendix-shear-b0}
\begin{equation}\begin{aligned}
		\eta_s^{RA}(E)=&\frac{\hbar^3v_f^2}{8\pi^2}\int dkk^3\left(\frac{1}{E-\hbar v_fk-\Sigma^R}+\frac{1}{E+\hbar v_fk-\Sigma^R}\right)\left(\frac{1}{E-\hbar v_fk-\Sigma^A}+\frac{1}{E+\hbar v_fk-\Sigma^A}\right)\\
		=&\frac{\hbar^3v_f^2}{2\pi^2}\int dkk^3\frac{E-\Sigma^R}{(E-\Sigma^R)^2-(\hbar v_fk)^2}\frac{E-\Sigma^A}{(E-\Sigma^A)^2-(\hbar v_fk)^2}\\
		=&\frac{\hbar}{2\pi^2}(E-\Sigma^R)(E-\Sigma^A)\int dkk
		\left[\frac{(\hbar v_fk)^2}{(E-\Sigma^R)^2-(\hbar v_fk)^2}-\frac{(\hbar v_fk)^2}{(E-\Sigma^A)^2-(\hbar v_fk)^2}\right]\frac{1}{(E-\Sigma^A)^2-(E-\Sigma^R)^2}\\
		=&\frac{\hbar}{2\pi^2}\frac{(E-{\rm Re}\Sigma^R)^2+({\rm Im}\Sigma)^2}{2{\rm Im}\Sigma(E-{\rm Re}\Sigma^R)}\int dkk{\rm Im}\left[\frac{(E-\Sigma^R)^2}{(E-\Sigma^R)^2-(\hbar v_fk)^2}\right]\\
		=&\frac{\hbar}{2\pi^2}\frac{(E-{\rm Re}\Sigma^R)^2+({\rm Im}\Sigma)^2}{2{\rm Im}\Sigma(E-{\rm Re}\Sigma^R)}\left\{(E-{\rm Re}\Sigma)\int dkk{\rm Im}\left[\frac{E-\Sigma^R}{(E-\Sigma^R)^2-(\hbar v_fk)^2}\right]\right.
		\left.-{\rm Im}\Sigma\int dkk{\rm Re}\left[\frac{E-\Sigma^R}{(E-\Sigma^R)^2-(\hbar v_fk)^2}\right]\right\}\\
		\approx&\frac{\hbar}{2\pi^2}\frac{(E-{\rm Re}\Sigma^R)^2+({\rm Im}\Sigma)^2}{2{\rm Im}\Sigma(E-{\rm Re}\Sigma^R)}\left\{(E-{\rm Re}\Sigma)(-\frac{\pi^2}{4}\rho)-{\rm Im}\Sigma\frac{A}{2(\hbar v_f)^2}{\rm Re}\Sigma\right\}\\
		\approx&\hbar\frac{E^2+({\rm Im}\Sigma)^2}{-16{\rm Im}\Sigma}\rho(E)\\
		=&\frac{1}{8}E^2\rho\tau+\frac{\hbar^2}{32\tau}\rho
		%	{\color{red}\approx}&\frac{2}{\pi^2}\frac{A}{2(\hbar v_f)^2}\frac{(E-\Sigma^R)(E-\Sigma^A)}{(E-\Sigma^A)^2-(E-\Sigma^R)^2}[(E-\Sigma^R)\Sigma^R-(E-\Sigma^A)\Sigma^A]\\
		%	=&\frac{A}{\pi^2(\hbar v_f)^2}\frac{(E-{\rm Re}\Sigma)^2+({\rm Im}\Sigma)^2}{4i(E-{\rm Re}\Sigma){\rm Im}\Sigma}2i(E-2{\rm Re}\Sigma){\rm Im}\Sigma\\
		%	{\color{red}\approx}&\frac{A}{2\pi^2(\hbar v_f)^2}\left[E^2+({\rm Im}\Sigma)^2\right]\\
\end{aligned}\end{equation}
\begin{equation}\begin{aligned}
		{\rm Re}\eta_s^{RR}(E)=&{\rm Re}\frac{\hbar^3v_f^2}{8\pi^2}\int dkk^3\left(\frac{1}{E-\hbar v_fk-\Sigma^R}+\frac{1}{E+\hbar v_fk-\Sigma^R}\right)^2
		=-{\rm Re}\frac{\hbar}{2\pi^2}(E-\Sigma^R)^2\int dkk\frac{(E-\Sigma^R)^2-(\hbar v_fk)^2-(E-\Sigma^R)^2}{\left[(E-\Sigma^R)^2-(\hbar v_fk)^2\right]^2}\\
		=&-{\rm Re}\frac{\hbar}{2\pi^2}(E-\Sigma^R)^2\left\{\int dkk\frac{1}{(E-\Sigma^R)^2-(\hbar v_fk)^2}-\int dkk\frac{(E-\Sigma^R)^2}{\left[(E-\Sigma^R)^2-(\hbar v_fk)^2\right]^2}\right\}\\
		=&-{\rm Re}\frac{\hbar}{2\pi^2}(E-\Sigma^R)\int dkk\left[\frac{E-\Sigma^R}{(E-\Sigma^R)^2-(\hbar v_fk)^2}\right]-{\rm Re}\frac{(E-\Sigma^R)^2}{4\pi^2\hbar v_f^2}\\
		\approx&-\frac{\hbar}{2\pi^2}{\rm Im}\Sigma\int dkk\left[\frac{E-\Sigma^R}{(E-\Sigma^R)^2-(\hbar v_fk)^2}\right]\\
		=&-\frac{\hbar^2}{16\tau}\rho\\
\end{aligned}\end{equation}
where we have used the SCBA of self-energy
\begin{equation}
	\int dkk\frac{E-\Sigma^{R/A}}{(E-\Sigma^{R/A})^2-(\hbar v_fk)^2}=\frac{A}{2(\hbar v_f)^2}\Sigma^{R/A},
\end{equation}
the relation between density of states and self-energy
\begin{equation}
	\rho(E)=-\frac{4}{\pi}{\rm Tr}[{\rm Im}G^R(E)]=-\frac{4}{\pi^2}\int kdk{\rm Im}\left[\frac{E-\Sigma^R}{(E-\Sigma^R)^2-(\hbar v_fk)^2}\right]
\end{equation}
and ignored the real part of self-energy ${\rm Re}\Sigma\to0$.

\newpage
\section{Derivation of static shear viscosity $\eta_s$ in the presence of $B\neq0$}\label{appendix-shear-bn0}
\begin{equation}\begin{aligned}
		\eta_s^{RA}(E)=&\frac{\hbar^3\omega_c^2}{4\pi^2l_B^2}\sum_{n}(n+1)(g^R_{n}g^A_{n+2}+g^R_{n+2}g^A_{n})\\
		&=\frac{\hbar^3\omega_c^2}{4\pi^2l_B^2}\sum_{n}(n+1)
		\left[\frac{E-\Sigma^R}{(E-\Sigma^R)^2-n(\hbar\omega_c)^2}\frac{E-\Sigma^A}{(E-\Sigma^A)^2-(n+2)(\hbar\omega_c)^2}+\frac{E-\Sigma^R}{(E-\Sigma^R)^2-(n+2)(\hbar\omega_c)^2}\frac{E-\Sigma^A}{(E-\Sigma^A)^2-n(\hbar\omega_c)^2}\right]\\
		=&\frac{\hbar^3\omega_c^2}{4\pi^2l_B^2}\sum_{n}(n+1)(E-\Sigma^R)(E-\Sigma^A)
		\left\{\frac{1}{(E-\Sigma^A)^2-(E-\Sigma^R)^2-2(\hbar\omega_c)^2}\left[\frac{1}{(E-\Sigma^R)^2-n(\hbar\omega_c)^2}-\frac{1}{(E-\Sigma^A)^2-(n+2)(\hbar\omega_c)^2}\right]\right.\\
		&+\left.\frac{1}{(E-\Sigma^R)^2-(E-\Sigma^A)^2-2(\hbar\omega_c)^2}\left[\frac{1}{(E-\Sigma^A)^2-n(\hbar\omega_c)^2}-\frac{1}{(E-\Sigma^R)^2-(n+2)(\hbar\omega_c)^2}\right]\right\}\\
		=&\frac{\hbar^3\omega_c^2}{8\pi^2l_B^2}
		\left\{\frac{-(\hbar\omega_c)^2}{4(E-{\rm Re}\Sigma)^2({\rm Im}\Sigma)^2+(\hbar\omega_c)^4}\left[(E-\Sigma^R)(g^A_0+2\sum_{n\geq1}g_n^A)+(E-\Sigma^A)(g^R_0+2\sum_{n\geq1}g_n^R)\right]\right.\\
		&+\left.\frac{2i(E-{\rm Re}\Sigma){\rm Im}\Sigma}{4(E-{\rm Re}\Sigma)^2({\rm Im}\Sigma)^2+(\hbar\omega_c)^4}\left[(E-\Sigma^R)(g^A_0+\sum_{n}2ng_n^A)-(E-\Sigma^A)(g^R_0+\sum_{n}2ng_n^R)\right]\right\}\\
		=&\frac{\hbar^3\omega_c^2}{8\pi^2l_B^2}
		\left\{\frac{-(\hbar\omega_c)^2}{4(E-{\rm Re}\Sigma)^2({\rm Im}\Sigma)^2+(\hbar\omega_c)^4}\left[2(E-{\rm Re}\Sigma)\sum_{ns}{\rm Re}G^R_{ns}-2{\rm Im}\Sigma\sum_{ns}{\rm Im}G^R_{ns}\right]\right.\\
		&+\frac{2i(E-{\rm Re}\Sigma){\rm Im}\Sigma}{4(E-{\rm Re}\Sigma)^2({\rm Im}\Sigma)^2+(\hbar\omega_c)^4}\left[\frac{-4i(E-{\rm Re}\Sigma){\rm Im}\Sigma}{(E-{\rm Re}\Sigma)^2+({\rm Im}\Sigma)^2}\right]\\
		&\left.+\frac{2i(E-{\rm Re}\Sigma){\rm Im}\Sigma}{4(E-{\rm Re}\Sigma)^2({\rm Im}\Sigma)^2+(\hbar\omega_c)^4}\frac{(E-{\rm Re}\Sigma)^2+({\rm Im}\Sigma)^2}{(\hbar\omega_c)^2}
		\left[2i{\rm Im}\Sigma\sum_{ns}{\rm Re}G_{ns}^R-2i(E-{\rm Re}\Sigma)\sum_{ns}{\rm Im}G_{ns}^R\right]\right\}\\
		\approx&\frac{\hbar}{8\pi^2l_B^2}
		\left\{\frac{(\hbar\omega_c)^4}{4E^2({\rm Im}\Sigma)^2+(\hbar\omega_c)^4}\left[2{\rm Im}\Sigma\sum_{ns}{\rm Im}G^R_{ns}\right]+\frac{2E{\rm Im}\Sigma(\hbar\omega_c)^2}{4E^2({\rm Im}\Sigma)^2+(\hbar\omega_c)^4}\left[\frac{4E{\rm Im}\Sigma}{E^2+({\rm Im}\Sigma)^2}\right]\right.\\
		&\left.+\frac{2E{\rm Im}\Sigma}{4E^2({\rm Im}\Sigma)^2+(\hbar\omega_c)^4}[E^2+({\rm Im}\Sigma)^2]
		\left[2E\sum_{ns}{\rm Im}G_{ns}^R\right]\right\}\\
		=&\frac{\hbar}{8\pi^2l_B^2}
		\left\{\frac{4\tilde{\omega}_c^2\tau^2}{1+4\tilde{\omega}_c^2\tau^2}\left[\frac{\hbar}{\tau}\frac{\pi^2l_B^2}{2}\rho(E)\right]+\frac{4\tilde{\omega}_c^2\tau^2}{1+4\tilde{\omega}_c^2\tau^2}\left[\frac{8E^2({\rm Im}\Sigma)^2}{(\hbar\omega_c)^2[({\rm Im}\Sigma)^2+E^2]}\right]+\frac{1}{1+4\tilde{\omega}_c^2\tau^2}[E^2+(\frac{\hbar}{2\tau})^2]\frac{2\tau}{\hbar}
		\left[\frac{\pi^2l_B^2}{2}\rho(E)\right]\right\}\\
		\approx&\frac{1}{8}\frac{E^2\rho\tau}{1+4\tilde{\omega}_c^2\tau^2}+\frac{\hbar^2}{32\tau}\rho\frac{1+8\tilde{\omega}_c^2\tau^2}{1+4\tilde{\omega}_c^2\tau^2}
\end{aligned}\end{equation}
\begin{equation}\begin{aligned}
		\eta_s^{RR}(E)=&\frac{\hbar^3\omega_c^2}{2\pi^2l_B^2}\sum_{n}(n+1)g^R_{n}g^R_{n+2}
		=\frac{\hbar^3\omega_c^2}{2\pi^2l_B^2}\sum_{n}(n+1)\frac{E-\Sigma^R}{(E-\Sigma^R)^2-n(\hbar\omega_c)^2}\frac{E-\Sigma^R}{(E-\Sigma^R)^2-(n+2)(\hbar\omega_c)^2}\\
		=&-\frac{\hbar}{4\pi^2l_B^2}\sum_{n}(n+1)(E-\Sigma^R)^2\left[\frac{1}{(E-\Sigma^R)^2-n(\hbar\omega_c)^2}-\frac{1}{(E-\Sigma^R)^2-(n+2)(\hbar\omega_c)^2}\right]\\
		=&-\frac{\hbar}{4\pi^2l_B^2}(E-\Sigma^R)(g_0^R+2\sum_{n\geq1}g_n^R)=-\frac{1}{\pi^2l_B^2}(E-\Sigma^R)\sum_{ns}G^R_{ns}\\
		\approx&-\frac{\hbar}{4\pi^2l_B^2}\left[{\rm Im}\Sigma\sum_{ns}{\rm Im}G^R_{ns}+i(E-{\rm Re}\Sigma)\sum_{ns}{\rm Im}G^R_{ns}\right]
		\approx-\frac{\hbar^2}{16\tau}\rho(E)\\
\end{aligned}\end{equation}
where we used the SCBA of self energy ${\rm Re}\Sigma(E)=\frac{(\hbar\omega_c)^2}{2A}\sum_{ns}{\rm Re}G_{ns}(E)$ and then assumed ${\rm Re}\Sigma\to0$, applied the expression of density of states $\rho(E)=-\frac{2}{\pi^2\l_B^2}\sum_{ns}{\rm Im}G_{ns}^R(E)$ and quasiparticle relaxation time $\tau(E)=-\frac{\hbar}{2{\rm Im}\Sigma}$, and introduced $\tilde{\omega}_c=\frac{\hbar\omega_c^2}{2E}$ which can transform to the cyclotron frequency in 2DEG by using the effective mass $m=\frac{2E}{v_f^2}$,
\begin{equation}
	\tilde{\omega}_c=\frac{\hbar\omega_c^2}{2E}\xrightarrow{m=\frac{2E}{v_f^2}}\frac{eB}{mc}; \ \ \ \ 
\end{equation}

\newpage
\section{Static Hall viscosity $\eta_H^{I,RA}$ and $\eta_H^{II}$}\label{appendix-hall}

\begin{equation}\begin{aligned}
		\eta_H^{I,RA}=&\frac{i\hbar^3\omega_c^2}{16\pi^2l_B^2}\sum_{n,s,s'}(n+1)\left[G^L_{n+2,s}(E)G^M_{n,s'}(E)-G^L_{n,s}(E)G^M_{n+2,s'}(E)\right]=\frac{i\hbar^3\omega_c^2}{4\pi^2l_B^2}\sum_{n}(n+1)\\
		&\left[\frac{E-\Sigma^R}{(E-\Sigma^R)^2-(n+2)(\hbar\omega_c)^2}\frac{E-\Sigma^A}{(E-\Sigma^A)^2-n(\hbar\omega_c)^2}-\frac{E-\Sigma^R}{(E-\Sigma^R)^2-n(\hbar\omega_c)^2}\frac{E-\Sigma^A}{(E-\Sigma^A)^2-(n+2)(\hbar\omega_c)^2}\right]\\
		=&\frac{i\hbar^3\omega_c^2}{4\pi^2l_B^2}\sum_{n}(n+1)
		\left\{\frac{(E-\Sigma^R)(E-\Sigma^A)}{(E-\Sigma^R)^2-(E-\Sigma^A)^2-2(\hbar\omega_c)^2}\left[\frac{1}{(E-\Sigma^A)^2-n(\hbar\omega_c)^2}-\frac{1}{(E-\Sigma^R)^2-(n+2)(\hbar\omega_c)^2}\right]\right.\\
		&-\left.\frac{(E-\Sigma^R)(E-\Sigma^A)}{(E-\Sigma^A)^2-(E-\Sigma^R)^2-2(\hbar\omega_c)^2}\left[\frac{1}{(E-\Sigma^R)^2-n(\hbar\omega_c)^2}-\frac{1}{(E-\Sigma^A)^2-(n+2)(\hbar\omega_c)^2}\right]\right\}\\
		=&\frac{i\hbar^3\omega_c^2}{4\pi^2l_B^2}\sum_{n}(n+1)
		\left\{\frac{1}{(E-\Sigma^R)^2-(E-\Sigma^A)^2-2(\hbar\omega_c)^2}\left[(E-\Sigma^R)g_n^A-(E-\Sigma^A)g_{n+2}^R\right]\right.\\
		&-\left.\frac{1}{(E-\Sigma^A)^2-(E-\Sigma^R)^2-2(\hbar\omega_c)^2}\left[(E-\Sigma^A)g_n^R-(E-\Sigma^R)g_{n+2}^A\right]\right\}\\
		=&\frac{i\hbar^3\omega_c^2}{8\pi^2l_B^2}\sum_{n}(n+1)
		\left\{\frac{2i(E-{\rm Re}\Sigma){\rm Im}\Sigma-(\hbar\omega_c)^2}{4(E-{\rm Re}\Sigma)^2({\rm Im}\Sigma)^2+(\hbar\omega_c)^4}\left[(E-\Sigma^R)g_n^A-(E-\Sigma^A)g_{n+2}^R\right]\right.\\
		&+\left.\frac{2i(E-{\rm Re}\Sigma){\rm Im}\Sigma+(\hbar\omega_c)^2}{4(E-{\rm Re}\Sigma)^2({\rm Im}\Sigma)^2+(\hbar\omega_c)^4}\left[(E-\Sigma^A)g_n^R-(E-\Sigma^R)g_{n+2}^A\right]\right\}\\
		=&\frac{i\hbar^3\omega_c^2}{8\pi^2l_B^2}
		\left\{\frac{2i(E-{\rm Re}\Sigma){\rm Im}\Sigma}{4(E-{\rm Re}\Sigma)^2({\rm Im}\Sigma)^2+(\hbar\omega_c)^4}\left[(E-\Sigma^A)(g_0^R+\sum_{n\geq1}2g_{n}^R)+(E-\Sigma^R)(g_0^A+\sum_{n\geq1}2g_{n}^A)\right]\right.\\
		&+\left.\frac{(\hbar\omega_c)^2}{4(E-{\rm Re}\Sigma)^2({\rm Im}\Sigma)^2+(\hbar\omega_c)^4}\left[(E-\Sigma^A)(g_0^R+\sum_{n\geq1}2ng_{n}^R)-(E-\Sigma^R)(g_0^A+\sum_{n\geq1}2ng_{n}^A)\right]\right\}\\
		=&\frac{i\hbar^3\omega_c^2}{8\pi^2l_B^2}
		\left\{\frac{2i(E-{\rm Re}\Sigma){\rm Im}\Sigma}{4(E-{\rm Re}\Sigma)^2({\rm Im}\Sigma)^2+(\hbar\omega_c)^4}\left[(E-\Sigma^A)\sum_{ns}G_{ns}^R+(E-\Sigma^R)\sum_{ns}G_{ns}^A\right]\right.\\
		&+\frac{(\hbar\omega_c)^2}{4(E-{\rm Re}\Sigma)^2({\rm Im}\Sigma)^2+(\hbar\omega_c)^4}\left[(E-\Sigma^A)G_0^R-(E-\Sigma^R)G_0^A\right]\\
		&+\left.\frac{(\hbar\omega_c)^2}{4(E-{\rm Re}\Sigma)^2({\rm Im}\Sigma)^2+(\hbar\omega_c)^4}\left[\frac{(E-\Sigma^A)(E-\Sigma^R)^2}{(\hbar\omega_c)^2}\sum_{ns}G_{ns}^R-\frac{(E-\Sigma^R)(E-\Sigma^A)^2}{(\hbar\omega_c)^2}\sum_{ns}G_{ns}^A\right]\right\}\\
		=&\frac{i\hbar^3\omega_c^2}{8\pi^2l_B^2}
		\left\{\frac{2i(E-{\rm Re}\Sigma){\rm Im}\Sigma}{4(E-{\rm Re}\Sigma)^2({\rm Im}\Sigma)^2+(\hbar\omega_c)^4}\left[2(E-{\rm Re}\Sigma)\sum_{ns}{\rm Re}G_{ns}^R-2{\rm Im}\Sigma\sum_{ns}{\rm Im}G_{ns}^R\right]\right.\\
		&+\frac{(\hbar\omega_c)^2}{4(E-{\rm Re}\Sigma)^2({\rm Im}\Sigma)^2+(\hbar\omega_c)^4}\left[\frac{4i(E-{\rm Re}\Sigma){\rm Im}\Sigma}{(E-{\rm Re}\Sigma)^2+({\rm Im}\Sigma)^2}\right]\\
		&+\left.\frac{(\hbar\omega_c)^2}{4(E-{\rm Re}\Sigma)^2({\rm Im}\Sigma)^2+(\hbar\omega_c)^4}\frac{(E-{\rm Re}\Sigma)^2+({\rm Im}\Sigma)^2}{(\hbar\omega_c)^2}\left[2i(E-{\rm Re}\Sigma)\sum_{ns}{\rm Im}G_{ns}^R-2i{\rm Im}\Sigma\sum_{ns}{\rm Re}G_{ns}^R\right]\right\}\\
		\approx&\frac{i\hbar^3\omega_c^2}{8\pi^2l_B^2}
		\left\{\frac{2iE{\rm Im}\Sigma}{4E^2({\rm Im}\Sigma)^2+(\hbar\omega_c)^4}\left[-2{\rm Im}\Sigma\sum_{ns}{\rm Im}G_{ns}^R\right]
		+\frac{(\hbar\omega_c)^2}{4E^2({\rm Im}\Sigma)^2+(\hbar\omega_c)^4}\left[\frac{4iE{\rm Im}\Sigma}{E^2+({\rm Im}\Sigma)^2}\right]\right.\\
		&+\left.\frac{E^2+({\rm Im}\Sigma)^2}{4E^2({\rm Im}\Sigma)^2+(\hbar\omega_c)^4}\left[2iE\sum_{ns}{\rm Im}G_{ns}^R\right]\right\}\\
		=&\frac{\hbar^3\omega_c^2}{4\pi^2l_B^2}
		\left\{\frac{(\hbar\omega_c)^2}{4E^2({\rm Im}\Sigma)^2+(\hbar\omega_c)^4}\left[\frac{-2E{\rm Im}\Sigma}{E^2+({\rm Im}\Sigma)^2}\right]+\frac{E^2-({\rm Im}\Sigma)^2}{4E^2({\rm Im}\Sigma)^2+(\hbar\omega_c)^4}\left[-E\sum_{ns}{\rm Im}G_{ns}^R\right]\right\}\\
		=&\frac{\hbar}{2\pi^2l_B^2}\frac{4\tilde{\omega}_c^2\tau^2}{1+4\tilde{\omega}_c^2\tau^2}\left[\frac{1}{\frac{2E\tau}{\hbar}+\frac{\hbar}{2E\tau}}\right]+\frac{1}{4}\frac{\tilde{\omega}_c\tau^2}{1+4\tilde{\omega}_c^2\tau^2}[E^2-(\frac{\hbar}{2\tau})^2]\rho\\
		\approx&\frac{1}{4}\frac{\tilde{\omega}_c\tau^2E^2}{1+4\tilde{\omega}_c^2\tau^2}\rho
\end{aligned}\end{equation}

\begin{equation}\begin{aligned}
		\eta^{II}_{H}(E)=&\frac{i\hbar^3\omega_c^2}{8\pi^2l_B^2}\int f_{\omega}d\omega\sum_{nss'}(n+1)\left[G^R_{n+2,s}(\omega)\frac{dG^R_{ns'}(\omega)}{d\omega}-G^R_{ns'}(\omega)\frac{dG^R_{n+2,s}(\omega)}{d\omega}\right]\\
		=&\frac{i\hbar^3\omega_c^2}{8\pi^2l_B^2}\int f(\omega)d\omega\sum_{nss'}(n+1)
		\left[\frac{1}{\omega-\Sigma^R-E_{n+2,s}}\frac{-(1-\partial_{\omega}\Sigma)}{(\omega-\Sigma^R-E_{ns'})^2}-\frac{1}{\omega-\Sigma^R-E_{ns'}}\frac{-(1-\partial_{\omega}\Sigma)}{(\omega-\Sigma^R-E_{n+2,s})^2}\right]\\
		=&\frac{i\hbar^3\omega_c^2}{8\pi^2l_B^2}\int d\omega f(\omega)[-(1-\partial_{\omega}\Sigma)]\sum_{nss'}(n+1)\frac{1}{E_{ns'}-E_{n+2,s}}\left[\frac{1}{\omega-\Sigma^R-E_{n+2,s}}-\frac{1}{\omega-\Sigma^R-E_{ns'}}\right]^2\\
		=&\frac{i\hbar^3\omega_c^2}{8\pi^2l_B^2}\int d\omega f(\omega)\sum_{nss'}(n+1)\left\{\frac{1}{E_{ns'}-E_{n+2,s}}\left[\frac{-(1-\partial_{\omega}\Sigma)}{(\omega-\Sigma^R-E_{n+2,s})^2}+\frac{-(1-\partial_{\omega}\Sigma)}{(\omega-\Sigma^R-E_{ns'})^2}\right]\right.\\
		&+\left.\frac{2(1-\partial_{\omega}\Sigma)}{(E_{ns'}-E_{n+2,s})^2}\left[\frac{1}{\omega-\Sigma^R-E_{ns'}}-\frac{1}{\omega-\Sigma^R-E_{n+2,s}}\right]\right\}\\
		=&\frac{i\hbar}{8\pi^2l_B^2}\int d\omega f(\omega)\sum_{ns}(n+1)\left\{\left[\frac{(1-\partial_{\omega}\Sigma)E_{n+2,s}}{(\omega-\Sigma^R-E_{n+2,s})^2}+\frac{(1-\partial_{\omega}\Sigma)E_{ns}}{(\omega-\Sigma^R-E_{ns})^2}\right]\right.
		+\left.2(1-\partial_{\omega}\Sigma)(n+1)\left[\frac{1}{\omega-\Sigma^R-E_{ns}}-\frac{1}{\omega-\Sigma^R-E_{n+2,s}}\right]\right\}\\
		=&-\frac{i\hbar}{8\pi^2l_B^2}\int d\omega [-f'(\omega)]\sum_{ns}(n+1)\left[\frac{E_{n+2,s}}{\omega-\Sigma^R-E_{n+2,s}}+\frac{E_{ns}}{\omega-\Sigma^R-E_{ns}}\right]\\
		&+\frac{i\hbar}{4\pi^2l_B^2}\int d\omega f(\omega)(1-\partial_{\omega}\Sigma)\sum_{ns}(n+1)^2\left[\frac{1}{\omega-\Sigma^R-E_{ns}}-\frac{1}{\omega-\Sigma^R-E_{n+2,s}}\right]\\
		=&-\frac{i\hbar}{8\pi^2l_B^2}\sum_{ns}(n+1)\left[\frac{E-\Sigma^R}{E-\Sigma^R-E_{n+2,s}}+\frac{E-\Sigma^R}{E-\Sigma^R-E_{ns}}-2\right]\\
		&+\frac{i\hbar}{4\pi^2l_B^2}\int d\omega f(\omega)(1-\partial_{\omega}\Sigma)\sum_{ns}(n+1)^2\left[\frac{1}{\omega-\Sigma^R-E_{ns}}-\frac{1}{\omega-\Sigma^R-E_{n+2,s}}\right]\\
		=&-\frac{i\hbar(E-\Sigma^R)}{8\pi^2l_B^2}\sum_{ns}(2n+\delta_{n,0})G_{ns}^R(E)+\frac{i}{\pi^2l_B^2}\sum_{ns}(n+1)
		%	&+\frac{i}{\pi^2l_B^2}\int d\omega f(\omega)(1-\partial_{\omega}\Sigma)\sum_{ns}(4n+\delta_{n,0})G_{ns}^R(\omega)\\
		+\frac{i\hbar}{4\pi^2l_B^2}\int d\omega f(\omega)(1-\partial_{\omega}\Sigma)\sum_{ns}(n+1)^2\left[G_{ns}^R(\omega)-G_{n+2,s}^R(\omega)\right]
\end{aligned}\end{equation}
where we have used
\begin{equation}
	\sum_{s'}\frac{1}{E_{ns'}-E_{n+2,s}}=\frac{-E_{n+2,s}}{(\hbar\omega_c)^2};\ \ \ \ \sum_{s}\frac{1}{E_{ns'}-E_{n+2,s}}=\frac{-E_{ns'}}{(\hbar\omega_c)^2};
\end{equation}
\begin{equation}
	\sum_{s'}\frac{1}{(E_{ns'}-E_{n+2,s})^2}=\sum_{s}\frac{1}{(E_{ns'}-E_{n+2,s})^2}=\frac{n+1}{(\hbar\omega_c)^2}
\end{equation}
We assume ${\rm Re}\Sigma\to0$, and the real part of $\eta_H^{II}$ is
\begin{equation}
	{\rm Re}\eta_H^{II}=\frac{\hbar}{8\pi^2l_B^2}\left\{E\sum_{ns}(2n+\delta_{n,0}){\rm Im}G_{ns}^R(E)-2\int d\omega f(\omega)(1-\partial_{\omega}\Sigma)\sum_{ns}(n+1)^2\left[{\rm Im}G_{ns}^R(\omega)-{\rm Im}G_{n+2,s}^R(\omega)\right]\right\}
\end{equation}
Since the broaden of the Landau level will strikingly affect the value of ${\rm Im}G_{ns}^R(\omega)-{\rm Im}G_{n+2,s}^R(\omega)$, the real part of $\eta^{II}_{H}$ is evaluated in the separated Landau levels region and overlapped Landau levels region, respectively. 

In separated Landau levels region,

\begin{equation}\begin{aligned}
		{\rm Re}\eta^{II}_{H}(E)\approx&\frac{\hbar E}{4\pi^2l_B^2}\sum_{ns}n{\rm Im}G_{ns}^R(E)+\frac{\hbar}{4\pi^2l_B^2}\int d\omega f(\omega)\sum_{ns}(n+1)^2\left[\delta(\omega-E_{ns})-\delta(\omega-E_{n+2,s})\right]\\
		=&\frac{\hbar}{4\pi^2l_B^2}\frac{E^3}{(\hbar\omega)^2}\sum_{ns}{\rm Im}G^R_{ns}(E)+{\rm sgn}(E)\frac{\hbar}{4\pi l^2_B}\left[\sum_{n=0}^N(n+1)^2-\sum_{n=0}^{N-1}(n+1)^2\right]\\
		=&-\frac{E^2\rho}{16\tilde{\omega}_c}+{\rm sgn}(E)\frac{\hbar}{4\pi l^2_B}(2N^2+2N+1)
\end{aligned}\end{equation}
where
\begin{equation}
	\sum_{ns}nG_{ns}=\frac{(E-\Sigma)^2}{(\hbar\omega_c)^2}\sum_{ns}G_{ns}-\frac{E-\Sigma}{(\hbar\omega_c)^2}(2N_c-1)
\end{equation}

In overlapped Landau levels region,
\begin{equation}\begin{aligned}
		{\rm Re}\eta^{II}_{H}(E)=&\frac{\hbar(E-\Sigma^R)}{4\pi^2l_B^2}\sum_{ns}n{\rm Im}G_{ns}^R(E)-\frac{\hbar}{4\pi^2l_B^2}\int d\omega f(\omega)\sum_{ns}(4n+\delta_{n,0}){\rm Im}G_{ns}^R(\omega)\\
		\approx&\frac{\hbar}{4\pi^2l_B^2}\frac{E^3}{(\hbar\omega)^2}\sum_{ns}{\rm Im}G^R_{ns}(E)-\frac{\hbar}{\pi^2l_B^2}\int d\omega f(\omega)\frac{\omega^2}{(\hbar\omega_c)^2}\sum_{ns}{\rm Im}G_{ns}^R(\omega)\\
		=&-\frac{\hbar}{2(\hbar\omega)^2}E^3\rho(E)+\frac{2\hbar}{(\hbar\omega_c)^2}\int_0^E d\omega\omega^2\rho(\omega)\\
\end{aligned}\end{equation}
Since $\rho(\omega)\propto{\rm Im}\Sigma(\omega)\propto \omega$ in this region, the above equation is approximately vanished,
\begin{equation}
	{\rm Re}\eta^{II}_{H}(E)\approx-\frac{\hbar}{2(\hbar\omega)^2}E^3\rho(E)+\frac{2\hbar}{(\hbar\omega_c)^2}\frac{E^3\rho(E)}{4}=0
\end{equation}

\newpage
\section{Derivation of dynamic shear viscosity $\eta_s(\Omega)$ in the presence $B=0$}\label{appendix-ac-shear-b0}
\begin{equation}\begin{aligned}
		{\rm Re}\eta_s^{LM}(\Omega)=&{\rm Re}\frac{\hbar^3v_f^2}{8\pi^2\Omega}\int^{E}_{E-\Omega}d\omega\int dkk^3\frac{2(\omega+\Omega-\Sigma^L_{\omega+\Omega})}{(\omega+\Omega-\Sigma^L_{\omega+\Omega})^2-(\hbar v_fk)^2}\frac{2(\omega-\Sigma^M_{\omega})}{(\omega-\Sigma^M_{\omega})^2-(\hbar v_fk)^2}\\
		=&{\rm Re}\frac{\hbar^3v_f^2}{2\pi^2\Omega}\int^{E}_{E-\Omega}\frac{(\omega+\Omega-\Sigma^L_{\omega+\Omega})(\omega-\Sigma^M_{\omega})}{(\omega+\Omega-\Sigma^L_{\omega+\Omega})^2-(\omega-\Sigma^M_{\omega})^2}d\omega\int dkk^3\frac{1}{(\omega-\Sigma^M_{\omega})^2-(\hbar v_fk)^2}-\frac{1}{(\omega+\Omega-\Sigma^L_{\omega+\Omega})^2-(\hbar v_fk)^2}\\
		=&{\rm Re}\frac{\hbar}{2\pi^2\Omega}\int^{E}_{E-\Omega}\frac{(\omega+\Omega-\Sigma^L_{\omega+\Omega})(\omega-\Sigma^M_{\omega})}{(\omega+\Omega-\Sigma^L_{\omega+\Omega})^2-(\omega-\Sigma^M_{\omega})^2}d\omega
		\int dkk\frac{(\omega-\Sigma^M_{\omega})^2}{(\omega-\Sigma^M_{\omega})^2-(\hbar v_fk)^2}-\frac{(\omega+\Omega-\Sigma^L_{\omega+\Omega})^2}{(\omega+\Omega-\Sigma^L_{\omega+\Omega})^2-(\hbar v_fk)^2}\\
		\stackrel{\text{SCBA}}{\approx}&{\rm Re}\frac{\hbar}{2\pi^2\Omega}\int^{E}_{E-\Omega}\frac{(\omega+\Omega-\Sigma^L_{\omega+\Omega})(\omega-\Sigma^M_{\omega})}{(\omega+\Omega-\Sigma^L_{\omega+\Omega})^2-(\omega-\Sigma^M_{\omega})^2}d\omega\frac{A}{2(\hbar v_f)^2}\left[(\omega-\Sigma^M_{\omega})\Sigma^M_{\omega}-(\omega+\Omega-\Sigma^L_{\omega+\Omega})\Sigma^L_{\omega+\Omega}\right]\\
		=&{\rm Re}\frac{A}{4\pi^2\hbar v_f^2\Omega}\int^{E}_{E-\Omega}d\omega\frac{(\omega+\Omega-\Sigma^L_{\omega+\Omega})(\omega-\Sigma^M_{\omega})}{(\omega+\Omega-\Sigma^L_{\omega+\Omega})^2-(\omega-\Sigma^M_{\omega})^2}\left[(\omega-\Sigma^M_{\omega})\Sigma^M_{\omega}-(\omega+\Omega-\Sigma^L_{\omega+\Omega})\Sigma^L_{\omega+\Omega}\right]\\
%		=&{\rm Re}\frac{A}{4\pi^2\hbar v_f^2\Omega}\int^{E}_{E-\Omega}d\omega\frac{1}{4}\frac{[(2\omega+\Omega-\Sigma^L_{\omega+\Omega}-\Sigma^M_{\omega})^2-(\Omega-\Sigma^L_{\omega+\Omega}+\Sigma^M_{\omega})^2]}{(2\omega+\Omega-\Sigma^L_{\omega+\Omega}-\Sigma^M_{\omega})(\Omega-\Sigma^L_{\omega+\Omega}+\Sigma^M_{\omega})}\\
%		&\left[(\omega-\Sigma^M_{\omega})\Sigma^M_{\omega}-(\omega+\Omega-\Sigma^L_{\omega+\Omega})\Sigma^L_{\omega+\Omega}\right]\\
		=&{\rm Re}\frac{A}{16\pi^2\hbar v_f^2\Omega}\int^{E}_{E-\Omega}d\omega\left[\frac{2\omega+\Omega-\Sigma^L_{\omega+\Omega}-\Sigma^M_{\omega}}{\Omega-\Sigma^L_{\omega+\Omega}+\Sigma^M_{\omega}}-\frac{\Omega-\Sigma^L_{\omega+\Omega}+\Sigma^M_{\omega}}{2\omega+\Omega-\Sigma^L_{\omega+\Omega}-\Sigma^M_{\omega}}\right]\\
		&\left[(\Sigma_{\omega+\Omega}^L+\Sigma_{\omega}^M)(\Sigma_{\omega+\Omega}^L-\Sigma_{\omega}^M)-\frac{\Omega}{2}(\Sigma_{\omega+\Omega}^L+\Sigma_{\omega}^M)-\frac{2\omega+\Omega}{2}(\Sigma_{\omega+\Omega}^L-\Sigma_{\omega}^M)\right]\\
		\stackrel{{\rm Re}\Sigma\to0}{\approx}&{\rm Re}\frac{A}{32\pi^2\hbar v_f^2\Omega}\int^{E}_{E-\Omega}d\omega\left(\frac{2\omega+\Omega+i\Gamma_+^{LM}}{\Omega+i\Gamma^{LM}_-}-\frac{\Omega+i\Gamma^{LM}_-}{2\omega+\Omega+i\Gamma_+^{LM}}\right)
		[-2\Gamma_+^{LM}\Gamma_-^{LM}+i((2\omega+\Omega)\Gamma_-^{LM}+\Omega\Gamma_+^{LM})]\\
		=&{\rm Re}\frac{A}{32\pi^2\hbar v_f^2\Omega}\int^{E}_{E-\Omega}d\omega\left\{-2\Gamma_+^{LM}\Gamma_-^{LM}+i[(2\omega+\Omega)\Gamma_-^{LM}+\Omega\Gamma_+^{LM}]\right\}\\
		&\left\{\frac{\Omega(2\omega+\Omega)+\Gamma_+^{LM}\Gamma_-^{LM}-i[(2\omega+\Omega)\Gamma_-^{LM}-\Omega\Gamma_+^{LM}]}{\Omega^2+(\Gamma^{LM}_-)^2}-\frac{\Omega(2\omega+\Omega)+\Gamma_+^{LM}\Gamma_-^{LM}+i[(2\omega+\Omega)\Gamma_-^{LM}-\Omega\Gamma_+^{LM}]}{(2\omega+\Omega)^2+(\Gamma^{LM}_+)^2}\right\}\\
		=&\frac{A}{32\pi^2\hbar v_f^2\Omega}\int^{E}_{E-\Omega}d\omega
		\left\{\frac{-2\Omega(2\omega+\Omega)\Gamma_+^{LM}\Gamma_-^{LM}-2(\Gamma_+^{LM})^2(\Gamma_-^{LM})^2+[(2\omega+\Omega)\Gamma_-^{LM}]^2-[\Omega\Gamma_+^{LM}]^2}{\Omega^2+(\Gamma^{LM}_-)^2}\right.\\
		&\left.-\frac{-2\Omega(2\omega+\Omega)\Gamma_+^{LM}\Gamma_-^{LM}-2(\Gamma_+^{LM})^2(\Gamma_-^{LM})^2-[(2\omega+\Omega)\Gamma_-^{LM}]^2+[\Omega\Gamma_+^{LM}]^2}{(2\omega+\Omega)^2+(\Gamma^{LM}_+)^2}\right\}\\
		=&\frac{A}{32\pi^2\hbar v_f^2\Omega}\int^{E}_{E-\Omega}d\omega
		\left\{\frac{[(2\omega+\Omega)\Gamma^{LM}_--\Omega\Gamma^{LM}_+]^2-2(\Gamma_+^{LM})^2[\Omega^2+(\Gamma^{LM}_-)^2]}{\Omega^2+(\Gamma^{LM}_-)^2}\right.\\
		&\left.-\frac{[(2\omega+\Omega)\Gamma^{LM}_--\Omega\Gamma^{LM}_+]^2-2(\Gamma_-^{LM})^2[(2\omega+\Omega)^2+(\Gamma^{LM}_+)^2]}{(2\omega+\Omega)^2+(\Gamma^{LM}_+)^2}\right\}\\
		=&\frac{A}{16\pi^2\hbar v_f^2\Omega}\int^{E}_{E-\Omega}d\omega
		\left\{(\Gamma_-^{LM})^2-(\Gamma_+^{LM})^2+\frac{[(2\omega+\Omega)\Gamma^{LM}_--\Omega\Gamma^{LM}_+]^2}{2}\left[\frac{1}{\Omega^2+(\Gamma^{LM}_-)^2}-\frac{1}{(2\omega+\Omega)^2+(\Gamma^{LM}_+)^2}\right]\right\}\\
\end{aligned}\end{equation}
where we have introduced $-i\Gamma^{LM}_+=\Sigma_{\omega+\Omega}^L+\Sigma_{\omega}^M$ and $-i\Gamma^{LM}_-=\Sigma_{\omega+\Omega}^L-\Sigma_{\omega}^M$.

At first, we consider the condition $E\approx0\ll\Omega$ so that $\omega<0<\omega+\Omega$. Combining the analytic solutions of self-energy, one can get
\begin{equation}\begin{aligned}
		-i\Gamma^{RA}_+=&\Sigma_{\omega+\Omega}^R+\Sigma_{\omega}^A=-i[\Gamma_0+\frac{\pi}{A}(\omega+\Omega)-\Gamma_0+\frac{\pi}{A}\omega]=-i\frac{\pi}{A}(2\omega+\Omega)\\
		-i\Gamma^{RA}_-=&\Sigma_{\omega+\Omega}^R-\Sigma_{\omega}^A=-i[\Gamma_0+\frac{\pi}{A}(\omega+\Omega)+\Gamma_0-\frac{\pi}{A}\omega]=-i(2\Gamma_0+\frac{\pi}{A}\Omega)\\
		-i\Gamma^{RR}_+=&\Sigma_{\omega+\Omega}^R+\Sigma_{\omega}^R=-i[\Gamma_0+\frac{\pi}{A}(\omega+\Omega)+\Gamma_0-\frac{\pi}{A}\omega]=-i(2\Gamma_0+\frac{\pi}{A}\Omega)\\
		-i\Gamma^{RR}_-=&\Sigma_{\omega+\Omega}^R-\Sigma_{\omega}^R=-i[\Gamma_0+\frac{\pi}{A}(\omega+\Omega)-\Gamma_0+\frac{\pi}{A}\omega]=-i\frac{\pi}{A}(2\omega+\Omega)
\end{aligned}\end{equation}
so that
\begin{equation}\begin{aligned}
		{\rm Re}\eta_s^{RA}(\Omega)=&\frac{A}{16\pi^2\hbar v_f^2\Omega}\int^{0}_{-\Omega}d\omega
		\left\{(\Gamma_-^{RA})^2-(\Gamma_+^{RA})^2+\frac{[(2\omega+\Omega)\Gamma^{RA}_--\Omega\Gamma^{RA}_+]^2}{2}\left[\frac{1}{\Omega^2+(\Gamma^{RA}_-)^2}-\frac{1}{(2\omega+\Omega)^2+(\Gamma^{RA}_+)^2}\right]\right\}\\
		=&\frac{A}{16\pi^2\hbar v_f^2\Omega}\int^{0}_{-\Omega}d\omega\left\{(2\Gamma_0+\frac{\pi}{A}\Omega)^2-\frac{\pi^2}{A^2}(2\omega+\Omega)^2+2\Gamma_0^2(2\omega+\Omega)^2\left[\frac{1}{\Omega^2+(2\Gamma_0+\frac{\pi}{A}\Omega)^2}-\frac{1}{(1+\frac{\pi^2}{A^2})(2\omega+\Omega)^2}\right]\right\}\\
		\stackrel{\Gamma_0\to0}{\approx}&\frac{1}{16\hbar v_f^2\Omega}\frac{1}{A}\int^{0}_{-\Omega}d\omega\left\{\Omega^2-(2\omega+\Omega)^2\right\}\\
		=&\frac{1}{16\hbar v_f^2}\frac{2\Omega^2}{3A}
\end{aligned}\end{equation}
and
\begin{equation}\begin{aligned}
		{\rm Re}\eta_s^{RR}(\Omega)\approx&\frac{A}{16\pi^2\hbar v_f^2\Omega}\int^{0}_{-\Omega}d\omega
		\left\{(\Gamma_-^{RR})^2-(\Gamma_+^{RR})^2+\frac{[(2\omega+\Omega)\Gamma^{RR}_--\Omega\Gamma^{RR}_+]^2}{2}\left[\frac{1}{\Omega^2}-\frac{\pi}{\Gamma^{RR}_+}\delta(2\omega+\Omega)\right]\right\}\\
		=&\frac{A}{16\pi^2\hbar v_f^2\Omega}\int^{0}_{-\Omega}d\omega
		\left\{\frac{\pi^2}{A^2}(2\omega+\Omega)^2-(2\Gamma_0+\frac{\pi}{A}\Omega)^2+\frac{[\frac{\pi}{A}(2\omega+\Omega)^2-\Omega(2\Gamma_0+\frac{\pi}{A}\Omega)]^2}{2\Omega^2}\right\}\\
		&-\frac{A}{16\pi^2\hbar v_f^2\Omega}\frac{\pi}{2}\Omega^2(2\Gamma_0+\frac{\pi}{A}\Omega)\\
		=&\frac{A}{16\pi^2\hbar v_f^2\Omega}\left[\frac{\pi^2}{A^2}\frac{\Omega^3}{3}-(2\Gamma_0+\frac{\pi}{A}\Omega)^2\Omega+\frac{\pi^2}{A^2}\frac{\Omega^3}{10}-\frac{\pi}{A}(2\Gamma_0+\frac{\pi}{A}\Omega)\frac{\Omega^2}{3}+(2\Gamma_0+\frac{\pi}{A}\Omega)^2\frac{\Omega}{2}-\frac{\pi}{2}\Omega^2(2\Gamma_0+\frac{\pi}{A}\Omega)\right]\\
		\stackrel{\Gamma_0\to0}{\approx}&\frac{\Omega^2}{16\hbar v_f^2}\left[-\frac{1}{2}-\frac{2}{5A}\right]
\end{aligned}\end{equation}
Thus the total real part of dynamic shear viscosity in the condition $E\approx0\ll\Omega$ is 
\begin{equation}
	\eta_s(\Omega)={\rm Re}\eta_s^{RA}(\Omega)-{\rm Re}\eta_s^{RR}(\Omega)=\frac{\Omega^2}{16\hbar v_f^2}(\frac{1}{2}+\frac{16}{15}\frac{1}{A})
\end{equation}

Then, we consider the condition $0<\Omega\ll E$ so that $0<\omega\lesssim\omega+\Omega$. Combining the analytic solutions of self-energy, one can get
\begin{equation}\begin{aligned}
		-i\Gamma^{RA}_+=&\Sigma_{\omega+\Omega}^R+\Sigma_{\omega}^A=-i[\Gamma_0+\frac{\pi}{A}(\omega+\Omega)-\Gamma_0-\frac{\pi}{A}\omega]=-i\frac{\pi}{A}\Omega\\
		-i\Gamma^{RA}_-=&\Sigma_{\omega+\Omega}^R-\Sigma_{\omega}^A=-i[\Gamma_0+\frac{\pi}{A}(\omega+\Omega)+\Gamma_0+\frac{\pi}{A}\omega]=-i[2\Gamma_0+\frac{\pi}{A}(2\omega+\Omega)]\\
		-i\Gamma^{RR}_+=&\Sigma_{\omega+\Omega}^R+\Sigma_{\omega}^R=-i[\Gamma_0+\frac{\pi}{A}(\omega+\Omega)+\Gamma_0+\frac{\pi}{A}\omega]=-i[2\Gamma_0+\frac{\pi}{A}(2\omega+\Omega)]\\
		-i\Gamma^{RR}_-=&\Sigma_{\omega+\Omega}^R-\Sigma_{\omega}^R=-i[\Gamma_0+\frac{\pi}{A}(\omega+\Omega)-\Gamma_0-\frac{\pi}{A}\omega]=-i\frac{\pi}{A}\Omega
\end{aligned}\end{equation}
so that
\begin{equation}\begin{aligned}
		{\rm Re}\eta_s^{RA}(\Omega)=&\frac{A}{16\pi^2\hbar v_f^2\Omega}\int^{E}_{E-\Omega}d\omega
		\left\{(\Gamma_-^{RA})^2-(\Gamma_+^{RA})^2+\frac{[(2\omega+\Omega)\Gamma^{RA}_--\Omega\Gamma^{RA}_+]^2}{2}\left[\frac{1}{\Omega^2+(\Gamma^{RA}_-)^2}-\frac{1}{(2\omega+\Omega)^2+(\Gamma^{RA}_+)^2}\right]\right\}\\
		=&\frac{A}{16\pi^2\hbar v_f^2\Omega}\int^{E}_{E-\Omega}d\omega
		\left\{[2\Gamma_0+\frac{\pi}{A}(2\omega+\Omega)]^2-\frac{\pi^2}{A^2}\Omega^2+\frac{[(2\omega+\Omega)[2\Gamma_0+\frac{\pi}{A}(2\omega+\Omega)]-\frac{\pi}{A}\Omega^2]^2}{2}\right.\\
		&\left.\left[\frac{1}{\Omega^2+[2\Gamma_0+\frac{\pi}{A}(2\omega+\Omega)]^2}-\frac{1}{(2\omega+\Omega)^2+\frac{\pi^2}{A^2}\Omega^2}\right]\right\}\\
		\approx&\frac{A}{16\pi^2\hbar v_f^2}
		\left\{[2\Gamma_0+\frac{\pi}{A}(2E+\Omega)]^2-\frac{\pi^2}{A^2}\Omega^2+\frac{[(2E+\Omega)[2\Gamma_0+\frac{\pi}{A}(2E+\Omega)]-\frac{\pi}{A}\Omega^2]^2}{2}\right.\\
		&\left.\left[\frac{1}{\Omega^2+[2\Gamma_0+\frac{\pi}{A}(2E+\Omega)]^2}-\frac{1}{(2E+\Omega)^2+\frac{\pi^2}{A^2}\Omega^2}\right]\right\}\\
		\approx&\frac{A}{16\pi^2\hbar v_f^2}
		\left(\frac{\pi^2}{A^2}4E^2+\frac{8E^4}{\frac{A^2}{\pi^2}\Omega^2+4E^2}\right)\\
\end{aligned}\end{equation}
and
\begin{equation}\begin{aligned}
		{\rm Re}\eta_s^{RR}(\Omega)=&\frac{A}{16\pi^2\hbar v_f^2\Omega}\int^{E}_{E-\Omega}d\omega
		\left\{(\Gamma_-^{RR})^2-(\Gamma_+^{RR})^2+\frac{[(2\omega+\Omega)\Gamma^{RR}_--\Omega\Gamma^{RR}_+]^2}{2}\left[\frac{1}{\Omega^2+(\Gamma^{RR}_-)^2}-\frac{1}{(2\omega+\Omega)^2+(\Gamma^{RR}_+)^2}\right]\right\}\\
		=&\frac{A}{16\pi^2\hbar v_f^2\Omega}\int^{E}_{E-\Omega}d\omega
		\left\{\frac{\pi^2}{A^2}\Omega^2-[2\Gamma_0+\frac{\pi}{A}(2\omega+\Omega)]^2+\frac{[(2\omega+\Omega)\frac{\pi}{A}\Omega-\Omega[2\Gamma_0+\frac{\pi}{A}(2\omega+\Omega)]]^2}{2}\right.\\
		&\left.\left[\frac{1}{\Omega^2+\frac{\pi^2}{A^2}\Omega^2}-\frac{1}{(2\omega+\Omega)^2+[2\Gamma_0+\frac{\pi}{A}(2\omega+\Omega)]^2}\right]\right\}\\
		\approx&\frac{A}{16\pi^2\hbar v_f^2}
		\left\{\frac{\pi^2}{A^2}\Omega^2-[2\Gamma_0+\frac{\pi}{A}(2E+\Omega)]^2+\frac{[(2E+\Omega)\frac{\pi}{A}\Omega-\Omega[2\Gamma_0+\frac{\pi}{A}(2E+\Omega)]]^2}{2}\right.\\
		&\left.\left[\frac{1}{\Omega^2+\frac{\pi^2}{A^2}\Omega^2}-\frac{1}{(2E+\Omega)^2+[2\Gamma_0+\frac{\pi}{A}(2E+\Omega)]^2}\right]\right\}\\
		\approx&-\frac{A}{16\pi^2\hbar v_f^2}\frac{\pi^2}{A^2}4E^2\\
\end{aligned}\end{equation}
where we have used on shell assumption $\int^E_{E-\Omega}d\omega=\Omega\int\delta(\omega-E)d\omega$. Thus the total real part of dynamic shear viscosity in the condition $0<\Omega\ll E$ is 
\begin{equation}
	\eta_s(\Omega)={\rm Re}\eta_s^{RA}(\Omega)-{\rm Re}\eta_s^{RR}(\Omega)=\frac{AE^2}{2\pi^2\hbar v_f^2}
	\left(\frac{\pi^2}{A^2}+\frac{E^2}{\frac{A^2}{\pi^2}\Omega^2+4E^2}\right)
\end{equation}

%\end{appendix}

%%%%%%%%%%%%%%%%%%%%%%%%%%%%%%%%%%%%%%%%%%%%%